\documentclass[a4paper,11pt]{article}
\pdfoutput=1 % if your are submitting a pdflatex (i.e. if you have
\usepackage{jheppub} % for details on the use of the package, please
\usepackage[T1]{fontenc} % if needed
\usepackage{ulem}
\usepackage[dvipsnames]{xcolor}
\usepackage{booktabs}
\usepackage{multicol}
\usepackage{multirow}
\usepackage{cancel}
\usepackage{pifont}
\usepackage{tikz}
\usepackage{comment}
\usepackage{subcaption}

\makeatletter
\newcommand\makebig[2]{%
\@xp\newcommand\@xp*\csname#1\endcsname{\bBigg@{#2}}%
\@xp\newcommand\@xp*\csname#1l\endcsname{\@xp\mathopen\csname#1\endcsname}%
\@xp\newcommand\@xp*\csname#1r\endcsname{\@xp\mathclose\csname#1\endcsname}%
}
\makeatother
\makebig{Bigggg}{4.5}

\usetikzlibrary{decorations.pathmorphing}
\usetikzlibrary{decorations.markings}

\title{\boldmath Walls, bubbles and doom - the cosmology of HEFT}

\author[a]{R. Alonso,}
\author[b]{J. C. Criado,}
\author[c]{R. Houtz}
\author[a]{and M. West}

\affiliation[a]{Institute for Particle Physics Phenomenology, Durham University, \\Durham, DH1 3LE, United Kingdom }

\affiliation[b]{Departamento de Física Teórica y del Cosmos, Universidad de Granada, \\E–18071 Granada, Spain}

\affiliation[c]{Department of Physics, University of Florida, Gainesville, FL 32611, USA}

\abstract{As experiment charts new territory at the electroweak scale, the enterprise to characterise all possible theories becomes all the more necessary. In the absence of new particles, this ambitious enterprise is attainable and has led to the Higgs Effective Field Theory (HEFT) as the most general characterising framework, containing the Standard Model Effective Field Theory (SMEFT) as a subspace.
The characterisation of this theory space led to the dichotomy SMEFT vs. HEFT$\backslash$SMEFT as the two possible realisations of symmetry breaking. The criterion to distinguish these two possibilities is non-local in field space, and phenomena which explore field space beyond the neighbourhood of the vacuum manifold are in a singular position to tell them apart. Cosmology allows for such phenomena, and this work focuses on HEFT$\backslash$SMEFT, the less explored of the two options, to find that first order phase transitions with detectable gravitational wave remnants, domain wall formation and vacuum decay in the far, far 
distant future can take place and single out HEFT$\backslash$SMEFT. Results in cosmology are put against LHC constraints, and the potential of future ground- and space-based experiments to cover parameter space is discussed.}

\begin{document}
\maketitle
\flushbottom

\clearpage

%%%%%%%%%%%%%%%%%%%%%%%%%%%%%%%%%%%%%%%%%%%%%%%%
\section{Introduction}
\label{sec:intro}

The direct experimental exploration of the electroweak (EW) scale is in full swing at the LHC, and we are closing in on the answers to questions historically central to particle physics: the hierarchy problem, the mechanism for electroweak symmetry breaking, and mass generation. Looking out into the cosmos, on the other hand, has proven to be a source of invaluable input to our theories of Nature, the latest potential window into early cosmology---possibly including electroweak transitions---being gravitational waves.

At the front line of this exploration is the Higgs scalar and its properties. Given the absence of particles beyond the Standard Model (SM) spectrum thus far, the formalism of Effective Field Theories (EFTs) provides a general, model independent framework to characterise the Higgs particle. This formalism presents a dichotomy in that the electroweak EFT might or might not admit a linear representation. These two options are dubbed SMEFT and HEFT$\backslash$SMEFT where backslash is the mathematical symbol for the difference of sets. The HEFT is the most general gauge and Lorentz invariant EFT we can write, and thus it includes both options. However, the literature often refers to the HEFT$\backslash$SMEFT simply as HEFT.
Here, we use $
\textrm{non-linear theory space}\equiv \textrm{HEFT$\backslash$SMEFT}
$ to make it clear that we are referring to EFT's that do not admit a linear representation.

\begin{figure}[h!]
    \centering
    \includegraphics[width=.5\textwidth]{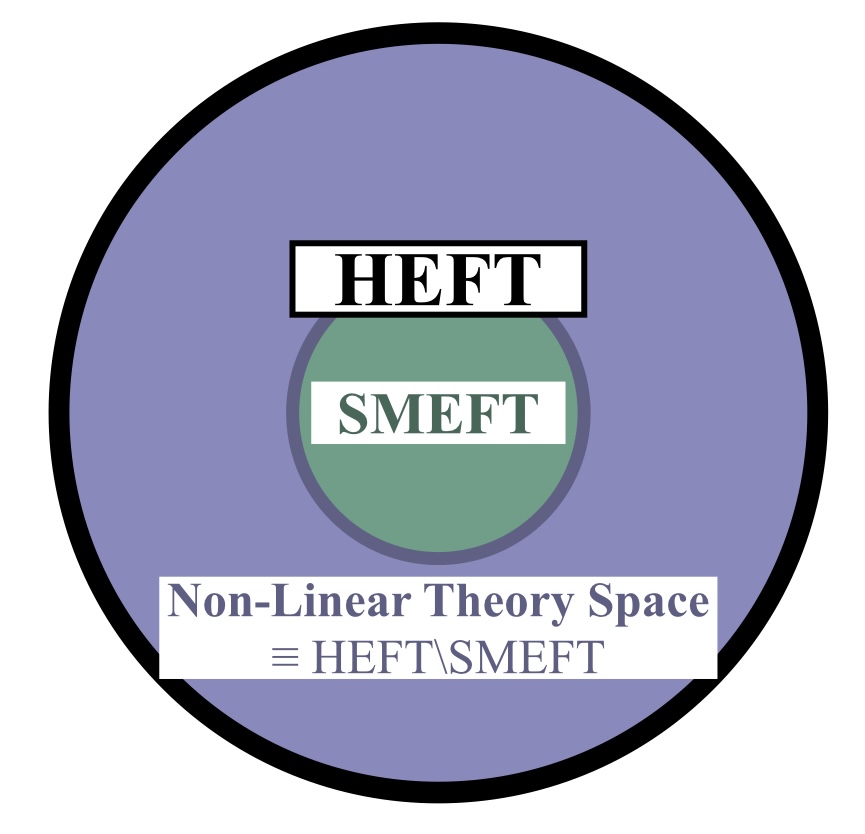}
    \caption{Visual depiction of the space of HEFT theories. There is some evidence for the boundary beetween non-linear and SMEFT theories belonging to the former, see e.g. Ref.~\cite{Alonso:2023upf}.}
    \label{fig:HEFTSpace}
\end{figure}

The characterisation of these theories has been laid out~\cite{Alonso:2015fsp,Alonso:2016oah,Cohen:2020xca}, examples of UV completions and general results for certain classes have been derived \cite{Banta:2021dek,Cohen:2021ucp,Gomez-Ambrosio:2022qsi,Gomez-Ambrosio:2022why,Finn:2019aip,Alonso:2021rac} and new features continue to emerge and be explored \cite{Cheung:2021yog,Cheung:2022vnd,Salas-Bernardez:2022hqv,Cohen:2022uuw,Alonso:2022ffe,Helset:2022pde,Helset:2022tlf,Assi:2023zid,Alminawi:2023qtf,Graf:2022rco,Sun:2022aag}. A pertinent remark is that the SMEFT vs HEFT$\backslash$SMEFT characterisation is an IR one, and there are known cases in which the UV description that completes a non-linear HEFT$\backslash$SMEFT theory presents linear scalar representations; what sets apart such linear UV theories is that they are non-decoupling with an upper bound on the mass of new states of $\sim4\pi v$. If such a contrast of UV and IR descriptions seems counter-intuitive, an example in nature is QCD and its natural low energy description using the chiral Lagrangian. Fully comprehensive characterisations are nonetheless elusive, in essence because the question is non-local in field space. Scattering experiments can only probe our theory around the vacuum to higher terms in our Taylor expansion of the Lagrangian in fields, incrementally improving our knowledge of the theory with more particles involved in the scattering process, yet still inherently local.
It is here that cosmology offers a global view of our theory 
and the possibility of testing it non-locally via phenomena such as phase transitions or topological defect formation. It is the aim of this work to examine the window that cosmology opens on HEFT$\backslash$SMEFT theories, especially those hardest to distinguish locally from the SM, and chart its complementarity with LHC data. 

The elementary result for EFT characterisation is the presence or absence of a point in scalar field space which is invariant, i.e. stays fixed, under the gauge group action. A useful visualisation in a two-dimensional field space with rotation as the gauge group has the fixed point at the origin. The extension to the electroweak theory led to naming it an $O(4)$ fixed point with $SU(2)_L\times U(1)_Y\subset O(4)$. If this fixed point is present, it is possible to cast the EFT as a SMEFT, and if absent the EFT is non-linear. This absence can arise in two broad forms: 
\begin{itemize}
    \item  type A, theories entirely without a fixed point,
    \item type B, theories with a singularity at the would-be fixed point in scalar space.
\end{itemize}
This distinction with SMEFT and within non-linear theories is one about the possible realisation of the symmetry, and therefore one can expect the electroweak phase transition epoch of the universe to shed light on the question---we do not consider here low energy inflation or other scenarios that bypass a universe at $T_{ew}$. The cosmology of ultraviolet completions giving rise to theories with a singularity, i.e. type B, has in fact been studied already in Ref.~\cite{Banta:2022rwg} (see also~\cite{Kanemura:2022txx}); this work focuses on type A. For a discussion of possible UV completions of type A theories see \cite{Alonso:2023upf}.

Non-perturbative dynamics depend generically on extended field configurations which hence probe the theory globally in field space. As such, these would be sensitive even to theories that locally resemble the SM. One of the prominent, phenomenologically relevant examples of such non-pertubative dynamics is B- and L-number violating sphaleron processes. The sphaleron energy is given by field configurations which will be modified in non-linear theories, and Sec.~\ref{sec:Sph} explores this modification.

\begin{figure}[t]
    \centering

    \begin{subfigure}[b]{0.425\textwidth}
         \centering
         \includegraphics[width=\textwidth]{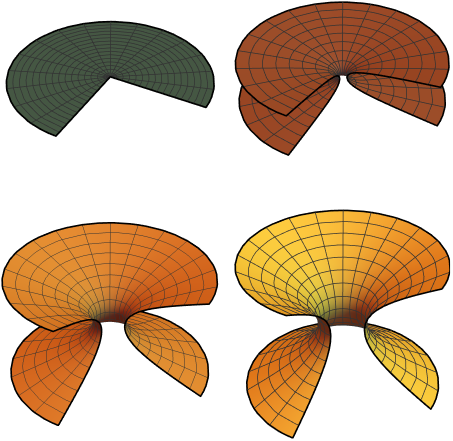}
         \caption{}
        \label{fig:Manifolds}
     \end{subfigure}
     \hfill
     \begin{subfigure}[b]{0.525\textwidth}
         \centering
         \includegraphics[width=\textwidth]{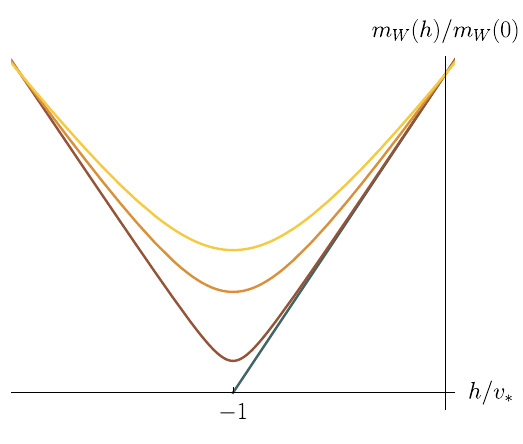}
         \caption{}
        \label{fig:WEffectiveMass}
     \end{subfigure}
    \caption{In Fig.~\ref{fig:Manifolds}, from top to bottom and left to right, 2D representations of the 4D scalar manifold for the SM case (top left) and type A theories with the metric of Eqs.~(\ref{LagFY}, \ref{vst}) and $\sin^2\chi=0.01$, 0.1, and 0.2. The Goldstone bosons correspond to the angle around the $z$ axis, while the Higgs $h$ parametrises the surface in the orthogonal direction to them. Fig.~\ref{fig:WEffectiveMass} shows $M_W(h)/M_W(0)$ for the SM and the type A theories in Fig.~\ref{fig:Manifolds} with the same colour coding.}
    \label{FigX}
    
\end{figure}

 The absence of a fixed point in type A theories requires, in turn,  a revision of the behaviour of the theory at high temperature. Indeed strictly speaking, electroweak symmetry cannot be restored at high temperature since there is not a point in the manifold for such restoration. One can still talk about discrete symmetry restoration nonetheless; a type A theory does not present a fixed point where the electroweak gauge boson masses would vanish, but it might contain a point where they reach their {\it minimum}. Let us suppose this point is a field distance $v_\star$ away from our current vacuum. One can then define reflection around this point: if $h$ is the Higgs fluctuation around the vacuum, let us define $\phi=h+v_\star$, and the reflection as $\phi\to-\phi$. The case of SMEFT has no such possible symmetry: $v_\star$ can be taken instead to be the distance to the fixed point, and by definition $\phi$, a radius, is always positive. The geometric approach helps visualise such theory. The possibility of extending a `radius' to negative values leads to a wormhole-like structure as shown in Fig.~\ref{fig:Manifolds}.

\begin{figure}[t]
\begin{subfigure}[b]{.25\textwidth}
 \begin{tikzpicture}
    \draw [->] (0,-1) --(0,1);
    \draw [->] (3,-1) --(3,1);
    \draw [thick] (-0.25,-1)  --(3.25,-1);
    \draw [thick,blue] (-0.25,-1) node[anchor=east] {$h_\star$} -- (0.5,-1);
    \draw [thick, ForestGreen] (2.5,-1) -- (3.25,-1) node[anchor=west] {$h_+$} ;
    \draw [thick, blue] (2.5,0.6) -- (3.25,0.6) node[anchor=west] {$h_0$} ;
    \filldraw [ultra thick,opacity=0.4] (0.5,-1) -- (2.5,-1) --(2.5,0.75) -- (0.5,0.75);
    \draw  (4.25,-0.2) node {\Bigggg\{} ;
    \draw [purple] (1.25,1) -- (1.75,1) node[anchor=south] {$t$};
    \draw [->,purple] (1.75,1) -- (2.25,1);
\end{tikzpicture}
\caption{}
\end{subfigure}
\hspace{1.25cm}
\begin{subfigure}[b]{.25\textwidth}
\begin{tikzpicture}
    \draw [->] (0,-1) --(0,1);
    \draw [->] (3,-1) --(3,1);
    \draw [thick] (-0.25,-1)  --(3.25,-1);
    \draw [thick,blue] (-0.25,-1) node[anchor=east] {$h_\star$} -- (1,-1);
    \draw [thick, ForestGreen] (1,-1) -- (3.25,-1) node[anchor=west] {$h_+$} ;
    \draw [thick, blue] (1,-1) .. controls (1.5,0.2) and (2.5,0.5)  ..(3.25,0.6) node[anchor=west] {$h_0$} ;
\end{tikzpicture}
\caption{}
\label{fig:breakno-split}
\end{subfigure}
\hspace{1cm}
\begin{subfigure}[b]{.25\textwidth}
\begin{tikzpicture}
    \draw [->] (0,-1) --(0,1);
    \draw [->] (3,-1) --(3,1);
    \draw [thick] (-0.25,-1)  --(3.25,-1);
    \draw [thick,blue] (-0.25,-1) node[anchor=east] {$h_\star$} -- (1.5,-1);
    \draw [thick, ForestGreen] (1.5,-1) -- (3.25,-1) node[anchor=west] {$h_+$} ;
    \draw [thick, blue] (1,0) .. controls (1.1,0.5) and (2.5,0.55)  ..(3.25,0.6) node[anchor=west] {$h_0$} ;
    \draw [thick, ForestGreen](1,0) .. controls (1.1,-0.5) and (1.4,-0.95)  ..(1.5,-1);
\end{tikzpicture}
\caption{}
\label{fig:breakno-form}
\end{subfigure}
 \caption{The possible extrema histories in SMEFT, minima in blue and maxima in green.}
    \label{fig:breakno}
\end{figure}
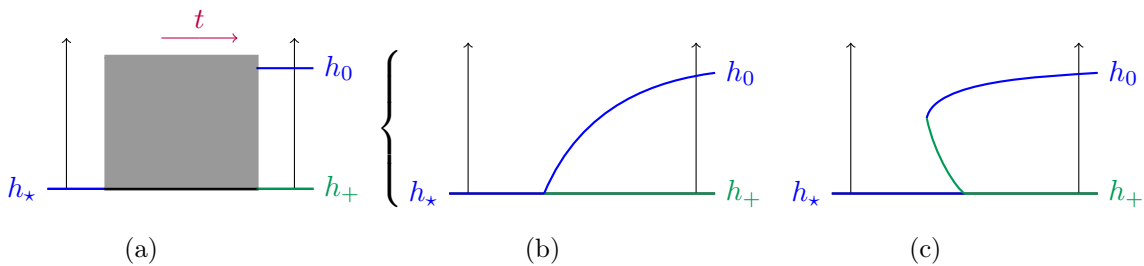

Even with this discrete parity introduction, one might not have symmetry restoration at high temperature. In fact, given the variety of possibilities for minima in these theories, it is convenient to consider the history of extrema of the finite temperature potential in this extended range for the Higgs field including `negative' values. We represent such histories as diagrams where time flows from left to right and the Higgs field value increases upwards, with lines showing the evolution, emergence and disappearance of extrema of the effective potential, see for example Fig.~\ref{fig:breakno}.
For reference, the characterisation of possibilities for the extrema history in the case of the Standard Model with variable  Higgs mass are well known but still useful to cast in this diagrammatic approach. One can have a transition from the symmetry-restored high temperature phase --with a potential of a single minimum at $h_\star$-- to the broken low temperature phase -- with a minimum at $h_0$ and a maximum $h_+$ sitting where $h_\star$ used to be-- via either (i) the splitting of the original minimum into a maximum and minimum (Fig.~\ref{fig:breakno-split}), or (ii) the formation of a maximum-minimum pair at a finite distance and with a barrier between  the new minimum and the symmetric minimum (Fig.~\ref{fig:breakno-form}). 
The latter gives rise to a first order phase transition while the former contains the case of the SM with its measured couplings, yielding a cross-over phase transition. The extension to SMEFT allows for changing the EW transition to possibility (ii), see e.g. Ref.~\cite{Chala:2018ari}, but still not changing the qualitative picture of possibilities as drawn above.

\begin{figure}[h]
    \centering
\begin{subfigure}[b]{.32\textwidth}
 \begin{tikzpicture}
    \draw [<->] (0,-1.5) --(0,1.5);
    \draw [<->] (3,-1.5) --(3,1.5);
    \draw [thick,blue] (-0.25,0) node[anchor=east] {$h_\star$} -- (0.5,0);
    \draw [thick, blue] (2.5,-1) -- (3.25,-1) node[anchor=west] {$h_-$} ;
    \draw [thick, ForestGreen] (2.5,0) -- (3.25,0) node[anchor=west] {$h_+$} ;
    \draw [thick, blue] (2.5,1) -- (3.25,1) node[anchor=west] {$h_0$} ;
    \filldraw [ultra thick,opacity=0.4] (0.5,-1.25) -- (2.5,-1.25) --(2.5, 1.25) -- (0.5,1.25);
    \draw  (4.15,0) node {\Bigggg\{} ;
\end{tikzpicture}
\caption{}
\end{subfigure}
\hspace{0cm}
\begin{subfigure}[b]{.32\textwidth}
 \begin{tikzpicture}
    \draw [<->] (0,-1.5) --(0,1.5);
    \draw [<->] (3,-1.5) --(3,1.5);
    \draw [thick,blue] (-0.25,0) node[anchor=east] {$h_\star$} -- (1,0);
    \draw [thick, blue] (1,0) ..controls (1.2,-0.7) and (2,-0.8) ..(3.25,-1)  node[anchor=west] {$h_-$} ;
    \draw [thick, ForestGreen] (1,0) -- (3.25,0) node[anchor=west] {$h_+$} ;
    \draw [thick, blue] (1,0) ..controls (1.2, 0.7) and (2 ,0.8) .. (3.25,1) node[anchor=west] {$h_0$} ;
\end{tikzpicture}
\caption{History P}
    \label{fig:breakit1to3-P}
\end{subfigure}
\hspace{0cm}
\begin{subfigure}[b]{.32\textwidth}
 \begin{tikzpicture}
    \draw [<->] (0,-1.5) --(0,1.5);
    \draw [<->] (3,-1.5) --(3,1.5);
    \draw [thick,blue] (-0.25,0) node[anchor=east] {$h_\star$} ..controls (0.5,-0.1) and (2,-0.8) .. (3.25,-1) node[anchor=west] {$h_-$} ;
    \draw [thick, ForestGreen] (1.5,0.5) ..controls (1.6,0.3) and (2,0.1) ..(3.25,0) node[anchor=west] {$h_+$} ;
    \draw [thick, blue] (1.5,0.5) ..controls (1.6,0.7) and (2,0.9) .. (3.25,1) node[anchor=west] {$h_0$} ;
 \end{tikzpicture}
\caption{History Q$_-$}
    \label{fig:breakit1to3-Qm}
\end{subfigure}
       \caption{Extrema histories we encounter in non-linear theories with one extrema at high temperature and three at low temperature.}
    \label{fig:breakit1to3}
\end{figure}
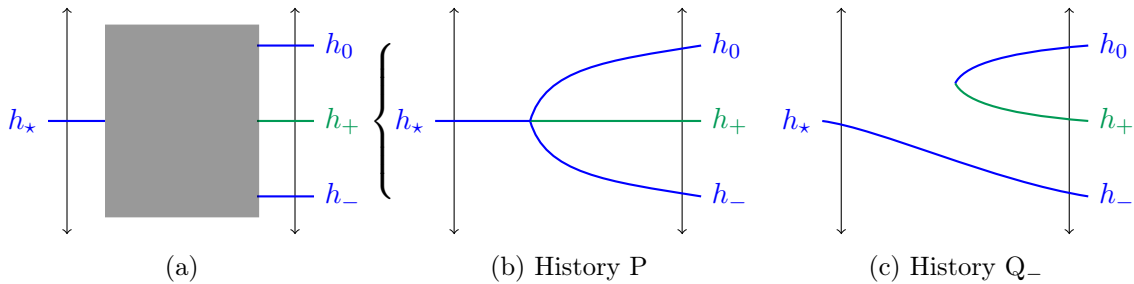

\begin{figure}
    \centering
    \begin{subfigure}[b]{.32\textwidth}
     \begin{tikzpicture}
    \draw [<->] (0,-1) --(0,1.5);
    \draw [<->] (3,-1) --(3,1.5);
    \draw [thick,blue] (-0.25,-0.5) node[anchor=east] {$h_\star$} -- (0.5,-0.5);
    \draw [thick, blue] (2.5,1) -- (3.25,1) node[anchor=west] {$h_0$} ;
    \filldraw [ultra thick,opacity=0.4] (0.5,-0.75) -- (2.5,-.75) --(2.5, 1.25) -- (0.5,1.25);
    \draw  (4,0.2) node {\Bigggg\{} ;
\end{tikzpicture}
    \caption{}
    \end{subfigure}
    \begin{subfigure}[b]{.32\textwidth}
     \begin{tikzpicture}
    \draw [<->] (0,-1) --(0,1.5);
    \draw [<->] (3,-1) --(3,1.5);
    \draw [thick,blue] (-0.25,-0.5) node[anchor=east] {$h_\star$} .. controls (1.5,-0.2) and (1.8,0.7)..(3.25,1) node[anchor=west] {$h_0$} ;
\end{tikzpicture}
    \caption{}
    \label{fig:breakit1to1-nomax}
    \end{subfigure}
    \begin{subfigure}[b]{.32\textwidth}
     \begin{tikzpicture}
    \draw [<->] (0,-1) --(0,1.5);
    \draw [<->] (3,-1) --(3,1.5);
    \draw [thick,blue] (-0.25,-0.5) node[anchor=east] {$h_\star$} .. controls (1,-0.49) and (1.45,-0.3)..(1.5,-0);
    \draw [thick, ForestGreen] (1,0.6) .. controls (1.05,0.3) and (1.4,0.2)..  (1.5,0);
    \draw [thick,blue] (1,0.6) ..controls (1.04,0.8) and (2,0.95).. (3.25,1)node[anchor=west] {$h_0$} ;
\end{tikzpicture}
    \caption{History R}
    \label{fig:breakit1to1-R}
    \end{subfigure}
    \caption{Possible extrema histories for a single minimum at high and low temperature}
    \label{fig:breakit1to1}
\end{figure}
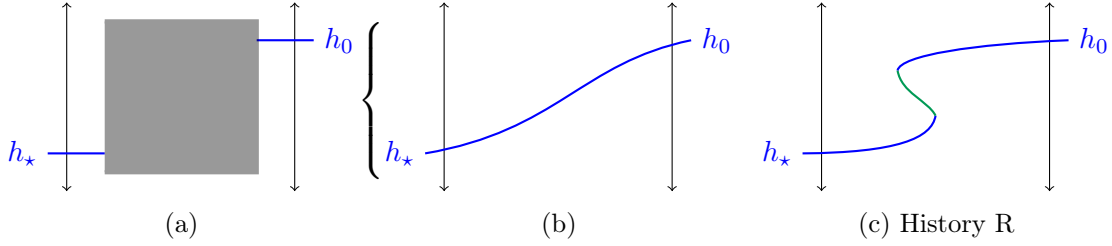

\begin{figure}[h]
    \centering  
\begin{subfigure}[b]{.32\textwidth}
 \begin{tikzpicture}
    \draw [<->] (0,-1.5) --(0,1.5);
    \draw [<->] (3,-1.5) --(3,1.5);
    \draw [thick,blue] (-0.25,0.75) node[anchor=east] {$h_\star$} -- (0.5,0.75);
    \draw [thick,ForestGreen] (-0.25,0) node[anchor=east] {$h_\star^+$} -- (0.5,0);
    \draw [thick,blue] (-0.25,-0.75) node[anchor=east] {$h_\star^-$} -- (0.5,-0.75);
    \draw [thick, blue] (2.5,-1) -- (3.25,-1) node[anchor=west] {$h_-$} ;
    \draw [thick, ForestGreen] (2.5,0) -- (3.25,0) node[anchor=west] {$h_+$} ;
    \draw [thick, blue] (2.5,1) -- (3.25,1) node[anchor=west] {$h_0$} ;
    \filldraw [ultra thick,opacity=0.4] (0.5,-1.25) -- (2.5,-1.25) --(2.5, 1.25) -- (0.5,1.25);
\end{tikzpicture}
    \caption{}
    \end{subfigure}
\hspace{1cm}
\begin{subfigure}[b]{.32\textwidth}
 \begin{tikzpicture}
    \draw [<->] (0,-1.5) --(0,1.5);
    \draw [<->] (3,-1.5) --(3,1.5);
    \draw [thick,blue] (-0.25,0.75) node[anchor=east] {$h_\star$} -- (0.5,0.75);
    \draw [thick,ForestGreen] (-0.25,0) node[anchor=east] {$h_\star^+$} -- (0.5,0);
    \draw [thick,blue] (-0.25,-0.75) node[anchor=east] {$h_\star^-$} -- (0.5,-0.75);
    \draw [thick, blue] (2.5,0.4) -- (3.25,0.4) node[anchor=west] {$h_0$} ;
    \filldraw [ultra thick,opacity=0.4] (0.5,-1.25) -- (2.5,-1.25) --(2.5, 1.25) -- (0.5,1.25);
\end{tikzpicture}
    \caption{}
    \end{subfigure}
\caption{Initial and  $T=0$ extrema we find as realisations in non-linear theories without high-temperature symmetry restoration.}
 \label{fig:breakitothers}
\end{figure}
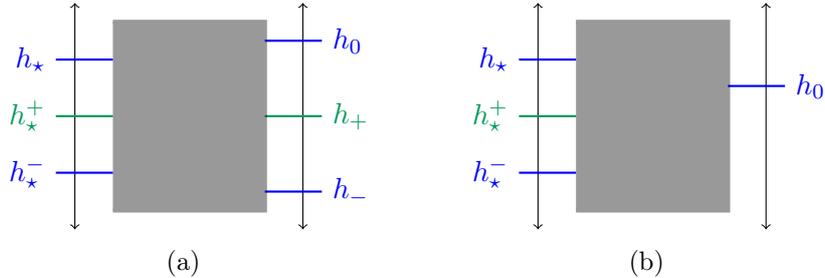

In non-linear theories by contrast, the range for $h$ need not have bounds, and the symmetric minimum might lose its natural extrema status (see Ref.~\cite{Alonso:2013nca}) but further, new minima arise from small deviations to SM couplings. The extension of the Higgs manifold shown in Fig.~\ref{FigX} is symmetric under the discrete reflection ($\phi\to-\phi$), which introduces a doubling of the minima, with a new one arising on the other side of the wormhole. 
This symmetric limit with Higgs parity restoration at high temperature -- i.e. a single minimum $h_\star$-- is shown in Fig.~\ref{fig:breakit1to3-P}, history P, and without parity restoration in Fig.~\ref{fig:breakit1to1}. For no high temperature symmetry restoration we can find high and low temperature configurations as in Fig.~\ref{fig:breakitothers}.

All the diagrams shown in Figs.~\ref{fig:breakit1to3}--\ref{fig:breakitothers}, are selected not by an artistic whim but rather because they do occur in the theories here considered. Generically one needs the further input of potential difference between extrema or barrier height to determine the phenomenology, but certain diagrams do however necessarily lead to processes markedly distinct from the SM. An example is
first instance in
 Fig.~\ref{fig:breakit1to1-nomax} with a single minimum for all history and hence no phase transition; while this in itself is a qualitative difference from the SM case, it is not one easily testable. History R in Fig.~\ref{fig:breakit1to1-R} by contrast does necessarily lead to a phase transition before the original minimum meets a sudden end encountering a maximum. Histories P,Q and R---Fig.~\ref{fig:breakit1to3} shows Q$_-$, while Q$_0$ is the parity-reflected case---will be the main cases of this study since they leave potentially observable traces or are partially ruled out.

In Figs.~\ref{fig:breakit1to3}--\ref{fig:breakitothers} we have also introduced  notation to label extrema; we will use $h_0$ for the minimum of the potential we find ourselves in today, $h_{+}$ and $h_-$ for possible maximum and minimum in the $T=0$ potential.
Note that our definition of $h_0$ as our vacuum today, --around which we have e.g. measured the mass of the Higgs to be $125$~GeV-- while conventional, does mean that any evolution of the universe that leads to the vacuum at $h_-$ today is discarded, without loss of generality.

%%%%%%%%%%%%%%%%%%%%%%%%%%%%%%%%%%%%%%%%%%%%%%%%%%%%%%%%%%%%%%%%%
  \section{Classical action}\label{sec:Classical}
  The tree level Lagrangian for the electroweak sector we use here to model non-linear theories reads
   \begin{align}\label{LagFY}
	\mathcal L&= \frac{1}2\partial h^2+\frac{F^2(h)}{2}v^2\frac{1}{2}\mbox{Tr}\left[D_\mu U D^\mu U^\dagger\right]-V(h)-\left[\frac{F_\psi(h)v}{\sqrt2}\bar\psi_L Y U \psi_R+\text{h.c.}\right]\\
 &\equiv\frac12 d_\mu\Phi^i G_{ij} d^\mu\Phi^j -V(h)-\left[\frac{F_\psi(h)v}{\sqrt2}\bar\psi_L Y U \psi_R+\text{h.c.}\right],
\end{align}
where $F(0)=F_\psi(0)=1$ and $v=246$~GeV. Here, $U$ is a special unitary $2\times 2$ matrix parameterised by 3 scalar degrees of freedom $\varphi^a$, the Nambu-Goldstone bosons, which, together with the Higgs degree of freedom $h$, span a four-dimensional manifold with coordinates $\Phi^i=(\varphi^a,h)$. We have used the latin letters $i,j,k$ for the components of $\Phi$, running from 1 to 4, and $a,b,c$ for Nambu-Goldstone bosons running from 1 to 3. The gauged covariant derivative $d_\mu\Phi=\partial_\mu \Phi+t_C A_\mu^C$, with electroweak bosons $A_\mu^C=\{W_\mu^I,B^\mu\}$, where $I=1,2,3$ and $t_C$ are the killing vectors~\cite{Alonso:2016oah} defined as, together with the metric
\begin{align}
\tilde g_{ab}=&\frac{v^2}{2}\mbox{Tr}\left(\frac{\partial U}{\partial\varphi^a} \frac{\partial U^\dagger}{\partial\varphi^b} \right)=v^2\frac{\partial u^T}{\partial\varphi^a}\frac{\partial u}{\partial\varphi^b}, & G_{ij}&=\left(\begin{array}{cc}
     F^2\tilde g_{ab}&  \\
     & 1
\end{array}\right),\\
    \frac{t_I^a}{v^2} &=\frac{ig\tilde g^{ab}}{4}\mbox{Tr}\left(U\frac{\partial U^\dagger}{\partial \varphi^b} \sigma_I\right) & \frac{t_Y^a}{v^2}&=\frac{ig_Y\tilde g^{ab}}{4}\mbox{Tr}\left(U^\dagger\frac{\partial U}{\partial\varphi^b} \sigma_3\right)\\&= \frac{g\tilde g^{ab}}{2}u^TT_I\frac{\partial u}{\partial\varphi^b},&
    &=\frac{g_Y\tilde g^{ab}}{2}u^TT_Y\frac{\partial u}{\partial \varphi^b},
\end{align}
where $u$ is a real four vector of norm one, $u^Tu=1$, and $\sigma$ are the Pauli matrices. We use $T$ to stand for a subset of the generators of $SO(4)$ such that $T^T=-T$, $Tr[TT]=-4$. There are a number of relations these objects obey due to unitarity such as $(\partial U)U^\dagger=- U\partial U^\dagger$.
For simplicity we will take a universal $F$, $F_\psi=F$ with no flavour structure, which for concreteness will read, as a function of the Higgs field,
\begin{align}
F(h)&=\sqrt{s_\chi^2+c_\chi^2(1+h/v_\star)^2}=\sqrt{s_\chi^2+c_\chi^2(1+\gamma_ah/v)^2}\,,
\label{vst}
\end{align}
where one has $\gamma_a=v/v_\star$. In the general formulation of the HEFT, $F^2(h)$ can be an arbitrary analytic function $F^2(h) = 1 + \sum_{n=1}^\infty f_n h^n$. Our choice corresponds to setting
\begin{equation}
    f_1 = \frac{2 \cos^2\chi}{v_\star},
    \quad
    f_2 = \frac{\cos^2\chi}{v_\star^2},
    \quad
    f_{n > 2} = 0\,,
\end{equation}
which just amounts to imposing that $F^2(h)$ is quadratic in $h$ with generic coefficients.
This is enough to model the main feature of our non-linear theory: a scalar manifold with a double sheet structure, connected by a throat at a distance $v_\star$ from our vacuum $h=0$. It is at the narrowest point of this throat where $F$ reaches its minimum,  and so do electroweak particle masses. For $\chi>0$, Min$[F(h)]>0$ which we maintain throughout, a limit of decreasing $\chi$ is depicted in Fig.~\ref{FigX} for both $F$ and the shape of the manifold.
 The field-dependent mass term for gauge bosons then reads
\begin{align}\label{WZMss}
    m_W(h)&=\frac{gv}{2}F(h)\,, & m_Z(h)&=\frac v2F(h)\sqrt{g^2+g_Y^2}\,,
\end{align}
with $F\geq  \sin \chi$ over the full range of $h$, and the minimum is achieved for
\begin{align}
    \textrm{Min}[m_W(h)]\equiv\frac{gv}{2}F(-v_\star)=\frac{gv\sin\chi}{2}\,.\label{mWmin}
\end{align}

We dub the reflection transformation introduced in the previous section Higgs parity---not to be confused with other acceptations~\cite{Dunsky:2019api}---whose action on $\phi$, the distance to the throat of the manifold, reads
\begin{align}
    \phi=h+v_\star&\equiv h+v\gamma_a^{-1}\,,  \label{eq:phidef}\\
    \boxed{\textrm{Higgs Parity}}\quad& \phi\to -\phi\,, \label{eq:HPar}
\end{align}
which as remarked in the introduction is not a symmetry applicable to SMEFT.

A scalar manifold and metric differing from the SM flat case will imply different couplings and deviations in observables. Scattering matrix elements are given in terms of tensors in field space projected on the different directions with vierbeins; in particular, the tensor that dictates scalar couplings to order $p^2$ is the curvature tensor and its covariant derivatives~\cite{Alonso:2015fsp,Cohen:2020xca}. One has in our case a Riemann tensor $\mathcal R$ that follows from the metric $G$ as
\begin{align}
    \mathcal R_{a\,\,c\,\,}^{\,\,b\,\,d}&= R_\varphi (\tilde g_{ac}\tilde g^{bd}-\delta_a^d\delta_c^b)\,, & \mathcal R_{a h}{}^{bh}&=R_h\delta_a^b\,,
\end{align}
and the only two independent functions
	\begin{align} v^2R_\varphi(h)&\equiv\frac{1}{F^2}-v^2[(\log F)']^2=\frac{c_\chi^2(1+h/v_\star)^2(1-c_\chi^2\gamma_a^2)+s_\chi^2}{(s_\chi^2+c_\chi^2(1+h/v_\star)^2)^2}\,, & v^2R_\varphi(0)&=1-c_\chi^4\gamma_a^2\,,\label{eq:Rphi}
 \\ v^2R_h(h)&\equiv-\frac{v^2F''}{F}=-\frac{\gamma_a^2s_\chi^2c_\chi^2}{(s_\chi^2+c_\chi^2(1+h/v_\star)^2)^2}\,, & v^2R_h(0)&=-\gamma_a^2s_\chi^2c_\chi^2\,,\label{eq:Rh}
\end{align}
 where we also present curvature evaluated at the origin because this is what LHC is sensitive to and can set bounds on, as in Ref.~\cite{Alonso:2021rac}.

  Lastly the tree level potential, which might generate new extrema, is a quartic potential accommodating the Higgs mass but with free cubic and quartic terms  
\begin{align} V(h)&=
\frac{m_h^2}{2}h^2+\frac{m_h\sqrt{\lambda}}{2}\gamma_4(1-\epsilon)h^3+\frac{\lambda}{8}\gamma_4^2h^4\,,\label{eq:VtreeDef}
\end{align}
with the SM limit $\gamma_4\to 1,\epsilon\to0$. We have defined $\sqrt{\lambda}=m_h/v\simeq125/246$ as a way to parametrise our system, but note this does not imply an SM-like quartic coupling.

\begin{figure}[h]
    \centering

    \begin{subfigure}[b]{0.475\textwidth}
         \centering
         \includegraphics[width=\textwidth]{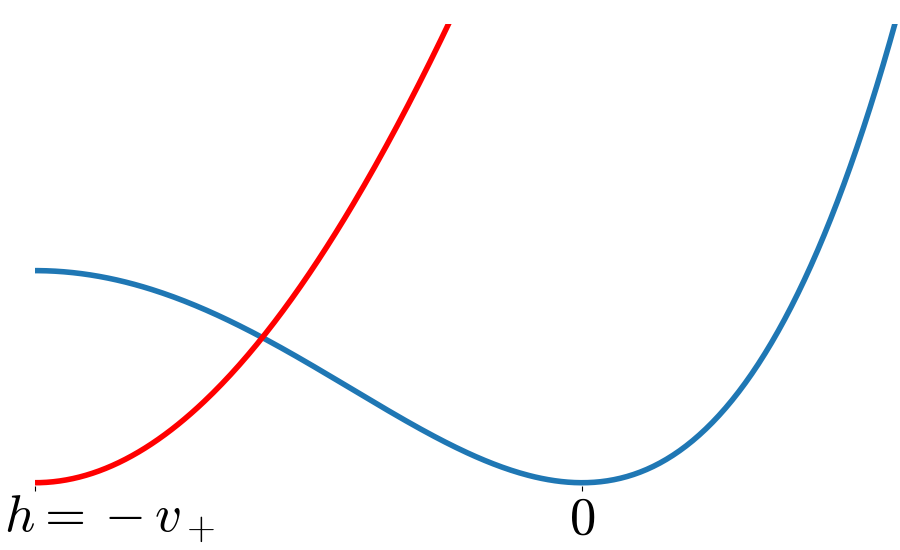}
         \caption{}
        \label{fig:SMEFTPotential}
     \end{subfigure}
     \hfill
     \begin{subfigure}[b]{0.475\textwidth}
         \centering
         \includegraphics[width=\textwidth]{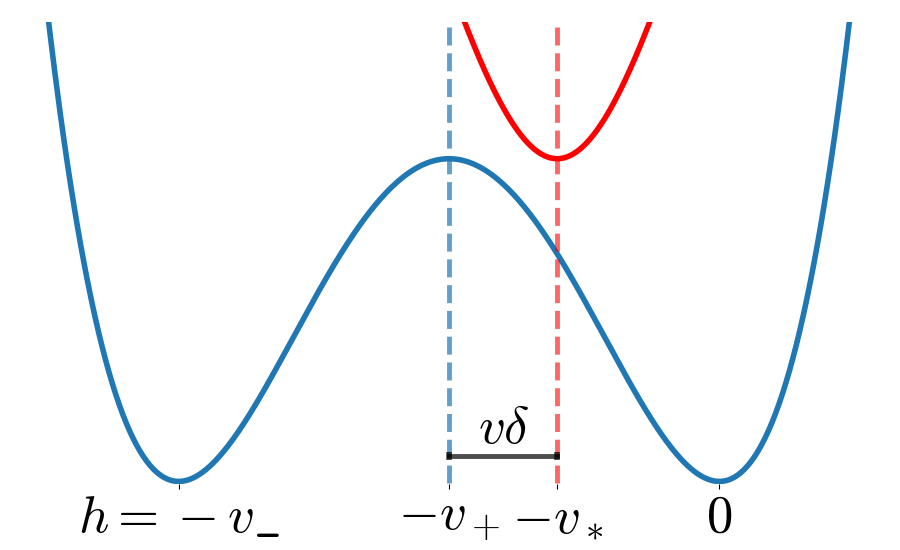}
         \caption{}
        \label{fig:NLPotential}
     \end{subfigure}
    \caption{Schematic diagram of the tree level Higgs potential $V(h)$ in blue and in red the Higgs-dependent W mass, $v^2m_W^2(h)$, in arbitrary units to set our conventions. Along the horizontal axis runs the Higgs field $h$. Fig.~\ref{fig:SMEFTPotential} shows the SM and SMEFT case, with $h\in[\,-v_+,\infty)\,$, and Fig.~\ref{fig:NLPotential} shows the non-linear case where $h\in(\,-\infty,\infty)\,$. The quantities $v_\pm,v_*$ and $\delta$ are defined in Eqs.~\eqref{vpmhpm}, \eqref{vst} and \eqref{fullde} respectively.}
    \label{FigY}
 
\end{figure}

The tree-level potential has a minimum by design at $h=0$, where we define the vacuum we inhabit today. There are two other possible extrema at a distance $v_{\pm}$ which read, in terms of $\gamma_4$, $\epsilon$:
\begin{align}\label{vpmhpm}
h_{\mp}(T=0)\equiv-v_{\mp}\equiv-v\gamma_4^{-1}\frac{3(1-\epsilon)\pm\sqrt{1+9(\epsilon^2-2\epsilon)}}{2}\equiv\left\{\begin{array}{c}
-2v\gamma_4^{-1}\gamma_\epsilon^{-1}\\
-v\gamma_4^{-1}\gamma_\epsilon
\end{array}\right.,
\end{align}
where we have defined 
\begin{equation}
    \gamma_\epsilon \equiv
    \left\{\begin{array}{ll}
        \frac{3 (1 - \epsilon) - \sqrt{1 + 9(\epsilon^2 - 2\epsilon)}}{2}
        & \qquad \epsilon \leq 1 - \sqrt{8/9}
        \\
        \sqrt{2}
        & \qquad \epsilon > 1 - \sqrt{8/9}
    \end{array}\right..
\end{equation}
Only in the former range of $\epsilon$ are two other extrema present because for the solutions to be real one needs 
\begin{align}
9(1-\epsilon)^2&\geq 8\,.
\end{align}
That is, numerically, $\epsilon\leq 1-\sqrt{8/9}\simeq 0.057$. At the upper limit of this inequality $\gamma_\epsilon=\sqrt2$, and the two extrema meet at $h/v=-\sqrt{2}\gamma_4^{-1}$ to form an inflection point, which informs our definition of $\gamma_\epsilon$ for the upper interval.
It should be stressed that this small range for positive $\epsilon$ where two extrema are present does not guarantee an expansion in small epsilon is viable, in particular not when close to the upper limit.

The values of the potential at the different extrema are
\begin{align}
\! V(-v_{\mp})=\frac{\lambda v^4\gamma_4^{-2}}{4}\left(\frac{v_\mp}{v\gamma_4^{-1}}\right)^{\!2}\!\left(1+\frac{1-\epsilon}{2}\left(\frac{-v_\mp}{v\gamma_4^{-1}}\right)\right)=\left\{\begin{array}{c}
\gamma_4^{-2}\gamma_\epsilon^{-2}\lambda v^4(1-\gamma_\epsilon^{-1}(1-\epsilon))\\
\gamma_4^{-2}\gamma_\epsilon^2\lambda v^4(2-\gamma_\epsilon(1-\epsilon))/8
\end{array}\right..
\end{align}
Note that only differences in these values are significant, and so we have chosen $V(0) = 0$ in the definition of the potential in Eq.~\eqref{eq:VtreeDef}.
It is useful to define the distance between the local maximum of the potential and the minimum value of electroweak masses as
\begin{equation}\label{fullde}
    \delta \equiv
    \left\{\begin{array}{ll}
        \frac{v_+ - v_*}{v} =
        \gamma_4^{-1} \gamma_\epsilon - \gamma_a^{-1}
        & \qquad \epsilon \leq 1 - \sqrt{8 / 9}
        \\
        \gamma_4^{-1} \sqrt{2} - \gamma_a^{-1}
        & \qquad \epsilon > 1 - \sqrt{8 / 9}
    \end{array}\right.,
\end{equation}
and a visual definition of $\delta$ in the presence of two tree-level minima is given in Fig.~\ref{FigY}.

In the limit of small deviation from (local) SM couplings  $\epsilon,\delta \ll 1$, we have---noting that this expansion is in the $\epsilon<1-\sqrt{8/9}$ interval,
\begin{align}
\gamma_\epsilon&\simeq 1+3\epsilon,&V(h_{\mp})&
\simeq\left\{\begin{array}{c}
4\gamma_4^{-2}\lambda v^4\epsilon+\mathcal{O}(\epsilon^2)\\
\gamma_4^{-2}\lambda v^4/8+\mathcal{O}(\epsilon)
\end{array}\right..
\label{eqSUV}
\end{align}

\begin{figure}
    \centering
    \includegraphics[width=0.48\textwidth]{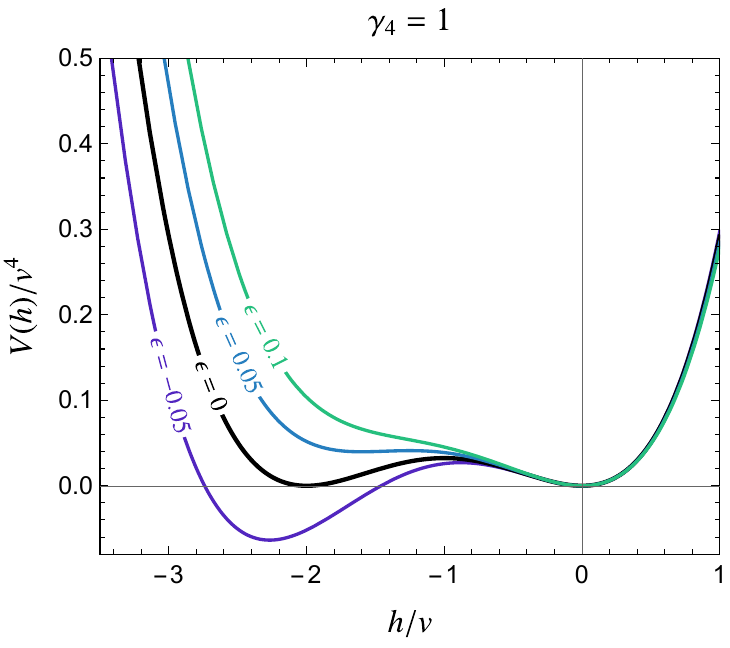}
   \hspace{.3cm} \includegraphics[width=0.48\textwidth]{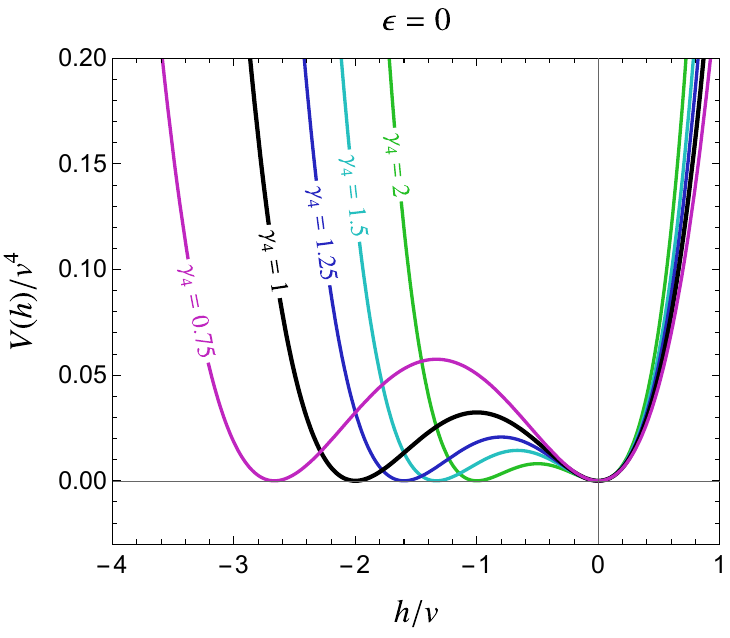}
   
   \includegraphics[width=0.48\textwidth]{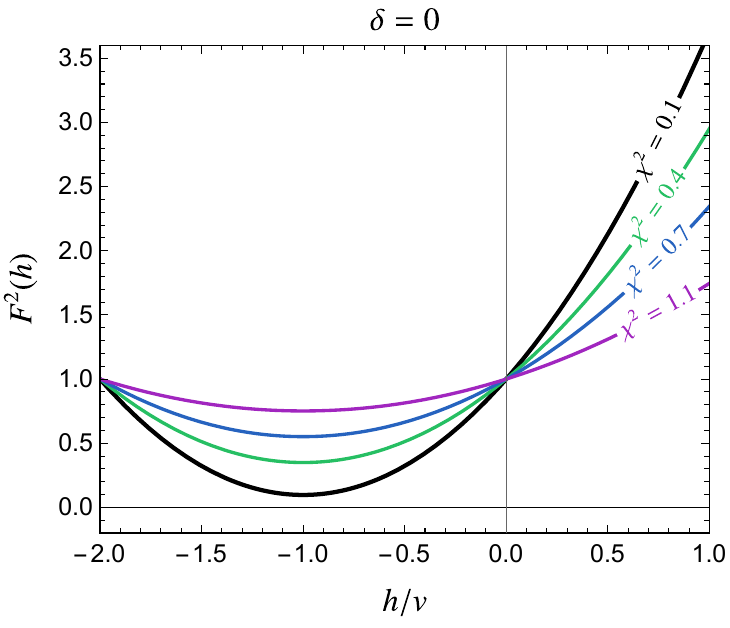}
   \hspace{.3cm} \includegraphics[width=0.48\textwidth]{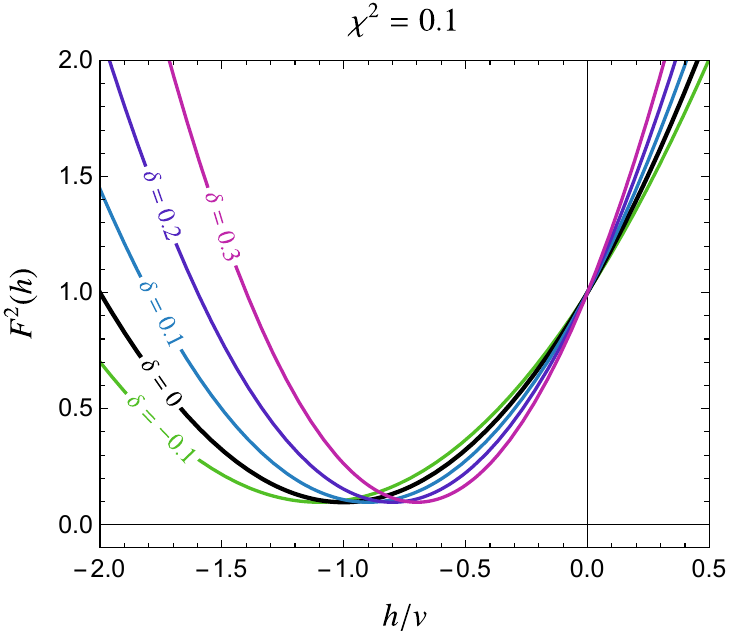}
    \caption{The effect of the 4 free parameters of our effective Lagrangian on the tree-level potential $V(h)$ and the $F(h)$ function. Top: variation of the potential $V(h)$ with $\epsilon$ (left) and $\gamma_4$ (right). Bottom: variation of the $F(h)$ function with $\chi$ (left) and $\delta$ (right).}
    \label{fig:changepar}
\end{figure}

Now through the definition of $v_\star$ in Eq.~\eqref{eq:phidef}, those of $v_{\pm}$ in Eq.~\eqref{vpmhpm} and the above, we can write the theory in terms of 4 free parameters $\delta,\chi,\epsilon$ and $\gamma_4$ which control respectively,

\begin{itemize}
    \item[$\chi$] the width of the throat, defined in Eq.~\eqref{vst} and depicted in Figs.~\ref{FigX} and~\ref{fig:changepar}.
    \item [$\epsilon$] the energy difference between the two tree-level minima and the Higgs triple coupling; this is defined in Eq.~\eqref{eq:VtreeDef}, and its effect depicted in Fig.~\ref{fig:changepar}.
    \item[$\gamma_4$] the distance between the two tree level minima and the triple and quartic Higgs coupling,  defined in Eq.~\eqref{eq:VtreeDef}, and depicted in Fig.~\ref{fig:changepar}.
    \item[$\delta$] the distance between the throat and the potential tree level maximum, defined in Eqs.~\eqref{vst} and \eqref{fullde} and depicted in Figs.~\ref{FigY} and~\ref{fig:changepar}.
\end{itemize}
As we shall see, this modelling of non-linear theories is versatile enough to unveil a rich phenomenology of which here we purport to show what we believe to be a representative sample.

\section{Modification of sphaleron solutions}
\label{sec:Sph}

The absence of a $O(4)$-symmetric point in the HEFT$\backslash$SMEFT leads to manifolds topologically different
from the SMEFT case.
Thus, one may be able to differentiate both theories, even when they appear to be the same locally (and hence inseparable by processes described by the perturbative $S$-matrix), by studying the physics of field configurations that depend on the global properties of this manifold.
In this section, we focus on the sphaleron configurations which, in principle, rely on the existence of $O(4)$-symmetric point to exist.

In the SM, a sphaleron is the maximum-energy point in a minimal-energy non-contractible loop through scalar and gauge field space.
This loop can be approximated by a family of static field configurations depending on a parameter $\alpha \in [0, \pi]$ as 
\begin{align}\label{eq:sph-ansatz}
    h &= h(r),
    &
    U &=
    e^{i \alpha \sigma_3 / 2}
    \left(
        \cos(\alpha) + i \sin(\alpha) \, \hat{x} \cdot \vec{\sigma}
    \right)
    e^{i \alpha \sigma_3 / 2},
    \\\
    W_0 &= 0, 
    &
    W_j &= -\frac{i}{g} a(r) U \partial_j U^\dagger
\end{align}
where $\hat x =\vec x/|x|$ is the unit vector in space, $U$ and $W$ are the $SU(2)$ Goldstone and gauge fields\footnote{The effects of the $U(1)$ sector are neglected here which is an approximation valid up to corrections of order $g'/g$, see Ref.~\cite{Klinkhamer:1984di}.}, and $h(r)$ and $a(r)$ are radial functions that minimise the energy and satisfy the boundary conditions
\begin{equation}
h(0) = h_\odot, \qquad a(0) = 0, \qquad
\lim_{r \to \infty} h(r) = 0, \qquad
\lim_{r\to\infty} a(r) = 1,
\end{equation}
with $h = h_\odot$ being the $O(4)$-invariant point.
The fact that $U|_{\alpha = 0} = U|_{\alpha = \pi} = 1$ ensures that the curve described in field space by varying $\alpha$ is indeed a loop.
Furthermore, $U$ can be viewed as a topologically non-trivial mapping from the 3-sphere into itself, and therefore the loop is not contractible (see Ref.~\cite{Manton:1983nd} for the original proof).

The sphaleron is the configuration at the midpoint of the loop, with $\alpha = \pi / 2$.
In order for it to be well-defined everywhere in space, it is crucial that $h(0) = h_\odot$.
This is because $U|_{\alpha\neq0,\pi}$ is singular at the origin.
Indeed, taking the limit $r \to 0$ from different spatial directions will yield different $U$ matrices.
This makes the field configuration undefined at $r = 0$, unless all values of $U$ can be identified and collapsed into a single point, making it single-valued.
This can only be done if there is a $O(4)$-invariant point $h_\odot$ where $F(h_\odot) = 0$.\footnote{See Ref.~\cite{Spannowsky:2016ile} for a study of sphalerons in non-SM theories with this property.}
Therefore, fully well-defined sphaleron configurations cannot exist in HEFT$\backslash$SMEFT theories which have $F(h) > 0$ for all $h$.

One can nevertheless work with the definition we have given for $U$ away from the origin, and see if the singularity there has physical consequences.
In particular, in order for it to be physically meaningful, the sphaleron solution must have finite energy, using an ansatz as that of Eqs.~\eqref{eq:sph-ansatz} with $\alpha=\pi/2$:
\begin{align}
    E &=
    \int d^3x \left[
        \frac{4}{g^2 r^2} \left(\frac{da}{dr}\right)^2
        + \frac{8 a^2 (1-a)^2}{g^2 r^4}
        + \frac{1}{2} \left(\frac{dh}{dr}\right)^2
        + \frac{v^2 F(h)^2}{r^2}(1 - a)^2
        + V(h,\Phi)
    \right].
\end{align}
While the energy density around $r = 0$ would diverge as $1 / r^2$ if $F(h) \neq 0$, the volume integral cancels this divergence for a finite result. Another road to reach the same finiteness conclusion is to remember that the sphaleron energy remains finite when $\lambda\to \infty$ in the SM, i.e. we freeze the Higgs field value on its vacuum and take the particle out of the spectrum. It should be noted nonetheless that had we been dealing with a 2-dimensional theory, the solutions would have indeed disappeared.

The question that follows is then how is the sphaleron energy, which we now know to be finite, modified. This energy does know about the Higgs manifold non-locally and is, as such, qualitatively different from, say, scattering experiments. In particular, the relevant question (even if just theoretical) as posed in Ref.~\cite{Alonso:2021rac} is whether a non-linear theory locally SM-like around the vacuum but globally different can be told apart by the sphaleron energy.

A smooth set of functions $F_n$ for non-linear theories that look locally ever more SM like for increasing integer $n$ were proposed in Ref.~\cite{Alonso:2021rac}, here we rewrite them as
\begin{align}
    F_n&=1+\frac{h}{v}+c_n\left(\frac{h}{v}\right)^n, & c_n&=(-1)^n\frac{(n-1)^{n-1}}{n^n(1-F_\star)^{n-1}}.
\end{align}
These functions have a minimum $F_\star$ at $f_\star=-v_\star/v=-n(1-F_\star)/(n-1)$, we take $0<F_\star<1$ with the lower limit in particular to keep a non-linear type A theory.
We use them as a probe for approaching the SM while in non-linear theory space.
As the origin boundary condition for the Higgs, we take as $h(0)=-v_\star$ where $F'(-v_\star)=0$.

\begin{figure}
    \centering

    \begin{subfigure}[b]{0.49\textwidth}
         \centering
         \includegraphics[width=\textwidth]{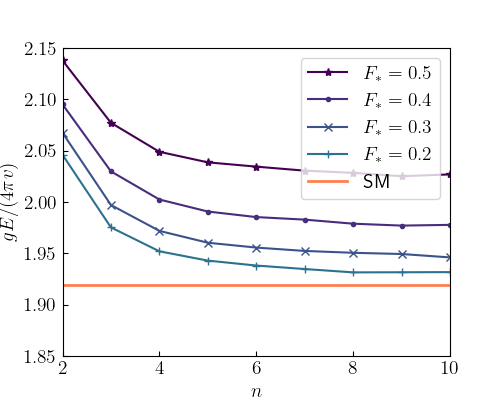}
         \caption{}
        \label{fig:SphEnergy}
     \end{subfigure}
     \hfill
     \begin{subfigure}[b]{0.49\textwidth}
         \centering
         \includegraphics[width=\textwidth]{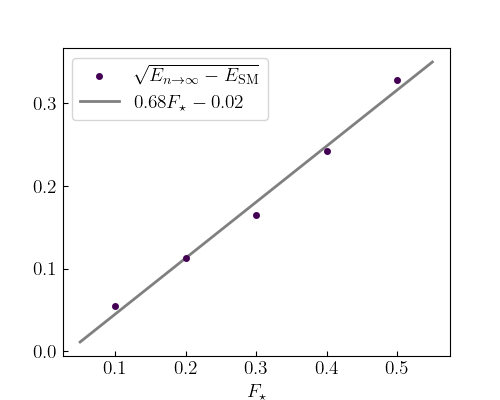}
         \caption{}
        \label{fig:LimSphEnergy}
     \end{subfigure}
    \caption{Fig.~\ref{fig:SphEnergy}: sphaleron energy as a function of $n$ for different values of $F_\star$. Fig.~\ref{fig:LimSphEnergy}: square root of the difference between the limit value of the energy as $n \to \infty$ (taken from $n = 10$) and the SM energy as a function of $F_\star$, together with a linear fit.}
    \label{fig:sphaleron-energies}
 
\end{figure}

\begin{table}
    \centering
\begin{tabular}{cc@{\quad\qquad}ccccccc}
    \toprule
     \multicolumn{2}{c}{\multirow{2}{*}{Ansatz}} & & 
     \multicolumn{6}{c}{$n$}
     \\ \cmidrule{4-9}
     & & & 1(SM) & 2&3 &4 &5&6   \\ \midrule
     &&$E_{Sph}$ &1.948   &2.079   &1.991   &1.971   &1.966   &1.966\\
     \textbf{A)} & Eqs.~\eqref{eq:a-ansatz},~\eqref{eq:f-ansatz} &$\zeta_a$    &1.4    &1.2    &1.4    &1.4    &1.4    &1.4 \\
     &&$\zeta_h$    &1.6    &2.4    &2.0    &1.8    &1.7    &1.6\\
     \midrule
     \textbf{B)} & Neural network  &$E_{Sph}$
     &1.919 &2.053 &1.976 &1.952 &1.943 &1.938 \\
     \bottomrule
\end{tabular}
\caption{Values of the sphaleron energy (and corresponding radii $\zeta_a$ and $\zeta_h$) at $F_\star=0.2$ and varying $n$, as computed with methods \textbf{A)} and \textbf{B)}.}
    \label{fig:TabEsph}
\end{table}

The dimensionless radius $\zeta=gvr$ simplifies the energy to read
\begin{align}
    E=&\frac{4\pi v}{g}\int d\zeta \left[4\dot a^2+\frac{8a^2(1-a)^2}{\zeta^2}+\frac{\zeta^2}2\dot f^2 +F(vf)^2(1-a)^2+\zeta^2\frac{\lambda}{8g^2}(f^2+2f)^2\right],
\end{align}
with $f=h/v$.
The sphaleron is obtained by minimising this energy as a function of the profiles $a(\zeta)$ and $f(\zeta)$.
We use two different approaches to minimize it for different values of $n$ in the $F_n$ functions defined above:
\begin{description}
\item[A)] \textbf{Analytical ansatz.} While, due to their non-linear character, one cannot solve the Euler-Lagrange equations that follow from the variational principle for the energy $E$:
    \begin{align}\label{eq:sph-energy}
        &-8 \ddot a+16\frac{a(1-1)(1-2a)}{\zeta^2}-2(1-a)F(f)^2=0,\\
        &-\zeta^2 \ddot  f-2\zeta \dot f+2FF'(f)(1-a)^2+\zeta^2\frac{\lambda}{g^2}f(1+f/2)(1+f)=0,
    \end{align}
     one can solve their asymptotic form in the limits $\zeta\to 0,\infty$. 
    Following Ref.~\cite{Klinkhamer:1984di}, we devise an ansatz as a piece wise function with the solutions to the asymptotic equations which we join at a radius $\zeta_a$ ($\zeta_h$ for the Higgs function). Such a function has therefore 4 integration constants, two of which are fixed by the boundary conditions, while the other two are found by imposing the same value for the function and its first derivative at $\zeta_a$ ($\zeta_h$). This leads to
    \begin{align}
        \hat a(\zeta)&=\left\{\begin{array}{lcr}
             \zeta^2\left(\frac{\zeta_a F_\star^2+6}{6\zeta_a(4+\zeta_a)}-\frac{F_\star^2}{12}\log(\zeta/\zeta_a)\right)&& \zeta<\zeta_a  \\
             1-\frac{4-F_\star^2\zeta_a^2/6}{4+\zeta_a}e^{-(\zeta-\zeta_a)/2}&& \zeta>\zeta_a 
        \end{array}\right.,
        \label{eq:a-ansatz}\\
        \hat f(\zeta)&=\left\{\begin{array}{lcr}
             f_\star\left(1-\frac{p+\sigma \zeta_h}{2p+\sigma \zeta_h}\left(\frac{\zeta}{\zeta_h}\right)^p\right)&& \zeta<\zeta_h  \\
             f_\star \frac{p\zeta_h}{2p+\sigma\zeta_h}\frac{1}{\zeta} e^{-\sigma (\zeta-\zeta_h)}&& \zeta>\zeta_h 
        \end{array}\right.,
        \label{eq:f-ansatz}
    \end{align}
    where $\sigma=m_h/(2m_W)$, $2p=(\sqrt{1+8F''_\star F_\star}-1)$. Substituting this ansatz back in the energy expression Eq.~\eqref{eq:sph-energy} allows for minimisation in two variables $\zeta_a,\zeta_h$ which we do numerically. This semi-numerical method will not yield the true minimum energy since the ansatz are not solutions to the full equations, but it does have the advantage of treating boundary conditions analytically. Results are shown in Tab.~\ref{fig:TabEsph}.
\item[B)] \textbf{Numerical ansatz.} Directly minimise $E$ numerically, employing a small neural network as an ansatz for $f$ and $a$.
    We use the Elvet package~\cite{Araz:2021hpx} for this purpose.
    The neural network is densely connected, with 1 input ($\zeta$), two outputs ($f$ and $a$) and 2 hidden layers with 5 units each.
\end{description}
Both approaches lead to similar results, shown in Fig.~\ref{fig:sphaleron-energies}, with \textbf{B)} giving slightly lower energies than \textbf{A)} as expected.
From these results, it is clear that, as $n \to \infty$, the sphaleron energies tend to a value differing from the SM one, provided $F_\star\neq 0$.
We find that the difference between this value and the SM one is approximately proportional to $F_\star^2$, as displayed in Fig.~\ref{fig:LimSphEnergy}.
This non-pertubative phenomenon is hence sensitive to non-linear theories that would be indistinguishable for perturbative scattering.

The prospects for observation of zero-temperature sphaleron-mediated processes rely on $B+L$ violation and involve all three families of fermions. Such a process might be initiated by, e.g., the impact of an energetic neutrino and the production of a multiparticle flavourful final state on IceCube~\cite{Tye:2015tva,Ellis:2016dgb}, or by a high-energy  collision at LHC and future colliders~\cite{Ellis:2016ast,Papaefstathiou:2019djz}, although exponential suppression factors might make it unobservable in practice~\cite{Bezrukov:2003er, Khoze:2020paj}. At finite temperature, the sphaleron energy is to be computed with the temperature-dependent effective potential and processes that are sensitive to it include the decoupling temperature for $B+L$ violating effects in baryogenesis, see e.g.~\cite{Kanemura:2020yyr}.  
%The most relevant phenomenological role of Sphalerons is to be found in Cosmology in e.g. conversion of L to B violation for which purpose the potential here should be substituted by the finite temperature potential to be shown below. However as opposed to the type of phenomena to follow the modifications in HEFT for sphalerons do not change the qualitative picture and e.g. the ratio of B/L would be the same. As for the zero temperature Sph energy one could think of B violation by SM processes
%a scattering experiment.
In practice, however, finding percent deviations in sphaleron energy is not a feasible strategy to explore these theories. As we shall show, phase transitions on the other hand offer a much more promising avenue.

%%%%%%%%%%%%%%%%%%%%%%%%%%%%%%%%%%%%%%%%%%%%%%%%%%
\section{Finite temperature potential}\label{sec:FiniteT}

The key quantity to study phase transitions and defect formation is the effective potential at finite temperature. The computation of this potential is not without obstacles, long since identified and discussed in the literature extensively, see e.g. Ref.~\cite{Croon:2020cgk} for a review. In such context, here we opt for laying out our derivation with all assumptions and choices explicit while emphasising the limit of applicability of our results. In addition by facing old problems from a new theory, we believe this work might bring novel approaches and perspective.

The effective potential will be computed to one loop at finite temperature. This suffices to chart possible new phenomena that HEFT brings to the electroweak phase transition. Before proceeding to the computation, however, let us first discuss its limitations. The potential estimate here is not valid in all the field domain since perturbation theory breaks down in thermal field theory for gauge theories in the limit of vanishing transverse mode mass (i.e. the infrared problem \cite{Linde:1980ts})-- for the EW theory this occurs around the $O(4)$ fixed point. While the perturbative expansion can be supplemented with resummation techniques and dimensional reduction as detailed in Ref.~\cite{Croon:2020cgk,Croon:2023zay},\footnote{Another promising direction using nonperturbative methods is presented in~\cite{Croon:2021vtc}.} the deciding say lies in lattice. Rather than attempting to extend the potential computation to this domain beyond perturbativity, we will mark its limits and base our conclusions on results outside of it. It is worth noting already that the IR problem arises as the masses of gauge bosons approach zero and the presence of a lower bound on such masses in our theory ameliorates the problem. 

\begin{figure}
    \centering
    \includegraphics[width=0.8\textwidth]{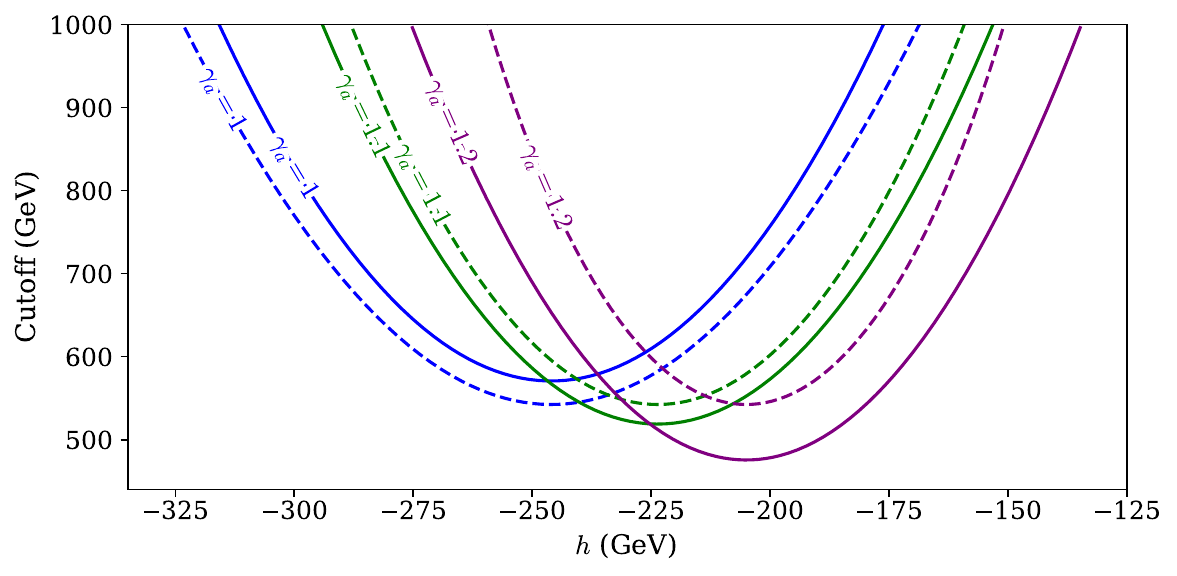}
    \caption{Field-dependent cut-off as estimated from perturbative unitarity to be $4\sqrt{\pi} \left|R_I(h)\right|^{-1/2}$ for sectional curvatures $R_\varphi(h)$ (solid) and $R_h(h)$ (dashed) as defined in Eqs.~\eqref{eq:Rphi} and~\eqref{eq:Rh}. The parameter $\chi=\sqrt{0.1}$ and the following $\gamma_a = v/v_\star$ values are chosen as those that lead to domain walls, $\gamma_a=1$ (see Sec.~\ref{sec:Walls}); the doom scenario, $\gamma_a=1.1$  (see Sec.~\ref{sec:vacuum-decay} and Fig.~\ref{DoomLHC} in Sec.~\ref{sec:LHC}); and $\gamma_a=1.2$ a conservative value for the observable bubble region shown in Fig.~\ref{fig:bubble_summary}. For comparison, the nucleation  temperature in phase transitions studied here is around $100$~GeV.}
    \label{fig:hDepCut}
\end{figure}

A second relevant consideration is that our potential, since it is computed in an EFT, is not valid to arbitrarily high temperatures. The amplitude for longitudinal boson scattering scales as $R\,s$, where $s$ is the square of the centre of mass energy. This points to a cutoff where new states would appear at $\Lambda\sim 4\sqrt{\pi} |R(0)|^{-1/2}$ \cite{Alonso:2021rac}, see also~\cite{Cohen:2021ucp} where unitarity leads to an estimate of $\Lambda\sim 4\pi v$. Experimental data from LHC constrains the curvature around the EW vacuum at zero temperature to be small so that the cut-off is above the TeV. In the evolution of the universe, however, the electroweak vacuum changes position in field space and scans over a range of values. Approximating as customary the thermally averaged cross-section in this environment as the zero temperature result convoluted with the thermal distributions, one can expect the cut-off estimation outlined above for $h=0$ to be extended to different manifold values  $\Lambda(h)\sim4\sqrt{\pi}  |R(h)|^{-1/2}$. This naive extension of the cut-off estimate to thermal field theory ignores the extra Boltzmann suppression that the contribution of heavy states receive (indeed UV divergences in thermal field theory are much softened with respect to zero temperature), but doing so is erring on the conservative side. We therefore consider the cut-off for the whole range that the Higgs field explores, and one has that it would be lowest at the highest curvature, i.e. the throat, see Fig.~\ref{fig:hDepCut}. At the throat one has  $\Lambda\sim4\sqrt{\pi} [R_\varphi(h=-v_\star)]^{-1/2}=4\sqrt{\pi} v \sin(\chi)=1.7\sin(\chi)$~TeV, imposing a minimum value for $\chi$ for our EFT to be applicable, and we impose a similar expression for $R_h$. In the following phenomenological study it is ensured that the cut-off is at least  $450$~GeV at the throat for the allowed parameter space.

\subsection{Calculation of the one loop finite temperature potential}

The geometric description presents relevant differences with the standard computation of one loop corrections for the effective potential, so let us sketch them here in some length. Path integral methods allow us to write one loop corrections in a few lines of algebra, while the notation is abstract enough to encompass both the finite and zero temperature case which we discuss simultaneously. In this functional computation one expands around a background value for the fields $\bar\Phi$, which satisfy the EoM at one loop order, along geodesics if the field space has curvature. This expansion reads, with $\delta\Phi$ being the field variation we will integrate over,
\begin{align}
    \Phi^i[\delta\Phi]&=\bar\Phi^i+\delta\Phi^i-\frac12\Gamma^i_{jk}(\bar\Phi)\delta\Phi^j\delta\Phi^k+\mathcal O (\delta\Phi)^3, & A^\mu&=\bar A^\mu+\delta A^\mu,
\end{align}
with $\Gamma$ the connection derived from the metric $G$. An expansion for fermions can be similarly derived to that for gauge bosons.

The path integral for the effective action $\Gamma_\textrm{eff}$ to one loop reads 
\begin{align}
    e^{i\Gamma_{\textrm{eff}}}=e^{iS[\bar\Phi,\bar A]}\int \sqrt{G} [\mathcal D\delta\Phi] [\mathcal D \delta A] e^{i \frac12\delta_\Phi^2S+i \frac12 \delta^2_AS+i\delta_\Phi\delta_A S},
\end{align}
where $\sqrt{G}$ is the scalar metric inserted for an invariant volume element and
the second variation around the background field for gauge bosons is a standard result, whereas for the scalars we have, when applied to the action in Eq.~(\ref{LagFY}),
\begin{align}
    \delta_\Phi^2S=&\int d^4x \delta\Phi^i\left(-G_{ij}D^2-\nabla_i\nabla_j V\right)\delta\Phi^j,
    & \nabla_i \nabla_j V&=\left(\begin{array}{cc}
        V'' & 0 \\
         0& FF'V'\tilde g_{ab} 
    \end{array}\right),
\end{align}
with $\nabla$ the covariant derivative w.r.t. field coordinates and $D$ the covariant spacetime derivative acting in the scalar manifold tensor space \cite{Alonso:2016oah}, which hereafter setting the background field to a constant equals the ordinary derivative. The extension to higher loop order can be found in Ref.~\cite{Alonso:2022ffe}.
It is well worth remarking that the mass term for Goldstones, $FF'V'$, (which vanishes at the vacuum) will be altogether missed in this coordinate system if not doing a covariant treatment. This non-covariant treatment would lead to inconsistency as one can realise by resorting to the SM with linear scalar coordinates where all 4 d.o.f. have a mass term away from the vacuum; in essence the point is that even in flat space, when in spherical coordinates, one has to use a covariant formulation.
The mixed term in gauge and scalar fields one can borrow from Ref.~\cite{Alonso:2017tdy}:
\begin{align}
    \delta_\Phi\delta_A S=\int d^4x [-\delta A_B D_\mu t_{B,i} \delta \Phi^j].
\end{align}
This  term can be absorbed in a redefinition of the path-integral fields without changing the measure as done in Ref.~\cite{Alonso:2017tdy}, but here instead as is more common practice in thermal field theory we cancel it out with the introduction of a gauge fixing term
\begin{align}
    \mathcal L_{g.f.}=-\frac{1}{2\xi}\sum_B \left(\xi t^i_B G_{ij}\delta\varphi^j+D_\mu \delta A^\mu_B\right)^2,
\end{align}
which puts our gauge bosons in the $R_\xi$ gauge while producing an extra mass term for the scalars as
\begin{align}
    \delta^2_\Phi(S+S_{g.f.})=-\delta\Phi(D^2+\nabla^2V+\xi m_{gf}^2) \delta\Phi,
\end{align}
with the additional gauge dependent term
\begin{align}
     \left[m_{gf}^2\right]^{i}_{\,j}=&\sum_B  t_B^i t_{j,B}
     =\frac{(Fv)^2}{4}\left(\begin{array}{cc}
           g^2\delta_a^b+g_Y^2 \tilde g^{ac} v^2u^T T_Y\frac{\partial u}{\partial \varphi^c}\, u^T T_Y\frac{\partial u}{\partial\varphi^b} &0 \\
          0& 0
     \end{array}\right),
\end{align}
with Tr$[m_{gf}^2]=2m_W^2(h)+m_Z^2(h)$.
We note that the last term does depend on the angular degrees of freedom, the Goldstones $\varphi$; this does not mean they will feature in the effective potential (which could only come about via explicit breaking) but that the final result, when all tensors have had their indices contracted, will make any Goldstone boson dependence disappear.
The second variation of the action for gauge bosons on the other hand is
\begin{align}
    \delta^2_{A}(S+S_{g.f.})=\delta A^\mu\left(\eta_{\mu\nu}D^2-(1-\xi^{-1})D_\mu D_\nu+m_A^2\right)\delta A^\nu.
\end{align}
When it is only the effective potential one is after, derivatives of the background field can be neglected, and we obtain loop corrections as, for the scalar correction
\begin{align}
    -i\int d^4x V_{\hbar\Phi}&=-\frac12\mbox{Tr}\left[\log\left(G \partial^2+\nabla^2V+\xi m_{gf}^2\right)\right]+\frac12 \mbox{Tr}\log (G)\\
        &=-\frac T2\mbox{Tr}\left[\log\left( \delta_i^j \partial^2+\nabla_i\nabla^jV+\xi [m_{gf}^2]_i^j\right)\right].
\end{align}
The trace is over internal indexes, position and momentum, and for the full one loop corrections one adds over fermions and gauge bosons. This trace can now be made  specific to thermal (with periodic time of interval $1/T$) and $T=0$ corrections as (with momentum in Euclidean space)
\begin{align}
V_{1-\textrm{loop}}&=\frac12 \int \frac{d^4\ell}{(2\pi)^4}\mbox{tr}\left[\log\left(\ell^2+\nabla^2V+\xi m_{gf}^2\right)\right],\\
    V_{\textrm{Th}}&=\frac{T}2\sum_n \int\frac{d^3\ell}{(2\pi)^3}\mbox{tr}\left[\log\left(E_n^2+\ell^2+\nabla^2V+\xi m_{gf}^2\right)\right],
\end{align}
where the trace is now over scalar indexes, and the 0 components of the momenta are the Matsubara frequencies \cite{Quiros:1999jp} in $V_{\textrm{Th}}$. While the first term requires counterterms, the second only has a field independent divergence that one can leave aside. Let us renormalise the corrections in vacuo next by dividing our effective potential into
\begin{align}
V_{\textrm{eff}}(h,T)&=V(h)+\Delta V_{CW}(h)+\Delta V_{\textrm{Th}}(h,T),
\end{align}
where the one loop correction at $T=0$, $\Delta V_{CW}$ has the form, with cut off regularisation,
\begin{align}
\Delta V_{CW}&=\sum_i\left[\Delta V_{\textrm{1-loop},i}+\Delta V_{\textrm{1-loop}.i}^{c.t}\right],\\
    \Delta V_{\textrm{1-loop},i}&=\frac{1}{64\pi^2}n_i\left(m_i^4\left(\log(m_i^2/\Lambda^2)-\frac12\right)+2\Lambda^2 m_i^2\right),
\end{align}
where $i$ runs over $\Phi$ scalars, electroweak gauge bosons, ghosts and fermions, for the latter for practical purposes, as all other fermions contribute negligibly, only the top. There are three types of masses as far as the field dependence is concerned which will dictate the number of counterterm operators needed
\begin{align}
    %m_I^2&&=
    &F^2(h), & &V''(h), & &\frac{V'(h)F'(h)}{F(h)}.
\end{align}
The counterterms needed are of the form either of the three above or their square and due to functional dependence similarities they amount to 6 terms; 4 could be taken as the tree level terms present in the Higgs potential and the other two from $(VF'/F)$ are rational rather than polynomial and require an additional two counterterms. Equivalently here we take
\begin{align}
    V^{c.t}_{\textrm{1-loop},i}=&a_im_i^4(h)+b_im_i^2(h), & &\textrm{Counter-terms}
\end{align}
Which are fixed by the renormalisation conditions here imposed as
\begin{align}
&   \left.\frac{dV_{CW,h}}{dh}\right|_{h=0}=\left.\frac{d^2V_{CW,h}}{dh^2}\right|_{h=0}=
   \left.\frac{dV_{CW,\varphi}}{dh}\right|_{h=0}=\left.\frac{d^2V_{CW,\varphi}}{dh^2}\right|_{h=0}=0,\\
 &  \sum_{A_B,t}\left.\frac{dV_{CW,i}}{dh}\right|_{h=0}=\sum_{A_B,t}\left.\frac{d^2V_{CW,i}}{dh^2}\right|_{h=0}=0.&\label{RenCond}
\end{align}
These imply that $h=0$ will stay a minimum of the potential with mass $m_h$.
The vanishing of the field dependent mass for $\varphi$ conflicts with these renormalisation conditions in Landau's gauge. While one can, as in Ref.~\cite{Elias-Miro:2014pca}, approach this problem by resummation, here we avoid this gauge and rather in the following select for concreteness {\it Feynman's gauge}, $\xi=1$. 

The thermal contributions on the other hand are
\begin{align}
\Delta V_{\textrm{Th}}&=\frac{T^4}{2\pi^2}\sum_i n_iJ_{S_i}\left(\frac{m_i(h)^2}{T^2}\right),
\end{align}
with $n_i$ the number of degrees of freedom and $S_i$ identifying the statistics, Bose-Einstein ($b$),  or Pauli-Dirac ($f$); explicitly
\begin{align}
    J_{b/f}(x^2)=\int_0^\infty y^2 d y \log\left(1\mp\exp{\left[-\sqrt{x^2+y^2}\right]}\right).
\end{align}

All the above leads to the 1-loop finite temperature potential  
\begin{align}\label{VeffUs}
    V_{\textrm{eff}}=V(h)+\sum_i n_i\left\{\frac{m_i^2(h)}{64\pi^2}\left( m_i^2(h)\left[\log\left(\frac{m_i^2(h)}{m_i^2(0)}\right)-\frac 32\right] +2m_i^2(0)\right)+\frac{T^4}{2\pi^2}J_{S_i}\left(\frac{m_i^2(h)}{T^2}\right)\right\}.
\end{align}
Each term in this sum, in our Feynman gauge, is \begin{align}\label{eqn1}
   m_t^2(h)&= m^2_t(0) F(h)^2, & n_t&=-12, & m^2_h(h)&=V''(h), & n_h&=1, \\
   m^2_{W_T}(h)&=m^2_W(0)F(h)^2, & n_{W_T}&=4, &
   m^2_{W_L}(h)&=\frac{F'V'}{F} +m^2_{W_T}(h), & n_{W_L}&=2,\\
   m^2_{Z_T}(h)&=m^2_Z(0)F(h)^2, & n_{Z_T}&=2, & m^2_{Z_L}(h)&=\frac{F'V'}{F} +m^2_{Z_T}(h), & n_{Z_L}&=1.\label{eqn3}
\end{align}

In the following, to avoid cluttered notation, a mass with no argument ($m_W$) is to be understood as a constant, the mass measured at the vacuum; for a field-dependent mass the dependence will be made explicit, e.g. $m_W(\phi)$.
It is useful to extend the extrema defined in Sec.~\ref{sec:Classical} to be the extrema of the finite temperature effective potential, $h_{\pm}(T)$, $h_0(T)$ so that the end-point of the temperature `trajetory' returns $h_0(0)=0$,  $h_{\pm}(0)=-v_{\pm}$. On the other hand, potential differences and barriers read
\begin{align} \label{eq:DeV}
    \Delta V(T)&=V_{\textrm{eff}}(h_-(T),T)-V_{\textrm{eff}}(h_0(T),T), \\  \label{eq:UDef}
    U_0(T)&=V_{\textrm{eff}}(h_+(T),T)-V_{\textrm{eff}}(h_0(T),T), \\ U_-(T)&=V_{\textrm{eff}}(h_+(T),T)-V_{\textrm{eff}}(h_-(T),T).
\end{align}

At high temperatures, given the importance of electroweak particle corrections, the relevant point in field space is to be found around $h=-v_\star$ rather than our vacuum at $h=0$. For this reason, we will write in the following the potential as a function of $\phi$ as in Eq.~\eqref{eq:phidef} where e.g.
\begin{align}
\phi_{\pm}&=h_\pm+v_\star\,,& \phi_{0}&=h_0+v_\star\,.
\end{align}

\subsection{Limitations of the calculation}
\label{sec:ir-problem}

Were one to aim at more precision, the present approximation can be improved by higher zero temperature or thermal loop corrections. These however suffer from an infrared illness as outlined in the beginning of this section. This problem is identified when the expansion parameter controlling the IR divergences
\begin{align}\label{eq:IRparameter}
     \varepsilon_{IR}\equiv\frac{g^2T}{\pi m_{W}(\phi)},
\end{align}
 ceases to be small.\footnote{A similar effective expansion parameter to Eq.~\eqref{eq:IRparameter} exists for both the $Z$ and $h$ bosons.} While some amelioration can be provided by daisy resummation~\cite{Curtin:2016urg}, the `magnetic modes' stay massless in perturbation theory and make these divergences unavoidable~\cite{Linde:1980ts}. Whenever the expansion parameter is larger than one, we simply cannot trust perturbation theory and must resort instead to non-perturbative methods in the form of lattice.
 Luckily, we do not need access to this non-perturbative region of the parameter space to unveil a range of novel phenomenology.
We address the IR problem by identifying the region of the field values $\phi$ that possess a controlled expansion $\varepsilon_{IR}(\phi)<1$, and only there do we trust the analytically and perturbatively computed potential.
The expression that determines this region in $\phi$ reads
\begin{align}    \left(\frac{\phi}{T}\right)^2&>\gamma_a^{-2}\left[\left(\frac{2g}{c_\chi\pi}\right)^2-\tan^2\chi\frac{v^2}{T^2}\right],
    \qquad \boxed{\textrm{IR Bound}}\label{eq:IRB}
\end{align}
where
we note that for temperatures below $T_{IR}=\pi s_\chi v/(2g)$, the RHS is negative and this constraint disappears for a controlled expansion in the entire range of $\phi$.
Above $T_{IR}$ nonetheless the bound applies, and it implies that for small values of $\phi$ the computation of the potential is unreliable. 

Let us exemplify how this IR bound is used here in practice by taking the SM as a case study. The SM potential for high temperature is a 4th degree polynomial in the fields and the extrema can be solved for analytically, for a pedagogical exposition see Sec.~4.1 of~\cite{Quiros:1999jp}. One finds that a minimum-maximum pair appears at a temperature $T_1$ as an inflection point at a distance 
$\phi_0(T_1)/T_1=3(2m_W^3+m_Z^3)/(4\pi\lambda v^3) \propto 1/m_h^2$ from the origin. This minimum becomes degenerate in potential energy with the origin at the critical temperature $T_c$ and sits at a distance $\phi_0(T_c)/T_c=(4/3)\phi_0(T_1)/T_1$ from it. After $T_c$, we would naively expect to have a first order phase transition. For this perturbative argument to be trustworthy, however, one should require the new minimum to be in the region $\varepsilon_{IR}(\phi_0)<1$. The region $\varepsilon_{IR}(\phi)>1$ is an interval around the origin ($\phi=0$) which shrinks with time and disappears at $T_{IR}$, whereas the distance of $\phi_0$ to $\phi=0$ generally increases monotonically.
It suffices then to specify the earliest time/highest temperature at which this condition is imposed: if satisfied at a temperature $T$, lower temperatures will continue to respect the bound. We distinguish two possibilities:
 \begin{itemize}
     \item {\bf Weak IR constraint.} 
 The new minimum should be at a position where the IR expansion is under control at the critical temperature, that is
\begin{align}
\varepsilon_{IR}(\phi_0(T_c))<&1,\label{eq:ir-weak}\end{align}
which when applied to the SM returns
\begin{align}\textrm{SM:}\quad\frac{g}{\pi}\leq \frac{(2m_W^3+m_Z^3)}{2\pi\lambda v^3},\qquad\qquad m_h\leq\sqrt{\frac{(2m_W^3+m_Z^3)}{2 g v^3}}v\simeq  75\text{ GeV},
\end{align}
which translates into an upper bound on the Higgs mass for a reliable first order phase transition prediction, a well known qualitative condition here made quantitative as outlined. 
\item {\bf Strong IR constraint.} The new minimum should be at a position where the IR expansion is under control already at the temperature when it first appears, that is
\begin{align}
    \varepsilon(\phi_0(T_1))<&1,
    \label{eq:ir-strong}
\end{align}
which applied to the SM gives
\begin{align}
   \textrm{SM:}\quad \frac{g}{\pi}\leq \frac{3(2m_W^3+m_Z^3)}{8\pi\lambda v^3}, \qquad\qquad m_h\leq\sqrt{\frac{3(2m_W^3+m_Z^3)}{8 g v^3}}v\simeq  65\text{ GeV}.
\end{align}
This a stronger demand and as such demands a smaller allowed ranged of the Higgs field.
 \end{itemize}

None of these bounds, however, are satisfied in the SM given the measured Higgs mass, and one cannot affirm there is a first order phase transition; in fact  lattice computations reveal that the SM with the measured couplings presents instead a crossover transition~\cite{Csikor:1998eu,PhysRevD.93.025003}.
 The bounds we obtain are not far from the actual value for the first order phase transition (1OPT) endpoint as determined by lattice $72\pm 7$~GeV~\cite{Laine:1998jb}
 and are also in line with more explicit estimates of the loop expansion breakdown \cite{Arnold:1992rz}.

\begin{figure}[h!]
    \centering
    \begin{tikzpicture}
    \draw [<->] (0,-1.5) --(0,1.5);
    \draw [<->] (3,-1.5) --(3,1.5);
    \draw [thick,blue] (-0.25,0) node[anchor=east] {$h_\star$} ..controls (0.5,-0.1) and (2,-0.8) .. (3.25,-1) node[anchor=west] {$h_-$} ;
    \draw [thick, ForestGreen] (1.5,0.45) ..controls (1.6,0.25) and (2,0.15) ..(3.25,-0.05) node[anchor=west] {$h_+$} ;
    \draw [thick, blue] (1.5,0.45) ..controls (1.6,0.65) and (2,0.85) .. (3.25,0.95) node[anchor=west] {$h_0$} ;
    \draw (1.5,1.75) node {Strong IR constraint compliant};
    \draw [thick,red,fill=red,opacity=0.5] (-0.25,0.75) ..controls (3,0.1) and (3,-0.1) .. (-0.25,-0.75);
\end{tikzpicture}\qquad\qquad\quad
\begin{tikzpicture}
    \draw [<->] (0,-1.5) --(0,1.5);
    \draw [<->] (3,-1.5) --(3,1.5);
    \draw [thick,blue] (-0.25,0) node[anchor=east] {$h_\star$} ..controls (0.5,-0.1) and (2,-0.8) .. (3.25,-1) node[anchor=west] {$h_-$} ;
    \draw [thick, ForestGreen] (1.5,0.3) ..controls (1.6,0.1) and (2,0) ..(3.25,-0.2) node[anchor=west] {$h_+$} ;
    \draw [thick, blue] (1.5,0.3) ..controls (1.6,0.5) and (2,0.7) .. (3.25,0.8) node[anchor=west] {$h_0$} ;
    \draw (1.5,1.75) node {Strong IR constraint non-compliant};
    \draw [thick,red,fill=red,opacity=0.5] (-0.25,0.75) ..controls (3,0.1) and (3,-0.1) .. (-0.25,-0.75);
\end{tikzpicture}
    \caption{Schematic representation of thermal histories for the extrema of the effective potential satisfying (left) and violating (right) the strong IR constraint.}
    \label{fig:IRB}
\end{figure}
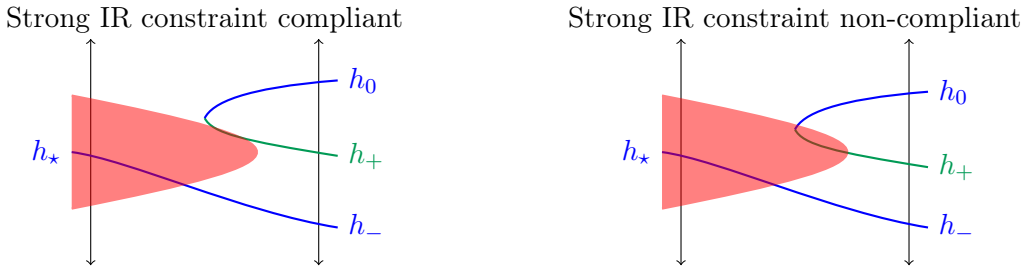

In our scenario with an extended range for $\phi$ the application is analogous, and we abstain from making predictions at temperatures where the minima fall in the $\varepsilon_{IR}>1$ region. To be explicit and consistent across cases, \textit{we demand that in a 1OPT, the minimum $\phi_0$ satisfies $\varepsilon_{IR}(\phi_0)<1$ at the critical temperature, $T_c$, (weak constraint) or the temperature when the minima first appears, $T_1$ (strong constraint).}
The diagrams for the histories of extrema allow for a visualisation of the strong IR constraint, as shown in Fig.~\ref{fig:IRB}.

Lastly, the large field regime of the $T=0$ potential might suffer from the same instability as the SM~\cite{Degrassi:2012ry} but in some instances is aggravated. One has, in the theory under study 
\begin{align} \nonumber
    V_{\text{eff}}(\phi\gg v)&\simeq \frac{\lambda\gamma_4^2}{8}\phi^4\left(1+\frac{3L(\phi)}{8\pi^2\gamma_4^2}\left[\frac{2m_W^4+m_Z^4-4m_t^4}{m_h^2v^2}\gamma_a^4c_\chi^4+\frac{m_h^2}{v^2}\gamma_4^4+\frac{2m_W^2+m_Z^2}{3v^2}\gamma_a^2c_\chi^2\gamma_4^2\right]\right),\\&\label{eq:BoundedfB} \equiv\frac{\lambda \gamma_4^2}{8}\phi^4\left(1+a_{\rm loop}L(\phi)\right)
\end{align}
with $L(\phi)=\log(\phi^2/v^2)$. In the SM case, top quark contributions drive the effective quartic coupling towards negative values and the electroweak minimum is metastable when field values are extended to the Planck mass. In the present case we will address the same problem with two prescriptions, {\it perturbativity} and {\it boundedness-from-below}. 

The constraint that we denote perturbativity follows from demanding a loop correction subleading to the tree level one;\footnote{Please note this constraint differs from unitarity bounds since we demand a convergent loop expansion rather than an effective coupling value below the unitarity limit, which in this study would be a subdominant constraint to perturbativity as here defined.} given the dependence in Eq.~\eqref{eq:BoundedfB} this demand is strongest at the edges of the field range, and it explicitly reads $a_{\rm loop}L(4\pi v)<1$. These edges should be chosen to capture a large enough domain of the potential to resolve the features of both minima, which roughly fall around $-v\leq \phi\leq v$. We require that this constraint holds instead over the field range $-4\pi v \leq \phi \leq 4\pi v$, about an order of magnitude larger in field space, which should safely capture the important features and ensure the probability of tunneling in some destabilized direction outside this range is considerably suppressed.

 A related but independent constraint arises from demanding the potential be bounded from below in the range we consider, which again we choose to be field values within $4\pi v$ of the origin $\phi=0$. To be conservative, we required that the potential does not start to turn downwards at high field values. We first checked whether the potential was concave up or concave down at the $\pm 4\pi v$ boundaries. A concave up potential passes the test. If the potential was concave down at the boundary, however, we required that the potential was increasing (decreasing) with $\phi$ at the positive (negative) $4\pi v$ boundary.
Lastly, we note that the 1-loop contribution to Eq.~\eqref{VeffUs} is complex for some values of the field value $\phi$ --e.g.  when $V''$ changes sign for a concave to convex tree-level potential. However it is the real part of the effective potential that is relevant for phase transitions \cite{Delaunay:2007wb, Weinberg:1987vp}, given  the imaginary component of $\Delta V_{CW}$ is very small relative to the real part, which is found to hold in all cases studied here. It is therefore the real part of the effective potential which will be considered in the following for numerical computations.

%%%%%%%%%%%%%%%%%%%%%%%%%%%%%%%%%%%%%%%%
\subsection{Symmetry (non)restoration and roads to the SM}
\label{sec:HighT}

The study of phenomenology will make use of numerical methods guided by analytical estimates; here we present some of the latter. Much in the way of the well studied electroweak phase transition in the SM for small Higgs mass, here a high temperature expansion helps draw the features of the problem and provide an understanding of the underlying dynamics. In addition, it also sheds some light on the question of limits in non-linear theories that might lead to the SM. 

At high temperatures, the thermal corrections can be approximated by polynomials in the masses provided we are in the regime~\cite{Laine:2016hma}:
\begin{align}\label{HTA}
 \frac{m_{i}(\phi)}{\pi T}&<1,
\end{align}
and hence this high temperature approximation is {\it valid} for a neighbourhood around the point $\phi=0$ which shrinks with decreasing temperature. We note that~(\ref{eq:IRB})  {\it excludes} a neighbourhood around the origin as opposed to the above; these two exclusion regions fortunately do not overlap.

The field coordinate $\phi=h+v_\star$ as outlined above is convenient to write thermal corrections in terms of which
\begin{align}
V_{\textrm{eff}}^{\textrm{HT}}=&G_o v^3\phi -D_o v^2\phi^2+E_ov\phi^3+\frac{\lambda \gamma_4^2}{8}\phi^4\\\nonumber
&+\sum_{B}n_i\left(\frac{m^2_i(\phi)}{8\pi^2}\left[\frac{m_{i}^2}{4}+\frac{\pi^2T^2}{3}\right]-\frac{T(m_i^2(\phi))^{3/2}}{12\pi}-\frac{m_i^4(\phi)}{64\pi^2}\log\left(\frac{m_{i}^2}{T^2A_B}\right)\right)\\ \nonumber
&+\sum_{F}n_i\left(\frac{m^2_i(\phi)}{16\pi^2}\left[\frac{m_{i}^2}{2}-\frac{\pi^2T^2}{3}\right]-\frac{m_i^4(\phi)}{64\pi^2}\log\left(\frac{m_{i}^2}{T^2A_F}\right)\right),
\end{align}
where  keeping up with our notation masses with no arguments are constants, the values measured at the vacuum, $n_i$ are given in Eqs.~\eqref{eqn1}--\eqref{eqn3} whereas
\begin{align}
    \log(A_B)&=5.4076-3/2, & \log(A_F)&=2.6351-3/2,
 \end{align}\begin{align}G_o   &=-\frac{\lambda \delta \gamma_\epsilon^2 }{2}
            (1-\gamma_\epsilon^{-1}\gamma_4\delta)
            (1+\gamma_\epsilon^{-1}\gamma_4\delta+3((1-\epsilon)\gamma_\epsilon^{-1}-1)) \label{eq:Go}\\ 
        &=-\frac{\lambda\gamma_4\gamma_a^{-1}\gamma_\epsilon^{-1}}{2}   
            (\gamma_4^{-1}\gamma_\epsilon-\gamma_a^{-1})(2-\gamma_\epsilon\gamma_a^{-1}\gamma_4),
\end{align}
\begin{align}
    D_o=&\frac{\lambda  }{4}\left(6(1-\epsilon)\gamma_4 \gamma_a^{-1}-3(\gamma_a^{-1}\gamma_4)^2-2\right),
\\
    E_o=&-\frac{\lambda}{2}\gamma_4\gamma_a^{-1}(\gamma_4-\gamma_a(1-\epsilon)).\label{eq:Eo}
\end{align}

It is useful for phenomenological purposes to highlight the terms that break Higgs parity. Couplings to the electroweak bosons and top quark are symmetric so that all asymmetry is sourced by the potential either at tree level, or through the effective masses of Goldstones and the Higgs.  Note that, as in the SM case, a cubic term would arise from thermal loops of gauge bosons, but this would be parity symmetric as $(\phi^2)^{3/2}$.  All sources of parity violation are turned off for $\epsilon$ and $\delta$ vanishing which aligns the maximum of the tree level potential with the throat of the manifold and sets the energy difference between minima to zero. 
This limit of vanishing $\delta$ and $\epsilon$ simplifies eqs.~\eqref{eq:Go}--\eqref{eq:Eo} so that the coefficients of parity violating terms are linear in $\epsilon$ and $\delta$ as $G_o\simeq -\lambda \delta/2$ and  $E_o\simeq\lambda\delta\gamma_4^2/2-2\lambda\gamma_4\epsilon$.
\begin{figure}
    \centering

    \begin{subfigure}[b]{0.475\textwidth}
         \centering
         \includegraphics[width=\textwidth]{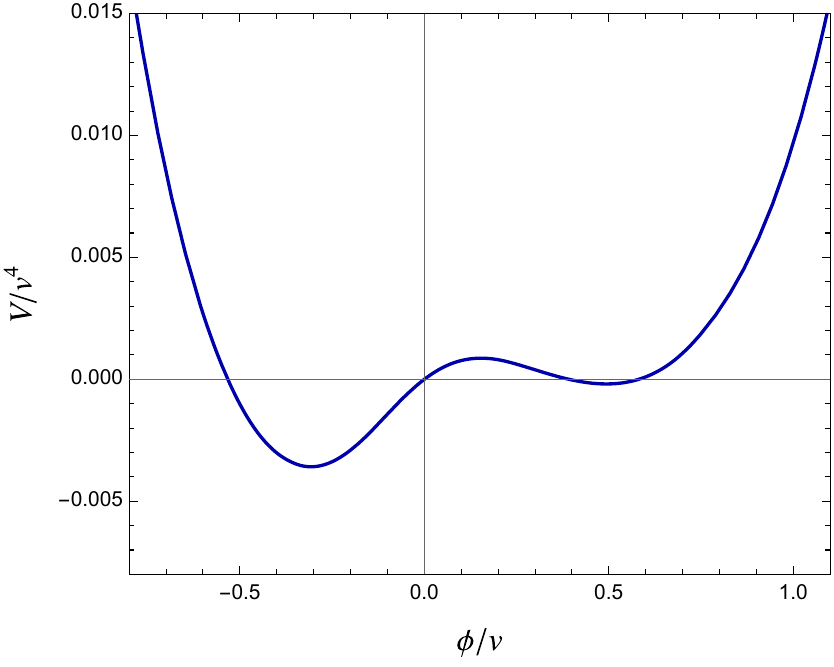}
         \caption{$T=246$ GeV}
        \label{fig:sym rest Tbig}
     \end{subfigure}
     \hfill
     \begin{subfigure}[b]{0.475\textwidth}
         \centering
         \includegraphics[width=\textwidth]{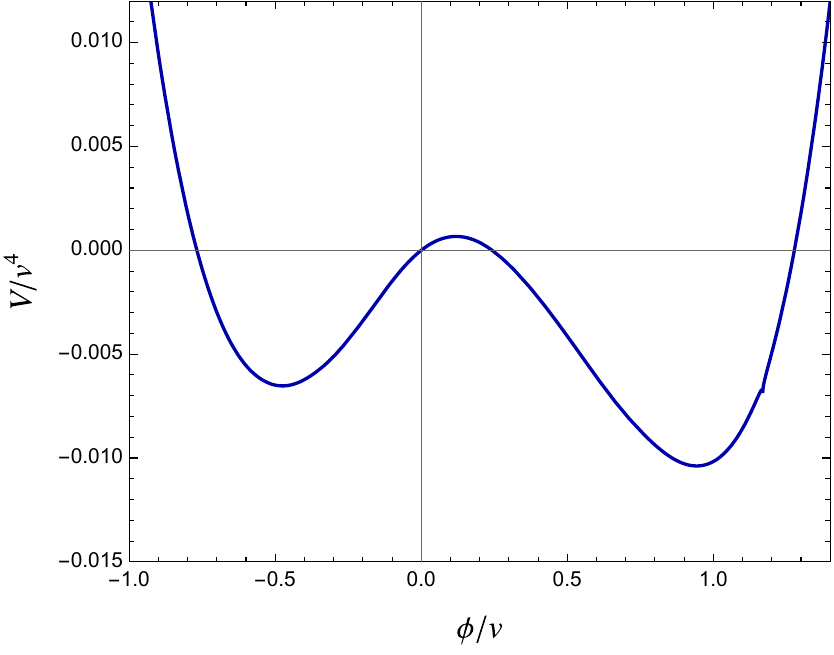}
         \caption{$T=190$ GeV}
        \label{fig:sym rest Tsmall}
     \end{subfigure}
    \caption{One-loop effective potential plotted for $\chi=\sqrt{0.1}$, obeying the high-temperature symmetry restoration bound in Eq.~\eqref{eq:betaL}. The benchmarks chosen here are $\gamma_4=1,\ \epsilon=0.02,\ \delta=-0.1$. Fig.~\ref{fig:sym rest Tbig} and Fig.~\ref{fig:sym rest Tsmall} show the potential evolution with temperature for $T=246$~GeV and $T=190$~GeV, respectively. Notice, in particular, the large $\chi$ smooths the potential around $\phi/v=0$ in contrast to Fig.~\ref{fig:Topo2}.
    }
        \label{fig:Topo1}
    \end{figure}

    \begin{figure}
    \centering

    \begin{subfigure}[b]{0.475\textwidth}
         \centering
         \includegraphics[width=\textwidth]
         {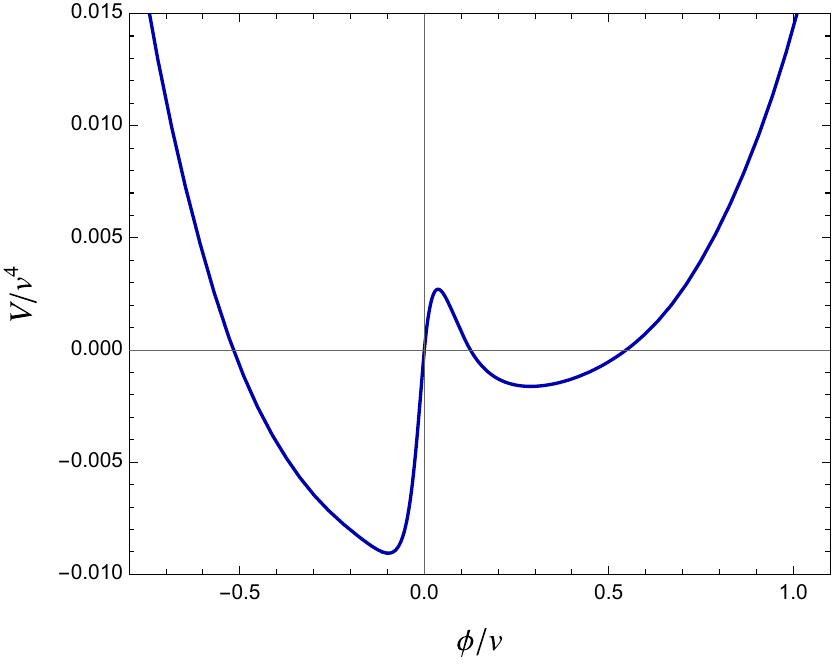}
         \caption{$T=246$ GeV}
        \label{fig:no sym rest Tbig}
     \end{subfigure}
     \hfill
     \begin{subfigure}[b]{0.475\textwidth}
         \centering
         \includegraphics[width=\textwidth]{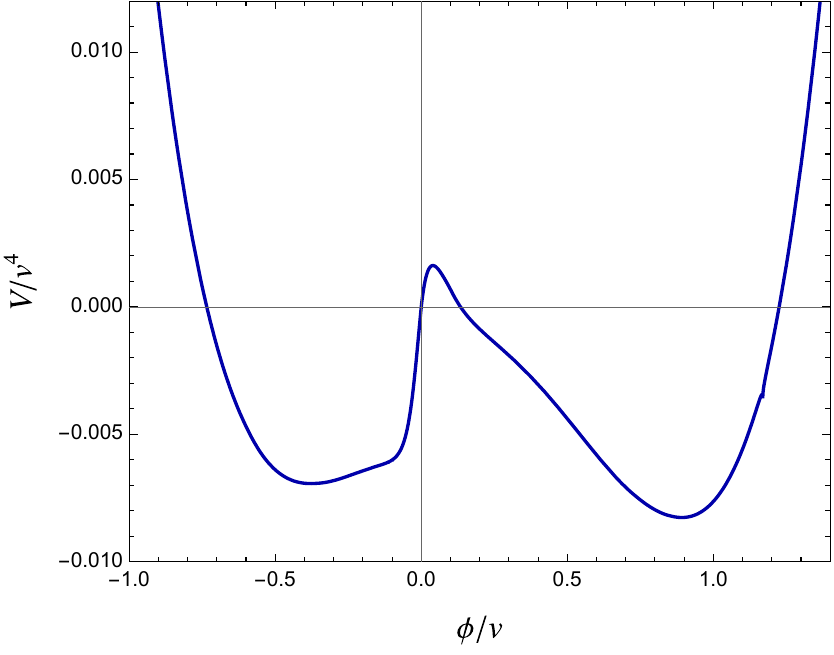}
         \caption{$T=190$ GeV}
        \label{fig:no sym rest Tsmall}
     \end{subfigure}
    \caption{One-loop effective potential plotted for $\chi=0.05$, which unlike Fig.~\ref{fig:Topo1}, does not satisfy the high-temperature symmetry restoration bound in Eq.~\eqref{eq:betaL}. The benchmarks chosen here are $\gamma_4=1,\ \epsilon=0.02,$ and $\delta=-0.1$. Fig.~\ref{fig:no sym rest Tbig} and Fig.~\ref{fig:no sym rest Tsmall} show the potential evolution with temperature for $T=246$~GeV and $T=190$~GeV respectively. For this choice of $\chi$, the potential is sharply peaked around $\phi/v=0$ in contrast to Fig.~\ref{fig:Topo1}.}
        \label{fig:Topo2}
    \end{figure}
    
While the high energy expansion yields a potential as a  polynomial in the fields and is amenable to analysis in SMEFT, in our non-linear theory the potential is instead a rational function of the fields due to the Goldstones contribution, $V'F'/F\propto (c_\chi^2\gamma_a^2\phi^2+s_\chi^2)^{-1}$.
Naively recovering a polynomial, as in the SM and SMEFT case, would require taking $\chi\to0$ while having the numerator polynomial start at $\phi^2$ to avoid poles, attainable by $\delta\to0$. Such a limit, in the direction of recovering the SM, is in fact a discontinuous one in what is a distinguishing feature of this non-linear theory. This feature is hinted at in the small $\chi$ limit with $\delta\neq0$, where plots of the potential in Fig.~\ref{fig:Topo2} show a sharp rise around the throat ($\phi=0$).\footnote{We have checked that the sharp features of this potential around the origin come primarily from the shape of the manifold. The imaginary parts of the potential are still much smaller than the real parts here.}  

This feature can be understood analytically if we inspect the very high temperature potential while still keeping in mind $T<4\pi v$:
\begin{align}
    V_{\textrm{eff}}^{\textrm{HT}}(\phi)& =\sum_{B} n_{i}\frac{T^2m_i^2(\phi)}{24}-\sum_F n_i\frac{T^2m_i^2(\phi)}{48}+\mathcal O(T^0)\\&=\frac{T^2}{24}\left((6 m_{W}^2+3m_{Z}^2+6 m_{t}^2)F^2(\phi)+V''(\phi)+3V'(\phi)\frac{F'(\phi)}{F(\phi)}\right)+\mathcal O(T^0)\\
    &=V_{\textrm{eff}}^{\textrm{HT}\,\prime}(0)\phi+\frac{V_{\textrm{eff}}^{\textrm{HT}\,\prime\prime}(0)}{2}\phi^2+\mathcal O(\phi^3,T^0) \label{eq:VeryHT},
\end{align}
with
\begin{align}\label{VpHT}
    V_{\textrm{eff}}^{\textrm{HT}\,\prime}(0)&=T^2 v \left(\frac{E_o}{4}+\frac{G_oc_\chi^2\gamma_a^2}{8s_\chi^2}\right),\\
    \frac{V_{\textrm{eff}}^{\textrm{HT}\,\prime\prime}(0)}{2}&=T^2\left(\frac{2m_W^2+m_Z^2+2m_t^2}{8v^2}c_\chi^2\gamma_a^2+\frac{\lambda\gamma_4^2}{16}-\frac{D_oc_\chi^2\gamma_a^2}{4s_\chi^2}\right).
\end{align}
 This expansion around the Higgs parity symmetric field value $\phi=0$ allows one to discuss symmetry restoration in analytic but approximate terms as follows. 
\begin{itemize}
    \item $V_{\textrm{eff}}^{\textrm{HT}\,\prime}(0)\neq 0$. 
The linear term in $\phi$ breaks Higgs parity and will prevent $\phi=0$ from being an extremum, its sign for small $\delta,\epsilon$  determining if this extremum has shifted to positive or negative values. 
 Assuming $V''_{\rm eff}>0$ and that the minimum is in a neighbourhood of the throat ($\phi=0$), one has a minimum shifted to negative $\phi$ values for positive slope, a condition expressed in terms of the model parameters as when subsituting in Eq.~\eqref{VpHT} as
\begin{align}
    -\frac{\lambda}{8}\gamma_4\gamma_a^{-1}(\gamma_4-\gamma_a(1-\epsilon))-\frac{\lambda c_\chi^2\gamma_4\gamma_a\gamma_\epsilon^{-1}}{16s_\chi^2}   
            (\gamma_4^{-1}\gamma_\epsilon-\gamma_a^{-1})(2-\gamma_\epsilon\gamma_a^{-1}\gamma_4)>0,
\end{align}
 and the opposite inequality for a minimum in the positive $\phi$ axis.
 In the limit of small $\epsilon$ and $\delta$ we can simplify the expression to derive
  \begin{align} \label{eq:HistParm}
    \delta\frac{c_\chi^2-2s_\chi^2}{16s_\chi^2}&<-\frac{\epsilon}{2\gamma_4} & &\begin{array}{c}
          \textrm{high T min at} \,\phi<0 \\
          \textrm{History Q}{}_-
    \end{array}\\
    \delta\frac{c_\chi^2-2s_\chi^2}{16s_\chi^2}&>-\frac{\epsilon}{2\gamma_4} & &\begin{array}{c}
         \textrm{high T min at}\, \phi>0  \\
          \textrm{History Q}{}_0
    \end{array}\\
    \delta\frac{c_\chi^2-2s_\chi^2}{16s_\chi^2}&\simeq-\frac{\epsilon}{2\gamma_4} & &\begin{array}{c}
         \textrm{high T min at}\, \phi\simeq0  \\
          \textrm{History P}
    \end{array}\,.
\end{align}
The equivalence with histories is not exact
given that this analytic approximation does only hold for small $\epsilon$ and $\delta$,  but it is useful in sketching the possibilities. This result divides the $(\epsilon, \delta)$ plane in half with a negative slope line that goes through the origin as in Fig.~\ref{TwoBinary}: above this line one would find history Q$_0$, below it Q$_-$ .
 \item  $V_{\textrm{eff}}^{\textrm{HT}\,\prime}(0)= 0$. 
 If the linear term vanishes, which is the case for $\epsilon,\delta\to 0$,
 $\phi=0$ is an extremum: a minimum for $V_{\textrm{eff}}^{\textrm{HT}\,\prime\prime}(0)>0$ and maximum for $V_{\textrm{eff}}^{\textrm{HT}\,\prime\prime}(0)<0$. The sign of the second term, $s_0$, will hence determine if Higgs parity is restored or not so
\begin{align}
    s_0\equiv\textrm{Sign}\big[&V_{\textrm{eff}}^{\textrm{HT}\,\prime\prime}(0)\big]\\=\textrm{Sign}\Big[&\frac{2m_W^2+m_Z^2+2m_t^2}{8}c_\chi^2\gamma_a^2\nonumber\\&-\frac{m_h^2\gamma_4^2}{16}\left(\gamma_a^2\gamma_4^{-2}t_\chi^{-2}\left(6(1-\epsilon)\gamma_4 \gamma_a^{-1}-3(\gamma_a^{-1}\gamma_4)^2-2\right)-1\right)\Big]\,,\label{eq:sign}
\end{align}
which to first order in an $\epsilon,\delta$ expansion returns 
\begin{align}\label{eq:betaL}
    s_0&\simeq\textrm{Sign}\left(\frac{(\sin 2\chi)^2}{\cos(2\chi)}-
    \frac{m_h^2}{m_W^2+m_t^2+m_Z^2/2}\right), \\ &\left\{\begin{array}{cl}
         s_0>0\Rightarrow\chi>0.3\,,& \textrm{Symmetry restoration at high }T \\ \nonumber
         s_0<0\Rightarrow\chi<0.3\,,& \textrm{Symmetry stays broken at high }T\,.
    \end{array}\right.
\end{align}
\end{itemize}
 The small $\chi$ limit leading to no symmetry restoration leads to an apparent contradiction: one expects this limit to yield SM-like couplings and with it the symmetry should naively be restored at high temperature. This finding sheds light on the question posed in Ref.~\cite{Alonso:2021rac} of whether non-linear theories have a limit in which the SM is recovered. Approaching SM-like couplings from our non-linear theory, we obtain a locally identical theory around $h=0$ which nonetheless does not have symmetry restoration at high temperature for any infinitesimally small value of $\chi$ since the theory is different at $\phi=0$. This discontinuous limit can be spotted opening up the term $V'F'/F$ in Eq.~(\ref{eq:VeryHT})
\begin{align}\label{finger}
       V_{\textrm{eff}}^{\textrm{HT}}&\supset V_{\rm peak} = - \frac{T^2D_o}{4}\frac{\phi^2}{\phi^2/v^2+s_\chi^2/(c_\chi^2\gamma_a^2)}
\end{align}

\begin{figure}[h]
    \centering
    \includegraphics[width=0.4\textwidth]{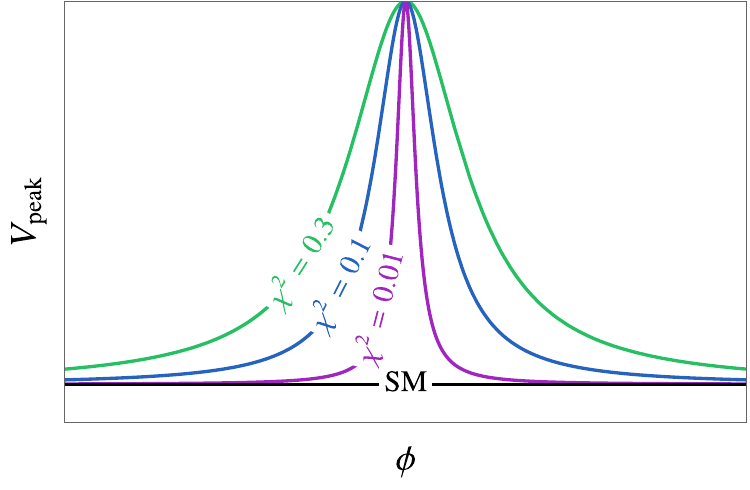}
    \caption{Contribution $V_{\rm peak}$ of the $V'F'/F$ term to the effective potential.}
    \label{fig:finger}
\end{figure}

This same limit of $\chi\to 0$ would give an ever smaller cut-off at the throat as the curvature increases as outlined in the discussion at the beginning of Sec.~\ref{sec:FiniteT}. New light states would be accessible in the thermal history which would change our EFT to a theory with extra degrees of freedom. This suggests that for a consistent theory, the functions in our manifold should be sufficiently smooth. The exploration of this interplay in the $\chi\to0$ limit and its phenomenology, however, we leave for future work on the connection of UV models and HEFT$\backslash$SMEFT.

Fig.~\ref{fig:finger} plots the contribution for different values of $\chi$ and the SM, and shows that at no finite stage is the SM limit (a constant for all field range, in black) reached. Sign of an anomalous limit is also present in the linear term of Eq.~\eqref{eq:VeryHT} since it reads in this expansion $G_o/s_\chi^2\sim \delta/\chi^2$, giving a different result depending on the order of limits that one takes to the SM. Unlike the behaviour in Eq.~\eqref{finger} however, the SM can be approached by first taking $\delta\to0$.

%%%%%%%%%%%%%%%%%%%%%%%%%%%%%%%%%%%%%%%%%%%%%%%
\section{Domain walls}\label{sec:Walls}

 The finite temperature effective potential is symmetric under Higgs parity in the limit $\epsilon,\delta\to0$. This limit contains couplings that are locally those of the SM yet the doubling of the Higgs domain implies two degenerate minima at low temperature. One has different possible phases for the discrete symmetry in the universe history and, although this theory might locally resemble the SM, the cosmological dynamics change drastically. As the most salient new possible feature, one has spontaneous breaking of Higgs parity and the creation of domain walls. The formation of domain walls in the Higgs field is, to the best of our knowledge, a unique feature of non-linear theories. In this section we describe and identify the conditions for this approximate symmetry to lead to walls with the assumption, validated below, that  $\delta$, $\epsilon$ are small enough to expand on.

Sec.~\ref{sec:HighT} showed that for $\chi>0.3$ and small $\epsilon$ and $\chi$ there will be symmetry restoration in the form of a single minimum at high energies, whereas the zero temperature potential presents two parity reflected minima. This well studied case of cosmological spontaneous breaking of discrete symmetry leads in general to the formation of walls and will be the focus of our study here.
The late time phenomenology of walls is quite insensitive to 
whether the transition is first or second order, but let us sketch how it occurs in both cases for illustration. For a first order PT, different patches in the universe will bubble to one of the two vacua, and after the bubbles collide and lose energy to the bath a network of walls will form, the characteristic size of this network dependent on the transition properties. 

For a second order PT, depicted in Fig.~\ref{fig:breakit1to3-P} (history P), shortly after the critical temperature the small separation and barrier between minima implies that
thermal fluctuations for correlation-volume ($l^3$ with $l^{-2}=V''(h_0)\simeq V''(h_-)$) patches allow for transitions from one vacuum to another. As temperature decreases, the probability of these jumps will decrease with the Boltzmann factor, $\sim e^{-l^3 U/T}$ where $U$ is the potential barrier given our approximate degeneracy assumption $U=U_0=U_-$ with $U_{0,-}$ in Eq.~\eqref{eq:UDef}, until this suppression is effective enough to in practice forbid the transitions. At this time patches of the correlation volume will be stuck in whichever vacuum they happened to be at, and the network of walls will form. Formation is marked by 
 Ginzburg's temperature defined as the temperature when the the negative of the exponent in Boltzmann's factor equals one,
\begin{align}
    T_G\equiv\left[V''_{\textrm{eff}}(h_0(T_G),T_G)\right]^{-3/2}U(T_G)\,.
\end{align}

This description allows for a qualitative picture of the symmetry non-restoration case for $\chi<0.3$. In this instance two minima survive until high temperature, but noting that the barrier scales like $v^2T^2$ and naively correlation the length is $T^{-1}$, one can expect that above $T\sim v$ the Boltzmann factor is large enough to allow for these jumps. If this is so, even without symmetry restoration, the Ginzburg temperature is well defined and one can expect formation of walls. This must be the case at least in a neighbourhood below $\chi=0.3$, but we leave its detailed study for future work.

 The definition of the Ginzburg temperature is implicit; in practice however it is well approximated by the zero temperature potential, in our case
\begin{align}
T_G\simeq \frac{\lambda v^4}{8\gamma_4^2m_h^3}\sim 60\gamma_4^{-2}\textrm{GeV},
\end{align}
where we have expanded on $\epsilon,\delta$ and kept only the first term. The probabilities $P_0$ and $P_-$ to find a patch of each vacua $h_0$ and $h_-$ at the Ginzburg temperature can then be estimated as
\begin{align}
    \frac{P_0}{P_-}\sim\exp \left(\frac{l^3 V_{\textrm{eff}}(h_-,T)}{T_G}-\frac{l^3 V_{\textrm{eff}}(h_0,T)}{T_G}\right)=
    \exp \left(\frac{\Delta V(T_G)}{U(T_G)}\right),
\end{align}
where $\Delta V $ is as defined in Eq.~\eqref{eq:DeV}. A large $\Delta V/U$ ratio would imply one vacuum is selected predominantly and very few walls form. We instead assume this ratio, which depends only on $\epsilon$, is small, which implies small $\epsilon$ according to Eq.~\eqref{eqSUV}. This translates into $\Delta V/U=32\epsilon\ll1$.

We note that regardless of how the network formed and the typical scale of structure, both types of phase transition have a large wall stretching out to the horizon. One can realise this in a 2D case by rolling a dice to fill each correlation patch inside a causal box with either of the two vacua. Once the filling is done, a zoom out to see the global structure will reveal the presence of this large wall.

Let us then continue taking $l$ as the typical scale of the network regardless of how it was formed.
 The small scale structure dynamics is ruled by the balance of pressure $p=\Delta V(T_G)$ and tension $\mu$.
The tension is well approximated by the zero temperature potential and to first non-vanishing order in small $\epsilon,\delta$
\begin{align}
	\mu=\int_{-\infty}^{\infty} d\ell \,T^{0}_{\,\,0}(h_{w}(\ell))=\int d\ell \left(\frac12(\nabla h_w)^2+V(h_w)\right)=\frac{2\sqrt{\lambda}v^3}{3\gamma_4^2},
\end{align}
where the profile function is found as a solution to the static field equations and reads $\gamma_4\sqrt{\lambda} h_w(\ell)/m_h=\tanh(m_h\ell)-1$.

The potential energy difference, Eq.~\eqref{eq:DeV}, vanishes in the parity symmetric limit and here it suffices to estimate it to linear order in $\delta$ and $\epsilon$. On top of this expansion, we also perform a loop expansion, with minima at $h_-=h_-^{(0)}+\kappa h_-^{(1)}$, $h_0=h_0^{(0)}+\kappa h_0^{(1)}$ and $\kappa=(4\pi)^{-2}$ to find the energy difference ($V_{\textrm{eff}}=V+V_\kappa$\,, $V_\kappa=V_{CW}+V_{\textrm{Th}}$)
\begin{align}
    \Delta V=&V(h_-^{(0)})+V_{\kappa}(h_-^{(0)})+\frac{d V (h_-^{(0)})}{dh}\kappa h_-^{(1)}+\mathcal O(\kappa^2)\\&-\left[V(h_0^{(0)})+V_{\kappa}(h_0^{(0)})+\frac{d V(h_0^{(0)})}{dh}\kappa h_0^{(1)}+\mathcal O(\kappa^2)\right]\\
    =&V(h_-^{(0)})+V_{\kappa}(h_-^{(0)})-[V(h_0^{(0)})+V_{\kappa}(h_0^{(0)})]+\mathcal{O}(\kappa^2),
\end{align}
where the derivative term cancels since it is evaluated at a minimum to the given approximation.
One can now use that $h_i^{(0)}$, the minima of the tree level potential, do not depend on $\delta$ and are the Higgs parity conjugate of one another to write
\begin{align}
  %  h_0^{(0)}&=h_s+\epsilon h_{a,0}^{(1)}+\mathcal{O}(\epsilon^2) & 
    h_-^{(0)}&=-2v\gamma_4^{-1}+\epsilon h_{a,-}^{(1)}+\mathcal{O}(\epsilon^2),
\end{align}
whereas we recall by definition and for all values of the tree level potential $h^{(0)}_0=0$. The expanded energy difference is then
\begin{align}
    \Delta V=&\epsilon\left.\left( \frac{\partial V(-2v\gamma_4^{-1})}{\partial\epsilon}-\frac{\partial V(0)}{\partial\epsilon}\right)\right|_{\epsilon=0}+\epsilon h_{a,-}^{(1)}\left.( \frac{d V(-2v\gamma_4^{-1})}{dh}\right|_{\epsilon=0}
    \\&
    +\delta\left.\left( \frac{\partial V_\kappa(-2v\gamma_4^{-1})}{\partial\delta}-\frac{\partial V_\kappa(0)}{\partial\delta}\right)\right|_{\epsilon,\delta=0},
\end{align}
where the potential derivative w.r.t. the field will cancel once more given it is evaluated at a minimum to the given approximation. The leading $\epsilon$ contribution comes at tree level and is straightforward to obtain. For $\delta$, a number of intermediate steps leads to
\begin{align}
 &\delta\left.\left(\frac{\partial}{\partial\delta}\left[ V_\kappa(-2v\gamma_4^{-1})-V_\kappa(0)\right]\right)\right|_{\delta,\epsilon=0}= \delta\sum_i\left[\frac{\partial m_i^2(-2v\gamma_4^{-1})}{\partial\delta}-\frac{\partial m_i^2(0)}{\partial\delta}\right]\frac{\partial V_{\kappa}}{\partial m_i^2}\\
=&\frac{T^2}{2\pi^2}\frac{\partial F^2(-2v\gamma_4^{-1})}{\partial\delta }\sum n_i m_i^2(0)  J'_i(m_i^2(0)/T^2)\\
=&\frac{6\gamma_4 \delta c_\chi^2T^2}{\pi^2}\left(2 m_W^2(0) J'_b\left(\frac{m_W^2(0)}{T^2}\right)+ m_Z^2(0) J'_b\left(\frac{m_Z^2(0)}{T^2}\right)-4 m_t^2(0) J'_f\left(\frac{m_t^2(0)}{T^2}\right)\right)\\
\equiv&\lambda v^4\gamma_4^{-2} \delta\left( b(x) x^{-2}\right),
 \end{align}
where $
x=v/T
$,
and we used that our renormalisation conditions of Eq.~\eqref{RenCond} imply the variation w.r.t. the mass vanishes so that the linear term in $\delta$ comes from the thermal $J$ functions, which are $F$-dependent (and $F$ is $\delta$-dependent).
The result is the potential energy difference  to leading order in $\epsilon,\delta$,
\begin{align}
    p_V(T)&=\Delta V(T)=
    4\lambda v^4\gamma_4^{-2}(\epsilon+b(x) x^{-2}\delta/4)+\mathcal{O}(\kappa\epsilon , \delta^2)\,.
\end{align}
 Wall dynamics will be determined by the balance of the volume pressure $p_V=\Delta V$ and an equivalent tension pressure $p_T=\mu/l$.
 Volume pressure $\Delta V$ domination at $T_G$ means the lowest energy vacuum patches will quickly expand against the other vacuum patches and make the $l$ size structure of walls disappear shortly after formation. The opposite case, approximately given by 
  $(\epsilon+b(x)x^{-2}\delta/4)<1/6$~\cite{Gelmini:1988sf},
 has that tension would drive dynamics together with the friction of the thermal bath. Indeed in vacuo, the tension would make the walls oscillate converting potential energy stored in tension into kinetic energy and back. The thermal bath however introduces friction with a pressure~$T^4v_w$ with $v_w$ the wall velocity which dissipates the energy of the walls efficiently back into the bath, some of it in the form of gravitational radiation. Both instances then lead to the disappearance of the $l$ size network shortly after formation.

 The dynamics of the large wall are dictated by the horizon rather than $l$ and one can translate the above discussion by $l^{-1}\to H$. As long as $ \mu H\geq p_V(T)$, structure of  size $H^{-1}$ will enter the horizon, oscillate and radiate as controlled by tension and friction leaving a potentially detectable trace. The opposite case wipes even this large wall away leaving virtually no trace. Focusing on detectable phenomenology, we restrict to the condition for long lived large walls, which is $p_T>p_V$ as outlined above, and reads explicitly  
 \begin{align}\label{HWallG}
 	p_T(H(T_G))=\mu H(T_G)&> p_V=\Delta V(T_{G}), \\ \left|\epsilon+b(x_G)x_G^{-2}\delta/4\right|&\lesssim\frac{x_G^{-2} v^2}{6m_h M_{\textrm{pl}}}\sqrt{\frac{4\pi^3g_\star}{45}},\label{HWallG2}
 \end{align}
where $x_G^{2}=v^2/T_G^2$. This condition can be met by a cancellation of $\epsilon$ and $\delta$, however, if they themselves are individually much larger than the RHS of the inequality above, the temperature dependence would mean that the condition will, shortly after $T_G$, not be satisfied.
For this reason we also impose a condition on the logarithmic derivative w.r.t. temperature of the LHS; explicitly the change after an e-fold variation in temperature should also be smaller than the RHS of Eq.~\eqref{HWallG2},
\begin{align}\left.\frac{d}{d\log x} \left(b(x) x^{-2}\delta/4\right)\right|_{x_G}\lesssim\frac{x_G^{-2} v^2}{6m_h M_{\textrm{pl}}}\sqrt{\frac{4\pi^3g_\star}{45}}.\label{eq:walldeB}
 \end{align}
 All the above conditions are met for $\epsilon=\delta=0$, a case in which the perfect degeneracy would make walls endure the history of the universe to be around today. If so, their effect must have been negligible in the cosmological evolution and in particular any anisotropic impact on the CMB less than one part in $10^5$. Given the energy density stored in walls is $\mu H$, the bound can be translated into the tension \cite{Gelmini:1988sf}
 \begin{align}\label{eq:tensionbound}
 	\mu H_0 < M_{\rm pl}^2 H_0^2 \quad\Rightarrow\quad \mu=\frac{2\sqrt{\lambda}vv_+^2}{3}=\frac{2\sqrt{\lambda}v^3}{3\gamma_4^2} \leq (0.1 \textrm{GeV})^3.
 \end{align} 
 Avoiding this bound would require smaller tension, attainable with larger $\gamma_4$, of order $\gamma_4\sim 10^{4}$, yet perturbativity from e.g. $h+h\to h+h$ scattering demands $\lambda\gamma_4^2<8\pi/3$, and this possibility is ruled out. The avenue of reducing the tension in our degenerate potential means bringing the vacua closer together, which is a modification at small field values only. As a result this conclusion is robust against higher dimensional operator insertions in the potential.

\begin{figure}[t]
    \centering

    \begin{subfigure}[b]{0.53\textwidth}
         \centering
          \includegraphics[width=\textwidth, clip=true, trim=.3cm .2cm .2cm 0cm]{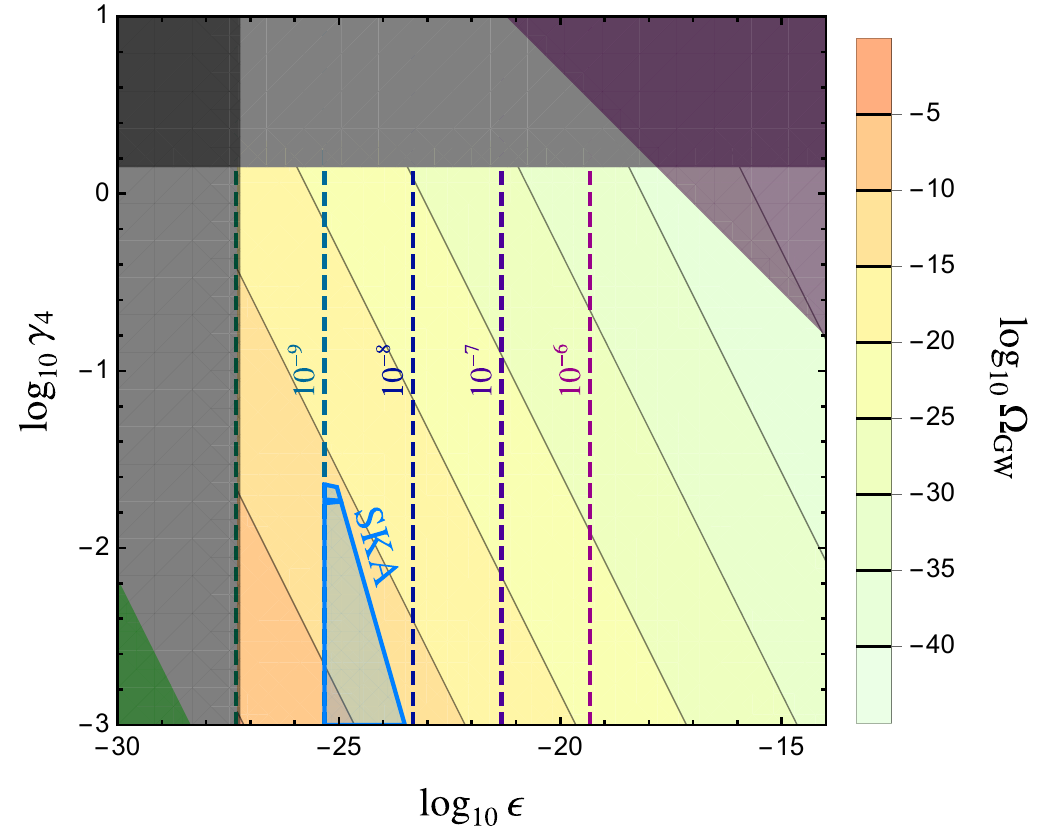}
         \caption{}
        \label{fig:DWPower}
     \end{subfigure}
     \hfill
     \begin{subfigure}[b]{0.44\textwidth}
         \centering
         \includegraphics[width=\textwidth, clip=true, trim=.1cm .1cm 0cm 0cm]{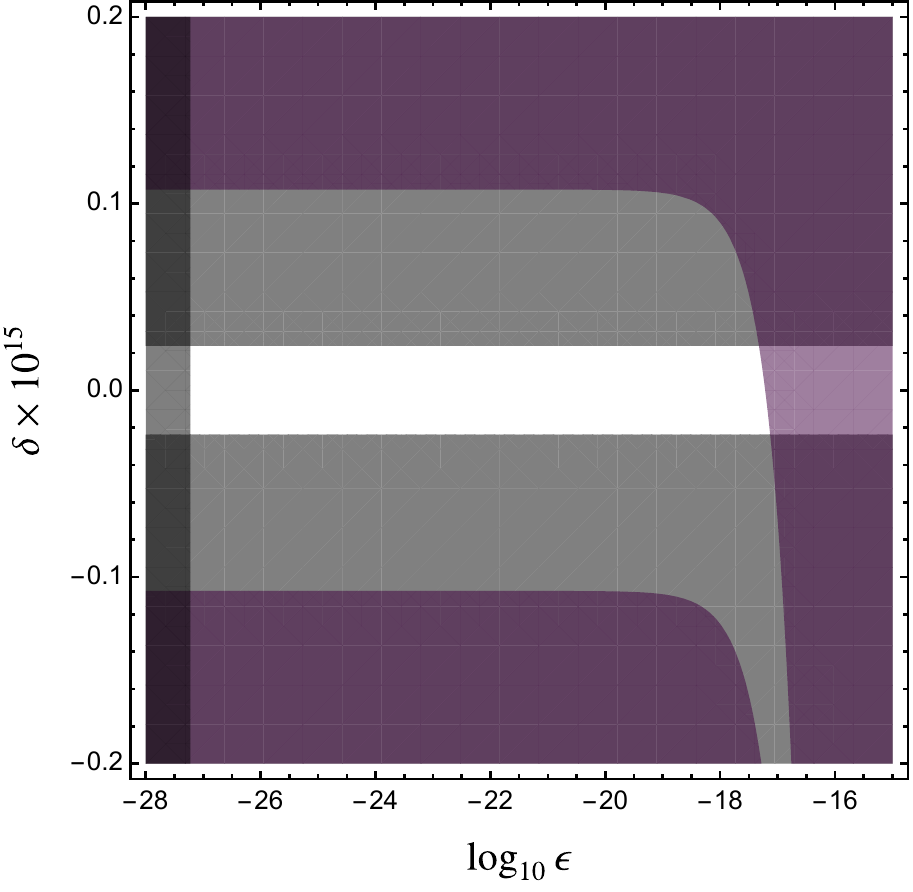}
         \caption{}
        \label{fig:DWdeeps}
     \end{subfigure}
    \caption{Fig.~\ref{fig:DWPower} shows the $(\epsilon, \gamma_4)$ plane with excluded regions in darker green (from the upper bound on walls energy density, Eq.~\eqref{eq:epsbd}), light gray horizontal (perturbativity, Eq.~\eqref{eq:BoundedfB}), dark gray (wall annihilation before BBN, \eqref{eq:WallBBN}). The region in dark purple is discarded if one is to have walls that survive long past the original phase transition (Eq.~\eqref{HWallG}). Also shown is the gravitational wave spectrum value at peak frequency $\Omega_{\rm GW}(f_{\rm peak})$ in the yellow gradient, see Eq.~\eqref{WallsGW} and Sec.~\ref{sec:GW}. Dashed lines for the peak frequencies run from $f_{\rm peak}=10^{-6}$~Hz on the far right to $10^{-10}$~Hz on the far left in intervals of $10$, and in blue the region SKA~\cite{Weltman:2018zrl} would be sensitive to (this region is however excluded by LHC bounds see Sec.~\ref{sec:LHC}). Fig.~\ref{fig:DWdeeps} shows the $(\epsilon, \delta)$ excluded by wall annihilation before BBN in dark gray vertical Eq.~\eqref{eq:WallBBN}, and the regions discarded for long lived walls following from Eq.~\eqref{HWallG2} (purple) and Eq.~\eqref{eq:walldeB} (gray horizontal).}
    \label{fig:wps}
 
\end{figure}

 Having ruled out the $\delta=\epsilon=0$ limit, we now turn to finite but small values of the Higgs parity breaking parameters. We note that the condition that ensures wall survival at the Ginzburg time, Eq.~\eqref{HWallG}, is more demanding if translated to later time and larger $x$;  for finite $\Delta V$ it will cease to be satisfied at  $T_w$, i.e.
 \begin{align}
 	\mu H_{w}\equiv p_V(T_w)=\Delta V(T_w)\quad \Rightarrow \quad T_w^2=\sqrt{\frac{45}{4\pi^3g_\star}}\frac{M_{\textrm{pl}}\Delta V}{\mu}=\sqrt{\frac{45}{4\pi^3g_\star}}6\epsilon m_hM_{\rm pl}.
 \end{align}
 Below this temperature, tension pressure gives way to vacuum pressure and we have that walls are swept away towards the horizon at a velocity close to the speed of light.
 
 A conservative bound~\cite{Lazanu:2015fua} is for this temperature to be higher than our earliest direct evidence of universe history, big  bang nucleo-synthesis. This constraint can be translated into a lower bound on $\epsilon$:
 \begin{align}\label{eq:WallBBN}
 	T_{w}>T_{BBN}\quad \Rightarrow \quad \epsilon \geq \frac{T_{BBN}^2}{6M_{\textrm{pl} }m_h} \sqrt{\frac{4\pi^3g_\star}{45}}.
 \end{align}
 In addition, the energy density of the walls should always be subdominant if one is to avoid entering an inflationary phase. Given its scaling, this constraint is strongest at the latest time, i.e. $T_w$ 
 \begin{align}\label{eq:epsbd}
 	\mu H(T_w)\leq g_\star T_w^4\quad \Rightarrow \quad \epsilon > \frac{m_h^2}{9 \lambda \gamma_4^2M_{\textrm{pl}}^2}.
 \end{align}
 These constraints are put together in Fig.~\ref{fig:wps} for an illustration of the parameter space region compatible with experimental data. We note that the lower bound in $\epsilon$ from Eq.~\eqref{eq:epsbd} means a locally SM-like potential (which has $\epsilon=0$) is excluded, whereas an upper bound follows from the LHC (non)measurement of the triple Higgs coupling, see Eq.~\eqref{eq:triplehbound}. Nevertheless the large disparity of scales involved leaves the possibility to close this window out of reach.

%%%%%%%%%%%%%%%%%%%%%%%%%%%%%%%%%%%%%%%%%%%%%%%%%%%%%%%%
\section{Past and future first order phase transitions}\label{sec:PT}

The presence of more than one minimum at a given temperature allows for tunnelling phenomena with potentially detectable imprints in our universe. This requires an evolution of the universe whose dynamics initially puts it in the false vacuum and a tunneling rate large enough to trigger a universe-wide transition, in the past or in the future. 

A well studied case that satisfies these conditions is the SM with small Higgs mass. The history for this case is depicted in Fig.~\ref{fig:breakno}; the high temperature symmetric minimum $\phi=0$ stops being the only extremum %turns into a false minimum 
after the appearance of a new minimum (and maximum) at a finite (and large enough, see Sec.~\ref{sec:HighT}) field distance. The potential energy difference between minima changes sign at a critical temperature $T_c$, after which tunneling is energetically viable but suppressed by the negative exponential of the bounce solution action. Given that the $\phi=0$ symmetric point turns eventually into a maximum, we have that the potential barrier between minima decreases and the energy difference increases, so sooner or later the transition will occur at what is dubbed the nucleation temperature $T_N$, where $T_N<T_c$.

Non-linear theories allow for first order phase transitions as that in the case of the light SM Higgs but also a set of qualitatively different ones. One could indeed attempt at the equivalent of the light Higgs SM in our non-linear theory; leaving every other parameter SM-like, decreasing $\gamma_4$ effectively increases field distances (i.e. decorrelates the quartic coupling and $m_h^2$, see Eq.~\eqref{eq:VtreeDef}) and could yield a first order phase transition. It is the case now, however, that there exist two degenerate true minima, $\phi_0,\phi_-$, given this limit is Higgs parity preserving, and uncorrelated patches of the universe will tunnel into one or the other minimum with equal likelihood.
This distribution implies that a network of domain walls would form in the boundary between different-vacua patches, the subsequent phenomenology of domain walls having been studied in Sec.~\ref{sec:Walls}.

This illustrates the variety of phenomena in non-linear theories. While the picture just described is more convoluted that typical case studies, one can break it down into staged transitions, each of which with a known theory formalism. For this reason we will study in this section scenarios that lead solely to a first order phase transition. Results derived here could be put together with those of Sec.~\ref{sec:Walls} for more intricate cases as the one described above with two transitions, however we leave this for future work. 
 Explicitly the cases studied will be histories Q of Fig.~\ref{fig:breakit1to3-Qm} and R of Fig.~\ref{fig:breakit1to1-R}, where we note that both are far from symmetric under Higgs parity and require sufficiently large $\epsilon$ and $\delta$ as shown below. For history Q  we have two possibilities by flipping the diagram vertically, these are Q$_0$ and Q$_-$ for a continuous connection of the high temperature with $h_0$ or $h_-$ respectively.  History R leads necessarily to a phase transition given the original minimum disappears, while history Q might not lead to a phase transition if the barrier and separation are never small enough. Following the discussion above we define the nucleation temperature as the temperature when the tunneling rate equals the Hubble 4-volume, which will signal the transition in the early universe
\begin{align}
    v^4 e^{-B(T_N)}\equiv H^4(T_N).
\end{align}
History Q does not specify which is the true minimum at low energies; this is dictated by the sign of $\Delta V(0)$ and in our parametrisation, $\epsilon$. The two options for the sign of $\epsilon$ then have to be put against the two histories Q$_{0,-}$ their selection being in turned mapped  to $\delta$. Small $\chi$ leads to more than one minimum at high temperature so we choose $\chi>0.3$ to ensure we remain over the limit of Eq.~\eqref{eq:betaL}.

For exposition purposes, here we use the small $\epsilon$ and $\delta$ limit to give a connection of history and parameter space through Eq.~\eqref{eq:HistParm}; although this expansion does not hold for the whole regime we explore (where we use the full expressions), it does identify possibilities and gives a rough outline of the regions. One has a contrast of high and low energy extrema dependent on the parameters as shown on Tab.~\ref{futuretable}.
Note that out of the four, one option is discarded since it never had $h_0$, where we are today, as a true vacuum.

\begin{table}
    \centering
    \begin{tabular}{ccccccc}
    \toprule
    && \multicolumn{2}{c}{$\epsilon < 0$}
    && \multicolumn{2}{c}{$\epsilon > 0$} \\
    && High $T$ & Low $T$
    && High $T$ & Low $T$ \\
    \midrule
    $(\mbox{ctg}^2_\chi - 2) \delta > -8\epsilon / \gamma_4$ 
    && $h_0(T)$ & $h_-(T)$
    && $h_0(T)$ & $h_0(T)$ \\
    $(\mbox{ctg}^2_\chi - 2) \delta < -8\epsilon / \gamma_4$
    && $h_-(T)$ & $h_-(T)$ 
    && $h_-(T)$ & $h_0(T)$ \\
    \bottomrule
\end{tabular}
\caption{True vacua at high and low temperatures for small $\epsilon$ and $\delta$.\label{futuretable}}
\end{table}

\begin{figure}
\begin{center}
  \includegraphics[width=0.7\textwidth]{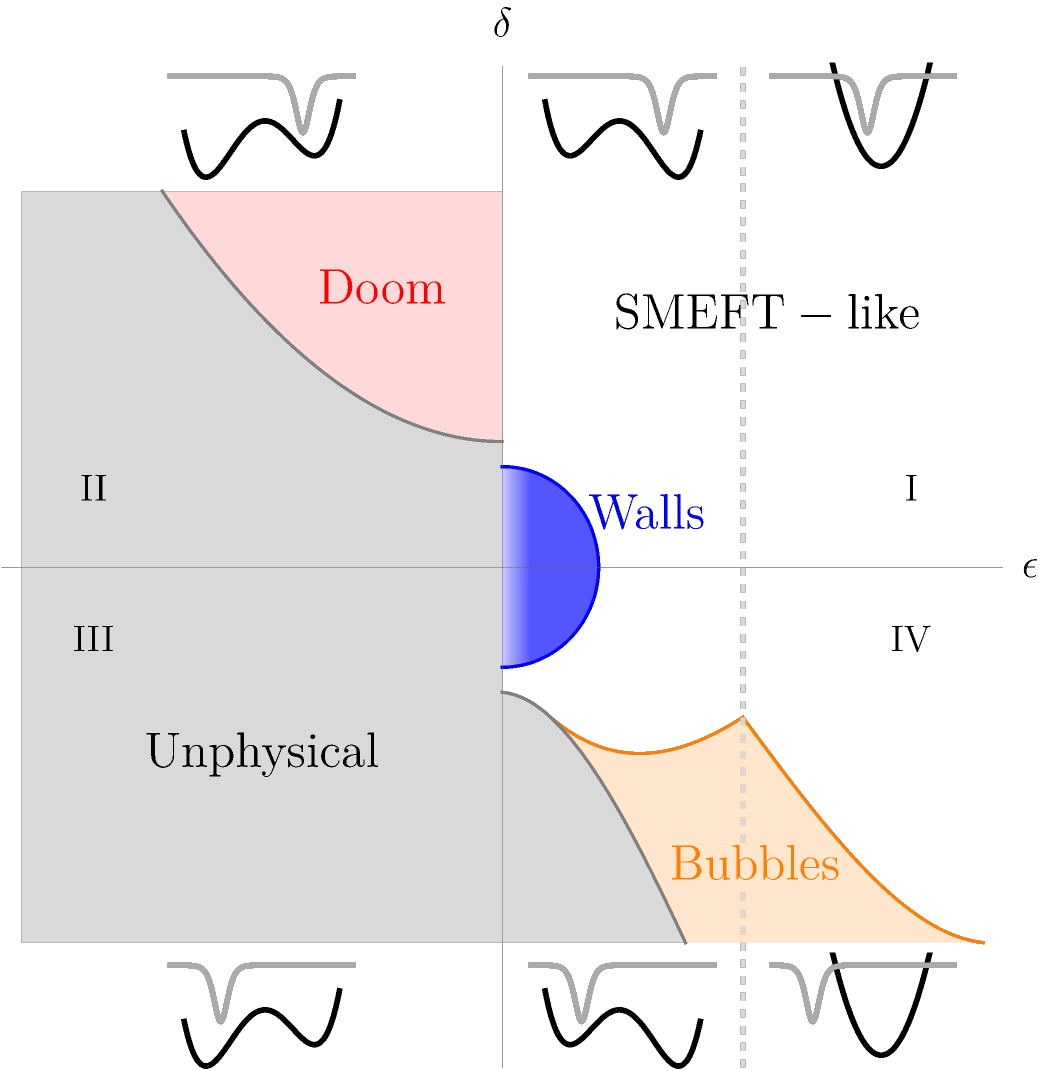}
  \caption{Schematic representation of the of the different cosmological phenomena arising in the $(\epsilon, \delta)$ plane. The small diagrams at the top and the bottom borders represent the tree-level potential in black, and the finite-temperature corrections in gray. }
  \label{TwoBinary}
\end{center}
\end{figure}

It is for a mismatch of high and low temperature minima that transitions are likely and that is the case for two out of remaining three options. These conditions are visualised in Fig.~\ref{TwoBinary}, which schematically displays the different phenomena that may be found in the $(\epsilon, \delta)$ plane:
\begin{itemize}
    \item In the first quadrant, both the tree-level potential and the finite-temperature corrections favor the $h_0$ SMEFT-like vacuum, and so we expect its cosmological phenomenology to be similar to that of the SMEFT. Note that a phase transition in this quadrant would require higher powers of $h/v$ (such as dim-6 or higher operators) in the tree-level potential.

    \item We mark as \emph{unphysical} the region where the thermal history selects the vacuum $h_-$ at present times, because, by definition, $h_-$ does not have the known measured masses and couplings of the SM particles. This is the case for the third quadrant, in which both the tree-level potential and the thermal corrections favor $h_-$; but also for part of the second quadrant, where $h_-$ is the tree-level vacuum; and part of the fourth quadrant, where the system may become trapped at $h_-$ and high temperatures, and not decay until after today.

    \item In the rest of the second quadrant, we find ourselves currently in the false vacuum $h_0$, a situation that we refer to as the \emph{doomsday} scenario, since a phase transition could be triggered at any time. However, as we will show in Sec.~\ref{sec:vacuum-decay}, the lifetime of the false vacuum we are in is much larger than the age of the universe. Notice that this possibility of the long-term survival of a false vacuum is in sharp contrast with the SM case.

    \item In the fourth quadrant, the true vacuum is $h_0$, but the finite-temperature corrections favor $h_-$. There will thus be a region where $h_-$ is selected at high temperatures, and a first-order phase transition happens before today. We have labelled this region \emph{bubbles}, since the phase transition will happen through bubble nucleation, as described in Sec.~\ref{sec:EWPT}.

    \item Finally, around the origin, when both $\epsilon$ and $\delta$ are small, long-lived domain walls are generated, as described in Sec.~\ref{sec:Walls}.
\end{itemize}
These considerations have to be put up against the IR problem; Sec.~\ref{sec:HighT} outlines our treatment of this issue. In summary we will discuss a phase transition {\it if the new minimum appears in the perturbative regime, $\varepsilon_{IR}(\phi_0(T_1))<1$ (strong IR constraint) or if the new minimum is in the perturbative region by the critical temperature $\varepsilon_{IR}(\phi_0(T_c))<1$ (weak IR constraint)}

 More complicated histories arise for small $\chi$ which we shall comment on but not study in detail. In this case two minima coexist at high temperature and we might have one or more minima at low energy. If the minima coexist above a temperature where thermal fluctuations of correlation-distance-size patches overcome the barrier height (above a Ginzburg temperature, see Sec.~\ref{sec:Walls}) one has a mixed spatial coexistence of the vacua. Above this temperature either the two minima are degenerate to a good approximation and the evolution is described in Sec.~\ref{sec:Walls}, or if they have a sizeable energy difference, thermal jumps from the true to the false vacuum will cease first to leave predominantly only the true vacuum. This vacuum might still not be the same as the zero temperature one, in which case a phase transition would be possible. We leave such cases for future study.
 
\subsection{First order electroweak phase transition} \label{sec:EWPT}

The possibility of nucleation in the past history of the universe is, given the considerations above, realised in the quadrant $\epsilon>0$, $\delta<0$ of the $(\epsilon, \delta)$ plane so that the true vacuum is today $\phi_0$ and at high temperatures $\phi_-$. In this quadrant we can find histories Q$_-$ and R, the latter for $\epsilon>1-\sqrt{8/9}$ where there is a single minimum at tree level. For $\phi_-$ to be the true vacuum at high temperature, however, not only negative but sufficiently sizeable $\delta$ is required; we found in a small $\epsilon,\delta$ expansion that this condition is  $\delta<- 8s^2_\chi/[\gamma_4(c_\chi^2-2s_\chi^2)]\epsilon $. This analytic result is validated around the origin but also extended to large values of the parameters (where one finds history R) by our numerical results in this section.

 The determining factor to characterise the transition, once one has arranged for the high temperature vacuum turning into a false one, is the bounce action. In particular it will determine whether the transition took place or whether the probability is too small for it to have occurred yet. At high temperatures this bounce action can be approximated by a 3-space-rotation symmetric action times the small time interval $1/T$
\begin{align}
    B=\frac{S_3}{T}.
\end{align}
There are no closed formulas for this action but rather a series of approximations of various accuracies and software for numerical solutions of the problem. Here we use both for a better understanding of the process. The software used is
\texttt{CosmoTransitions} \cite{Wainwright:2011kj} and \texttt{Anybubble} \cite{Masoumi:2016wot} whereas the analytic approximations are:

\begin{figure}
    \centering
    \includegraphics[width=0.65\textwidth]{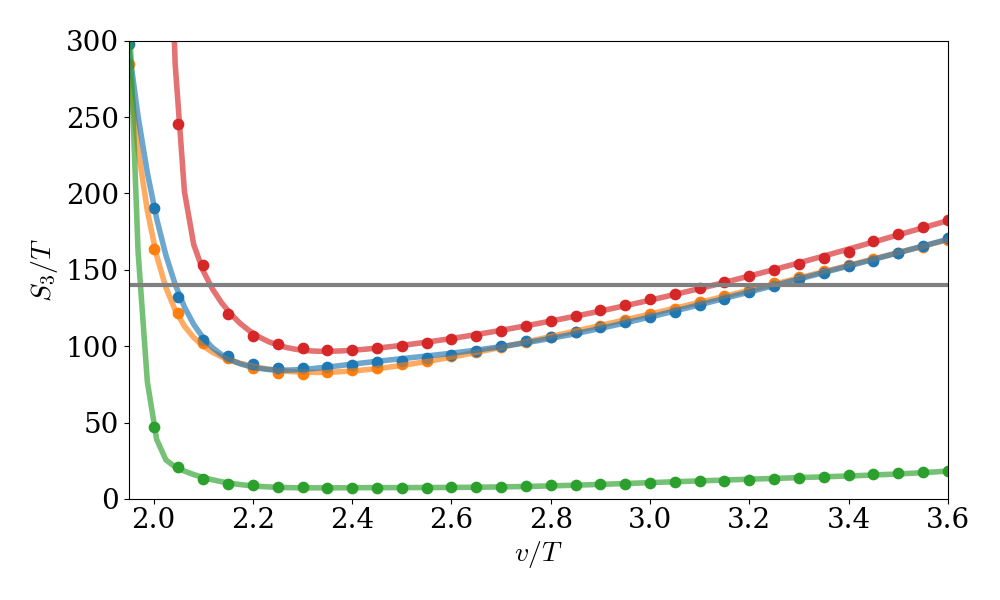}
    \caption{Comparison of the four methods to compute the tunnelling bounce action, as described in detail in the text of Sec.~\ref{sec:EWPT}. The displayed lines are evaluated for the parameters $\chi=\sqrt{0.1},\epsilon=0.04,\delta=-0.08$, with the results from \texttt{CosmoTransitions} in orange, \texttt{Anybubble} in blue, the quartic potential approximation in red and the thin wall approximation in green.}
    \label{fig:MethodsS3}
\end{figure}
\begin{itemize}
    \item 
\textit{Quartic potential.} The formula for the bounce action of a polynomial potential of degree four is known to a good approximation~\cite{LINDE1977306}. 
The formula gives, for a quartic potential with a local minimum at $\phi=0$, a true minimum at $\phi=\phi_m$, and maximum at $\phi_M$ as:
\begin{align}
 \bar V=\frac{\bar\lambda}{2}\left[\frac{\phi^4}{4}-\frac{\phi_m+\phi_M}{3}\phi^3+\frac{\phi_m\phi_M}{2} \phi^2\right],
\end{align}
and a bounce action as:
\begin{align}
\frac{S_3^L}{T}=&4.85 \sqrt{\frac{8\phi_m\phi_M}{\bar\lambda T^2}} \left(\alpha+\frac{\alpha^2}{4}\left[1+\frac{2.4}{1-\alpha}+\frac{0.26}{(1-\alpha)^2}\right]\right),\\
    \alpha=&\frac92\frac{\phi_m\phi_M}{(\phi_m+\phi_M)^2}.\label{eq:LindeB}
\end{align}
Our effective potential $V_{\textrm{eff}}$ is not a polynomial, but as an approximation it can be modelled as such. We do so by identifying the minimum $\phi_m=\phi_--\phi_0$ (if tunneling out of $\phi_-$) which leaves still two parameters in the potential, $\phi_M$ and $\bar \lambda$, which are not fixed to the maximum and quartic coupling in Eq.~\eqref{VeffUs}, but rather are implicitly defined by two conditions of equal energy difference between minima and barrier height, i.e.
\begin{align}
     \bar V(\phi_m;\bar\lambda,\phi_M)&\equiv \Delta V(T), & \bar V(\phi_M;\bar\lambda,\phi_M)&\equiv U_-(T).
\end{align}
These implicit definitions for $\bar\lambda$ and $\phi_M$ that ensure the model potential has the same energy difference and barrier height.

\begin{figure}
    \centering
    \includegraphics[width=\textwidth]{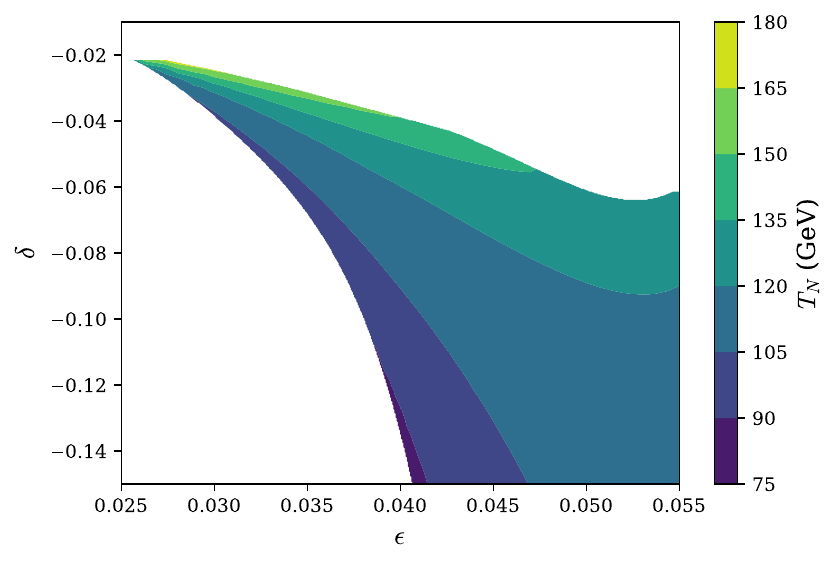}
    \caption{
    A slice of parameter space in $\epsilon,\delta$ for which the region in colour following history Q$_-$ meets the condition of Eq.~\eqref{eq:Tn} for some $T=T_N$, the nucleation temperature. For this plot, $\chi=\sqrt{0.1}$ and $\gamma_4=1.6$. The blank area in the upper region follows history Q$_0$, as the thermal corrections favour the $h_0$ vacuum, leading to a SMEFT-like cosmological history. In the lower blank region while still following history Q$_-$, the condition for nucleation, Eq.~\eqref{eq:Tn}, is not met for electroweak scale temperatures and as such we do not predict a first order phase transition in the early universe here. This and Fig.~\ref{TopoB} occupy roughly quadrant IV of Fig.~\ref{TwoBinary}.}
    \label{fig:daline}
\end{figure}

\item\textit{Thin wall and triangular approximation.} Shortly after the critical temperature $T_c$, one has that the thin wall approximation (which assumes a small energy difference between minima) holds and returns~\cite{Laine:2016hma}
 \begin{align}
     \frac{S_3^{tw}}{T}=&\frac{16\pi}{3}\frac{\mu^3}{T(\Delta V)^2},&
     \mu&=\int_{\phi_-}^{\phi_0} d\phi \sqrt{2(V_{\textrm{eff}}(\phi)-V_{\textrm{eff}}(\phi_-))}.
 \end{align}
Fitting the potential between minima to a triangle yields
 \begin{align}
\mu_{\triangle}=2\sqrt{2}\int_{\phi_-}^{\phi_+}\sqrt{(U_-)\phi/(\phi_+-\phi_-)}=\frac{4\sqrt{2}}{3}(\phi_+-\phi_-)\sqrt{U_{-}},
 \end{align}
and in this approximation
\begin{align}
    \frac{S_3^{tw}}{T}=\left(\frac{4\sqrt{2}}{3}\right)^3\frac{16\pi}{3} \frac{v}{T} \frac{U_{-}^{3/2}(\phi_+(T)-\phi_{-}(T))^3}{\Delta V^2 v}.
\end{align}
This expression is easy to evaluate, but is only valid for a short time after $T_c$ whereas Eq.~\eqref{eq:LindeB} continues to be valid at lower temperatures.
\end{itemize}

One can see a comparative plot for the two analytic approximations and the two numerical estimates of the bounce action for the 1-loop thermal effective potential defined in Eq.~\eqref{VeffUs} as a function of temperature in Fig.~\ref{fig:MethodsS3}. The numerical estimates are performed with
\texttt{Anybubble}~\cite{Masoumi:2016wot} and \texttt{CosmoTransitions}~\cite{Wainwright:2011kj} respectively. For  \texttt{Anybubble}, the potential has been first fitted to a 9th degree polynomial for technical reasons and then passed into the \texttt{Anybubble} code, for \texttt{CosmoTransitions}  we use the \texttt{tunneling1D} module to compute the bounce action.

\begin{figure}
    \centering
    \includegraphics[width=\textwidth]{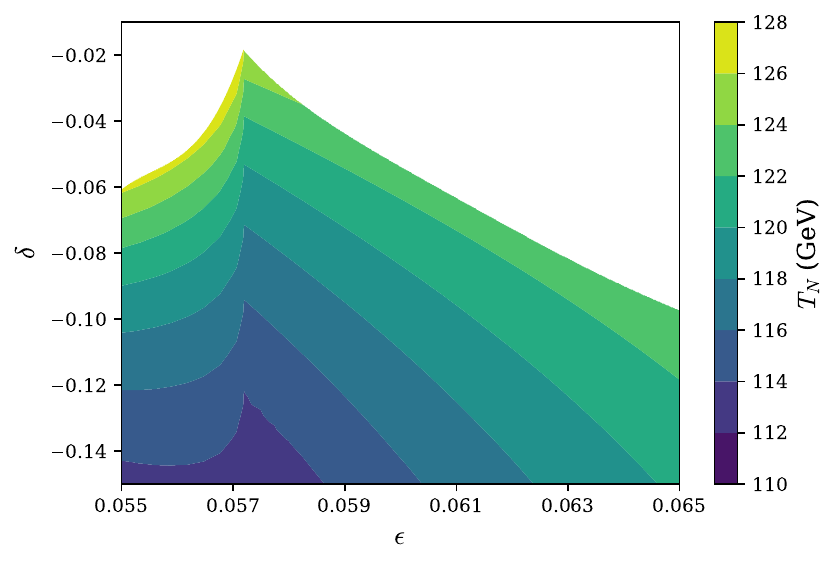}
    \caption{Extension of Fig.~\ref{fig:daline} to larger epsilon values and different histories, with $\gamma_4=1.6,\chi=\sqrt{0.1}$ as in the previous plot. The plot is discontinuous at $\epsilon=1-\sqrt{8/9}$ as a result of Eq.~\eqref{fullde}. For $\epsilon<1-\sqrt{8/9}$, the region in colour follows history Q$_-$ and the blank, upper region above this follows history Q$_0$ as in Fig.~\ref{fig:daline}. However when $\epsilon>1-\sqrt{8/9}$, the $T=0$ potential has a single minimum only and history R is followed for the region in colour. The blank region above only has a single minimum throughout its evolution. We note that the upper boundary from history R to a SMEFT-like history is slightly ambiguous as the potential becomes very flat and difficult to treat numerically.}
    \label{TopoB}
\end{figure}

The curves all start with a very large value of the
bounce action as right after the critical temperature, the energy difference between true and false minima is small and the thin wall approximation applies, giving a bounce action inversely proportional to the minima potential energy difference.
In the case of history Q$_-$, as temperature decreases the bounce action decreases and reaches a minimum value after which it grows again. For the case of history R however, the action decreases until it vanishes since the original minimum disappears and so does the barrier between minima. The condition for the transition to occur reads
\begin{align}\label{eq:Tn}
    \frac{S_3(T_N)}{T_N}
    \simeq
    -\log \frac{H^4(T_N)}{v^4}
    \simeq 140.
\end{align}
This condition will be met in history R but not necessarily in history Q$_-$, where it could be that the bounce action never decreases below 140 and one is stuck in the wrong vacuum. In Fig.~\ref{fig:daline} we show a slice of parameter space where the condition is met for history Q$_-$, where the gradient marks the nucleation temperature. Additionally Fig.~\ref{TopoB} extends this slice to encompass history R. As discussed in the next section, if one is stuck in $\phi_-$ below electroweak temperatures, one stays there until today. Since this conflicts with our definition of $\phi_0$ as our vacuum today these instances are excluded; visually that is the lower left white region in Fig.~\ref{fig:daline}. On the edge of this region we have a sizeable barrier, but just low enough so that the transition occurs, and it does so strongly. Further away from this lower edge of the wedge in Fig.~\ref{fig:daline}, we have smaller barriers and weaker first order phase transitions until at the upper edge one meets the numerical extension of the condition $(\mbox{ctg}_\chi^2-2) \delta < -  8 \epsilon/\gamma_4$ where the history changes to Q$_0$ and the universe is always in the true vacuum which is connected continuously at low and high temperatures. For history R with $\epsilon>1-\sqrt{8/9}$, this condition changes qualitatively and the edge of the first order phase transition region is found numerically.

%%%%%%%%%%%%%%%%%%%%%%

\subsection{Doom: future vacuum decay}
\label{sec:vacuum-decay}

The previous section showed that the tunneling rate out of a false vacuum might not be large enough to undergo a transition at temperatures near but below the critical temperature. At later times, the causal 4-volume increases as $H^{-4}$, and the bounce action is given by the 4d bounce action, well approximated by the potential for $T=0$. It could be that the growing Hubble volume exceeds both the inverse decay per time and volume at this later time, and vacuum decay would ensue; if so a conservative bound would be to require this to happen before BBN, 
\begin{align}
    B
    \leq 4 \log \frac{v}{H(T_{BBN})}
    \simeq 4 \log \frac{vM_{\textrm{pl}}}{T_{BBN}^2}
    \simeq 250 \,.
\end{align}
It is the case however, that for all studied cases of history Q$_-$, if the phase transition does not occur at $T_{ew}$, it will also not occur before BBN. This means that any point on Fig.~\ref{fig:daline} of Sec.~\ref{sec:EWPT} on the lower left white region, which marked instances where the universe remained in the false minimum $\phi_-$ past the electroweak epoch, the universe will remain in $\phi_-$ until today. For this reason these points are deemed unphysical.
 
\begin{figure}
    \centering    \includegraphics[width=0.6\textwidth, clip=true, trim=.1cm .2cm .0cm .0cm]{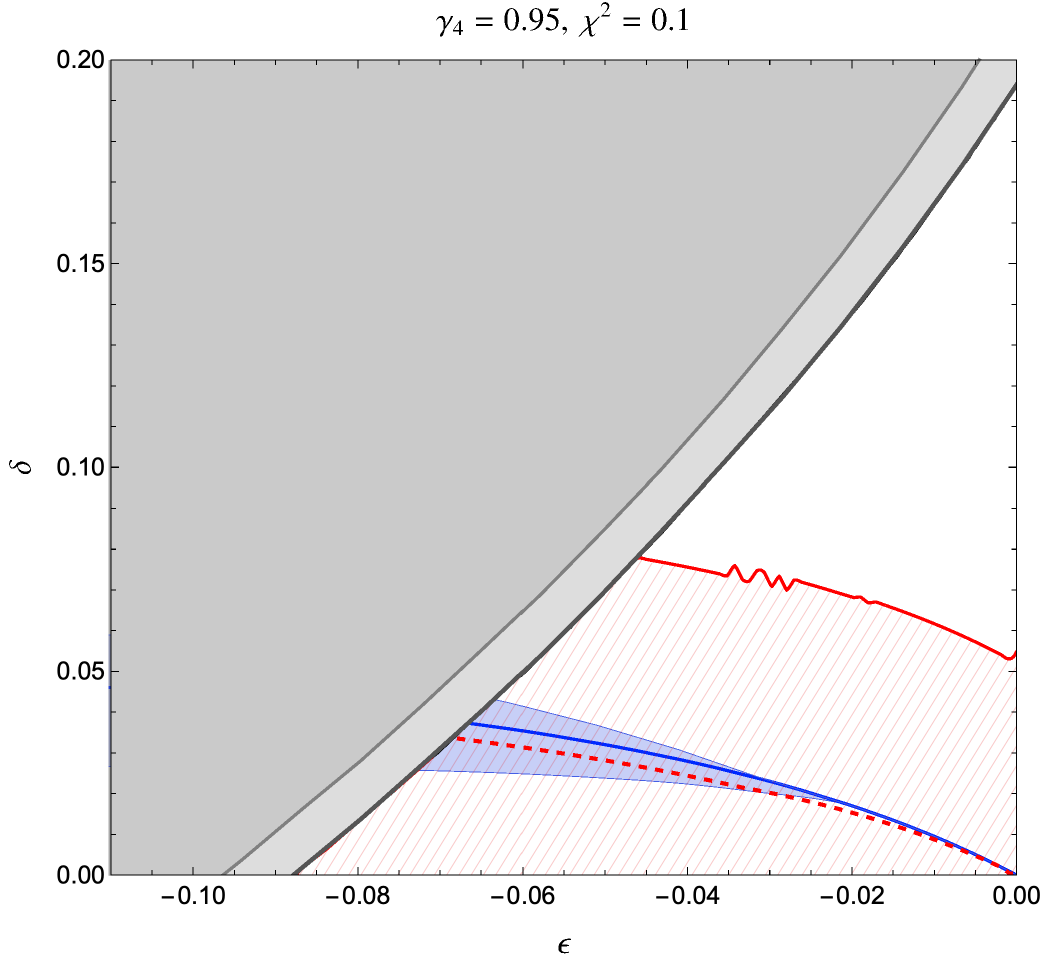}
    \caption{Bounds on the $(\epsilon, \delta)$ parameter space for the doom scenario at $\gamma_4 = 0.95$ and $\chi^2 = 0.1$. The upper-left gray region is excluded by the boundedness-from-below (dark) and perturbativity (light) criteria. The hatched region below the solid red line is excluded by the strong IR constraint, Eq.~\eqref{eq:ir-strong}, while the dashed red line shows the weak version, Eq.~\eqref{eq:ir-weak}. There is some noise associated with our method of numerically tracking the temperature evolution of the minima, and this noise is visible in the strong IR constraint curve. The blue region corresponds to the history displayed in Fig.~\ref{fig:breakitothers}, in which two minima exist in the high-temperature potential. The blue line separates the regions where the sign of the magnitude defined in Eq.~\eqref{eq:sign} is positive or negative and is an estimate for the regions for histories Q$_{-,0}$ respectively}
    \label{fig:Doom}
\end{figure}

 The situation of being stuck in a false minimum is realised in history Q$_0$ in the opposite quadrant, $\epsilon<0$, $\delta>0$, with the further approximate constraint of $(\mbox{ctg}_\chi^2 - 2) \delta \gtrsim -8 \epsilon /\gamma_4$ as sketched in Tab.~\ref{futuretable}. In practice, this approximate constraint has to be extended to a numerical search for history Q$_0$. In Fig.~\ref{fig:Doom}, we display in blue the region where the thermal history is that of Fig.~\ref{fig:breakitothers}, with two minima at high temperature. Above this region, we have history Q$_0$. The infrared problem as sketched in Sec.~\ref{sec:ir-problem} casts doubt on histories where extrema are to be found in regions with $\varepsilon_{IR}(\phi_{-})>1$; for such cases we do not trust our perturbative computation, and so we do not discuss the region below the red solid and dashed lines in Fig.~\ref{fig:Doom}.  
 We also show in gray colour in Fig.~\ref{fig:Doom} the regions excluded by the boundedness-from-below and perturbativity criteria described above.
 
 Even when history Q$_0$ can be perturbatively treated, the fact that at high temperature the only minimum is $h_0(T)$, which turns into a false vacuum at $T_c$, has to be reconciled with us being at $h_0(T)$ today; this demands that the decay rate be small enough so that the transition has not occurred yet
\begin{align}
    B\geq 4\log\frac{v}{H_0} \simeq 410.
\end{align}
This case with negative $\epsilon$ is however more restricted from the perturbativity and boundedness from below constraints. This can be understood since for $\epsilon<0$, $\gamma_\epsilon<1$ and given $\delta\gamma_4=\gamma_\epsilon-\gamma_4/\gamma_a$ for positive $\delta$ we require small $\gamma_4/\gamma_a$, yet one of the terms in the perturbative bound of Eq.~\eqref{eq:BoundedfB} scales with $(\gamma_4/\gamma_a)^{-2}$. This perturbative bound then translates into a lower bound on negative $\epsilon$ to obtain a bounded from below potential in a range of $\phi\sim 4\pi v$. On the other hand the limit of $\epsilon\to 0^-$ gives an infinite lifetime and hence is compatible with observation, the question then is if within this allowed window for $\epsilon$ decay lifetimes around the age of the universe can be found.

To answer this question we turn to the zero-temperature bounce action;  a (euclidean) space-time symmetric solution with the $T=0$ potential. 
A quick estimate of the thin wall approximation for our zero temperature potential, which holds if  
\begin{align}
    \frac{\Delta V}{3 \sigma m_h}=\frac{1}{2\gamma_\epsilon^2}(1-\gamma_\epsilon^{-1}(1-\epsilon))\ll 1,
\end{align}
gives the tension as in the bubble case expanding in $\epsilon$ 
\begin{align}
	\sigma = \int_{h_+}^0 dh \sqrt{2V(h)}=\mu+\mathcal O(\epsilon)=\frac{2\sqrt{\lambda}v^3\gamma_4^{-2}}{3},
\end{align}
and an expression for the bounce action as
\cite{Coleman:1977py}
\begin{align}
B=\frac{27\pi^2\sigma^4}{2\Delta V^3}=\frac{3^3\pi^2}{2}\frac{\lambda^2 2^4\gamma_4^{-8}}{3^4}\frac{1}{2^6\lambda^3 \epsilon^3\gamma_4^{-6}}=\frac{\pi^2}{24\lambda \gamma_4^2\epsilon^3}.
\end{align}
This analytic approximation returns, for $\gamma_4=1$, $\epsilon>-0.15$ decay lifetimes longer than today.
This is not a particularly accurate result: the use of
\texttt{CosmoTransitions} and  $\gamma_4=1$ returns
$\epsilon>-0.475$ for a decay after today.
All these values are however ruled out by the perturbativity and boundedness bound. These bounds are relaxed for smaller $\gamma_4$ allowing for larger $|\epsilon|$ values, so one might expect this limit to allow for shorter universe lifetimes. In this same limit however, the bounce action, while not well approximated by the thin wall, does factor out in the same way for $\gamma_4$, i.e. $B=\lambda^{-1} \gamma_4^{-2}f(\epsilon)$ and so reducing $\gamma_4$ actually increases the lifetime exponentially. One concludes then that the lifetime of the universe is in all instances well beyond the current age.

\begin{figure}
    \centering
    \includegraphics[width=0.6\textwidth]{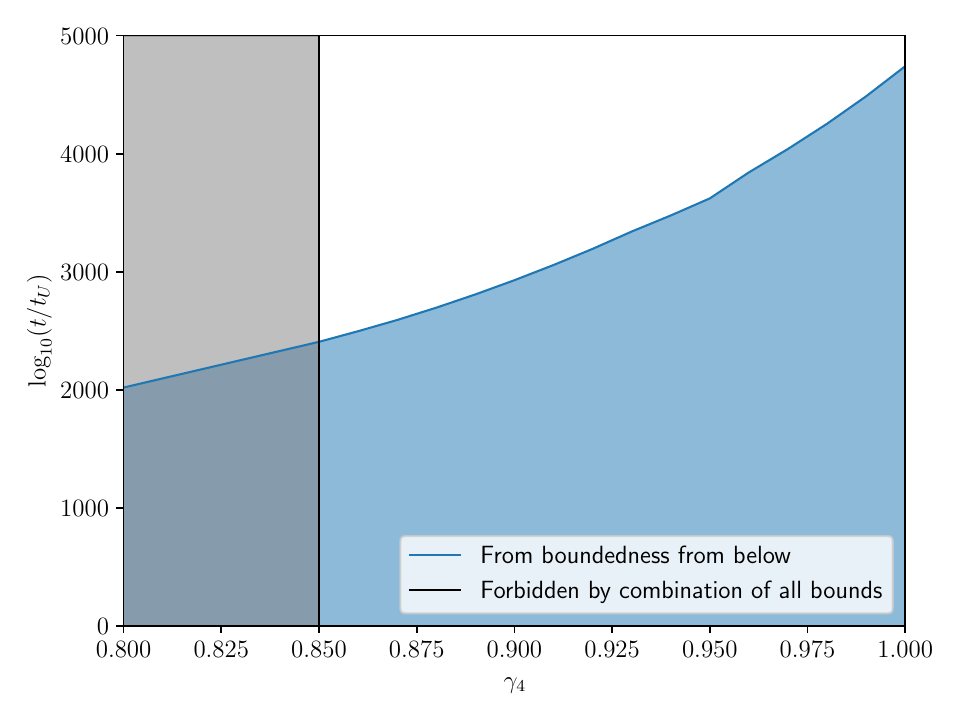} \\
    \includegraphics[width=0.32\textwidth]{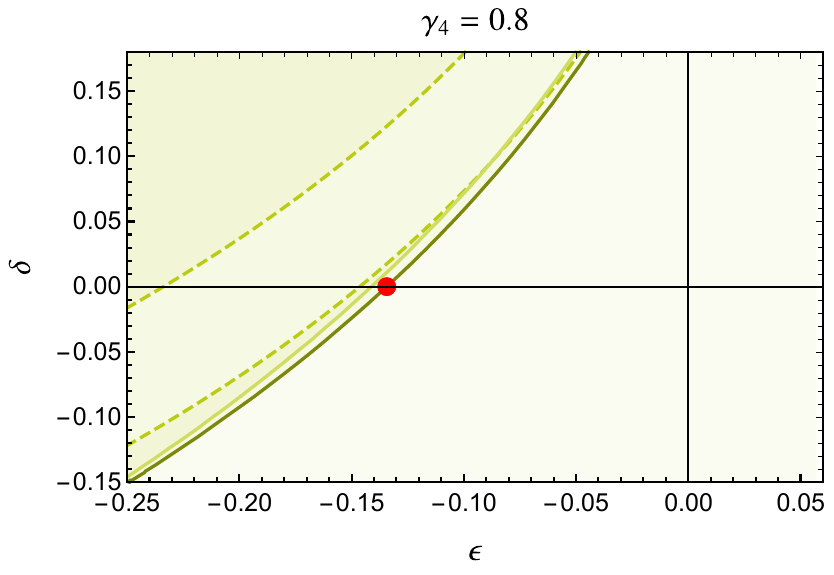}
    \includegraphics[width=0.32\textwidth, clip=true, trim=.1cm .2cm .1cm .1cm]{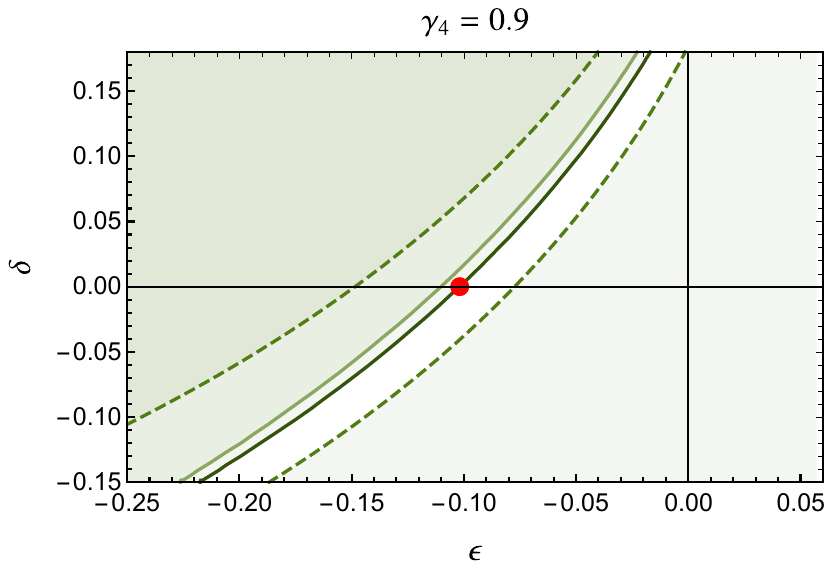}
    \includegraphics[width=0.32\textwidth, clip=true, trim=.1cm .2cm .1cm .1cm]{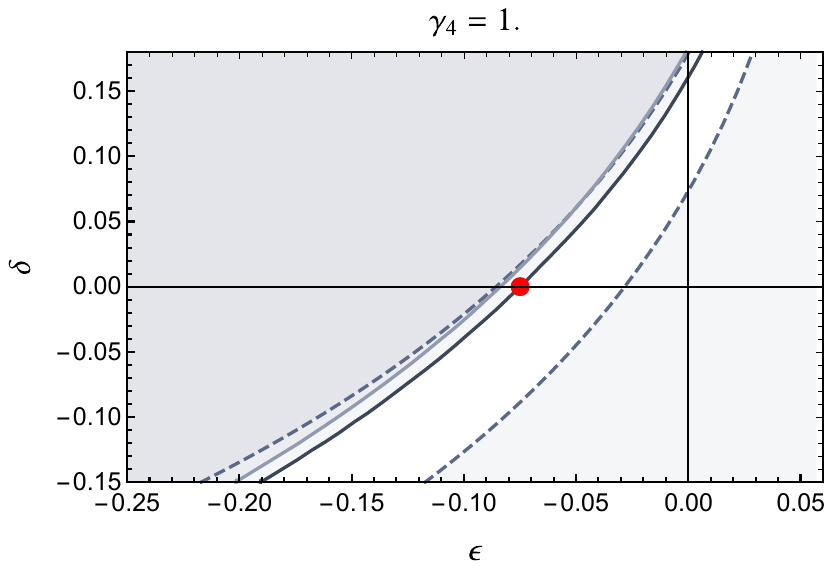}
    \caption{Top: excluded values of the lifetime of the universe as a function of $\gamma_4$. Bottom: example points with the minimal lifetime allowed by boundedness from below in the second quadrant of the $(\epsilon, \delta)$ plane, used to generate the blue line in the top plot. In these three plots, the shaded region is ruled out from various constraints: boundedness from below (darker, solid), perturbativity (lighter, solid), and LHC constraints (dashed). See Sec.~\ref{sec:LHC} and Fig.~\ref{DoomLHC} for more details on all the bounds shown in these plots.}
    \label{fig:Saved}
\end{figure}

The question that follows is then what are the allowed values for the lifetime. This turns our gaze to the future which changes the estimation qualitatively. Indeed the estimation of the Hubble 4-volume as the relevant spacetime factor is valid in a decelerated expansion with scale factor $a\sim t^p $  but not in an accelerated expansion. In this case revisiting the estimation for the relevant 4-volume, i.e. the past light cone of an observer at time $t_f$, one obtains
for $a\sim e^{H_0t}$
\begin{align}
\int_{\textrm{light cone}} &d^4x \sqrt{-\det(g_{\mu\nu})}=\frac{4\pi H_0^{-3} t_f}{3}.\\  \textrm{light cone eq.}&\quad -a(t)d\chi=dt,\qquad \chi(t_F)=0,
\end{align}
 i.e. an space-time volume which is a sphere of radius $H_0^{-1}$ times time, and one concludes that the accelerated expansion from this perspective is `living in a box', the longtime physicist dream. For a given value of potential parameters we find the lifetime $t_f$ by solving
\begin{align}
     \frac{4\pi H_0^{-3} t_f}{3} v^4 e^{-B}=1.
\end{align}
We have used this estimate to generate  Fig.~\ref{fig:Saved} which further illustrates the point of the lifetime of the universe being much larger the its age.
To generate it, we look for the point in the second quadrant of the $(\epsilon, \delta)$ space that gives the minimal lifetime allowed by the boundedness-from-below criterion, for different values of $\gamma_4$. Such points are shown in red in the 3 plots at the bottom of Fig.~\ref{fig:Saved}, corresponding to 3 values of $\gamma_4$. We then compute the corresponding lifetime and display it as the blue line in the top plot. The gray region on the left of the top plot is excluded by a combination of all bounds, as described in Sec.~\ref{sec:LHC}. Thus, the only allowed lifetimes are those in the white region, and are all larger than the age of the universe by a factor $>10^{2000}$.

\section{Gravitational Waves}\label{sec:GW}

The spectrum of gravitational waves is customarily given as 
\begin{align}
    \Omega_{\rm GW}=\frac{d \rho_{\rm GW}}{d\log k}=\frac{d \rho_{\rm GW}}{d\log f}\,, 
\end{align}
with $\rho_{\rm GW}$ the energy density in gravitational waves, $k$ the wave number and $f$ frequency, in natural units $2\pi f=k$. In our study two sources of gravitational waves have been identified: walls and first order phase transitions. Let us discuss each in turn.

 Gravitational emission occurs for domain walls as smaller curvature (larger radius) wall structure enters the causal horizon and starts a damped oscillation in the plasma. While most of the energy goes into the plasma, part of it is emitted as gravitational waves, with the contribution at time $t$ after the big bang for an interval $\Delta t$  estimated as $\Delta\rho_{\rm GW}/\Delta t\simeq G\mu^2/t$. Today the spectrum is redshifted and more so the earlier the emission, so that the peak frequency corresponds to the annihilation time $t_w$, and one has (see e.g. Ref.~\cite{Gelmini:2020bqg})
 \begin{align}
 	f_{\textrm{peak}}&= a_w H_w=\frac{g_0^{1/3}T_0}{g_{\bar w}^{1/3}T_w}  \sqrt{\frac{4\pi^3g_w}{45}}\frac{T_w^2}{M_{\rm pl}}=\frac{g_0^{1/3}}{g_{\bar w}^{1/3}}\left(\frac{4\pi^3g_w}{45}\right)^{1/4}T_0 \sqrt{\frac{6\epsilon m_h}{M_{\rm pl}}},\\
  &=\frac{g_0^{1/3}}{g_{\bar w}^{1/3}}\sqrt{\frac{4\pi^3g_w}{45}}\frac{T_0 T_{\rm BBN}}{M_{\rm pl}}\frac{T_w}{T_{\rm BBN}}=1.1\times 10^{-10}\frac{T_w}{T_{\rm BBN}}\textrm{ Hz},
  \end{align}
  where $g_{\bar w}=g_w-g_\nu=g_w-21/4$ and in the last equality  we assumed that $T_w$ occurred while $g_w=10.75$ which is the latest allowed. Correspondingly, the power spectrum value at the peak  is given by:
  \begin{align}
 	\Omega_{\rm GW}(f_{\textrm{peak}})&=\frac{G_N\mu^2}{\rho_{cr}}\left(\frac{T_0}{T_w}\right)^4=\Omega_\gamma^0\frac{2^3\pi g_w}{3^5g_0}\frac{v^4}{M_{\rm pl}^4}\frac{1}{\epsilon^2\gamma_4^4}\\
  &=\Omega_{\gamma}^0 \frac{40}{3\pi^2g_0}\frac{m_h^2 v^4}{M_{\textrm{pl}}^2 T_{BBN}}\frac{T_{BBN}^4}{T_w^4} \gamma_4^{-4}=1.4\times 10^{-17}\frac{T_{BBN}^4}{T_w^4} \gamma_4^{-4}.\label{WallsGW}
 \end{align}
 One can note, when expressed in terms of the ratio of temperatures for the BBN and wall annihilation, that the frequency is bounded to be greater than $0.1$~nHz, and the peak in the spectrum increases with decreasing $\gamma_4$ and decreasing frequency. These features are shown in Fig.~\ref{fig:wps}; given that the strength of the gravitational wave signal is greater for lower frequencies, one finds that the low frequency Square Kilometer Array (SKA) experiment~\cite{Weltman:2018zrl} would probe part of the parameter space.\footnote{We estimate that SKA is sensitive to the GW signal when its power spectrum evaluated at the peak frequency lies above the sensitivity curve.} As we shall see in Sec.~\ref{sec:LHC}, when put against LHC bounds for our model, however, the parameter region that would yield a signal at SKA is already excluded. The shape of the full spectrum can be approximated by $k^{-1}$, and $k^{3}$ right and left of the peak respectively~\cite{Gelmini:2020bqg}, in qualitative agreement with simulations~\cite{Hiramatsu_2014}.

For a first order phase transition undergoing bubble nucleation, the gravitational wave signal can be estimated using the thermal parameters derived from the tunnelling action: the speed of the transition $\beta/H$ and the strength of the phase transition $\alpha$ related to the latent heat, 
both evaluated at the nucleation temperature $T_N$ given in Eq.~\eqref{eq:Tn}.\footnote{More precisely, the relevant temperature is the transition temperature $T_*$. Here we assume $T_* \approx T_N$ and do not distinguish between them, a safe assumption for fast phase transitions without significant reheating and in the absence of large supercooling~\cite{Caprini:2019egz, Caprini:2015zlo}
} The thermal parameters are explicitly:
\begin{align}
    \alpha &= \frac{1}{\rho_{rad}(T_N)}\left(\Delta V(T_N) - T_N \frac{d\Delta V}{dT}\big|_{T=T_N}\right) \,,\label{eq:alpha}\\[12pt]
    \frac{\beta}{H_*} &= T_N\frac{d(S_3/T)}{dT}\big|_{T=T_N} \label{eq:beta}\,,
\end{align}
where $H_*$ is the Hubble constant at the time of nucleation, $\Delta V(T)$ is defined in Eq.~\eqref{eq:DeV}, and $\rho_{rad}(T_N) = g_*\pi^2 T_N^4/30$ is the radiation energy density. The quantity $g_*$ is the effective number of relativistic degrees of freedom in the plasma at $T_N$, which was calculated by tracking particle decoupling following~\cite{Husdal:2016haj}.
The GW signal also depends on the velocity of the bubble wall in the rest-frame of the plasma, $v_w$,
which we calculate following the model-independent prescription in \cite{Ai:2023see}.

\begin{figure}[h!]
    \centering
    \includegraphics[width=\textwidth]{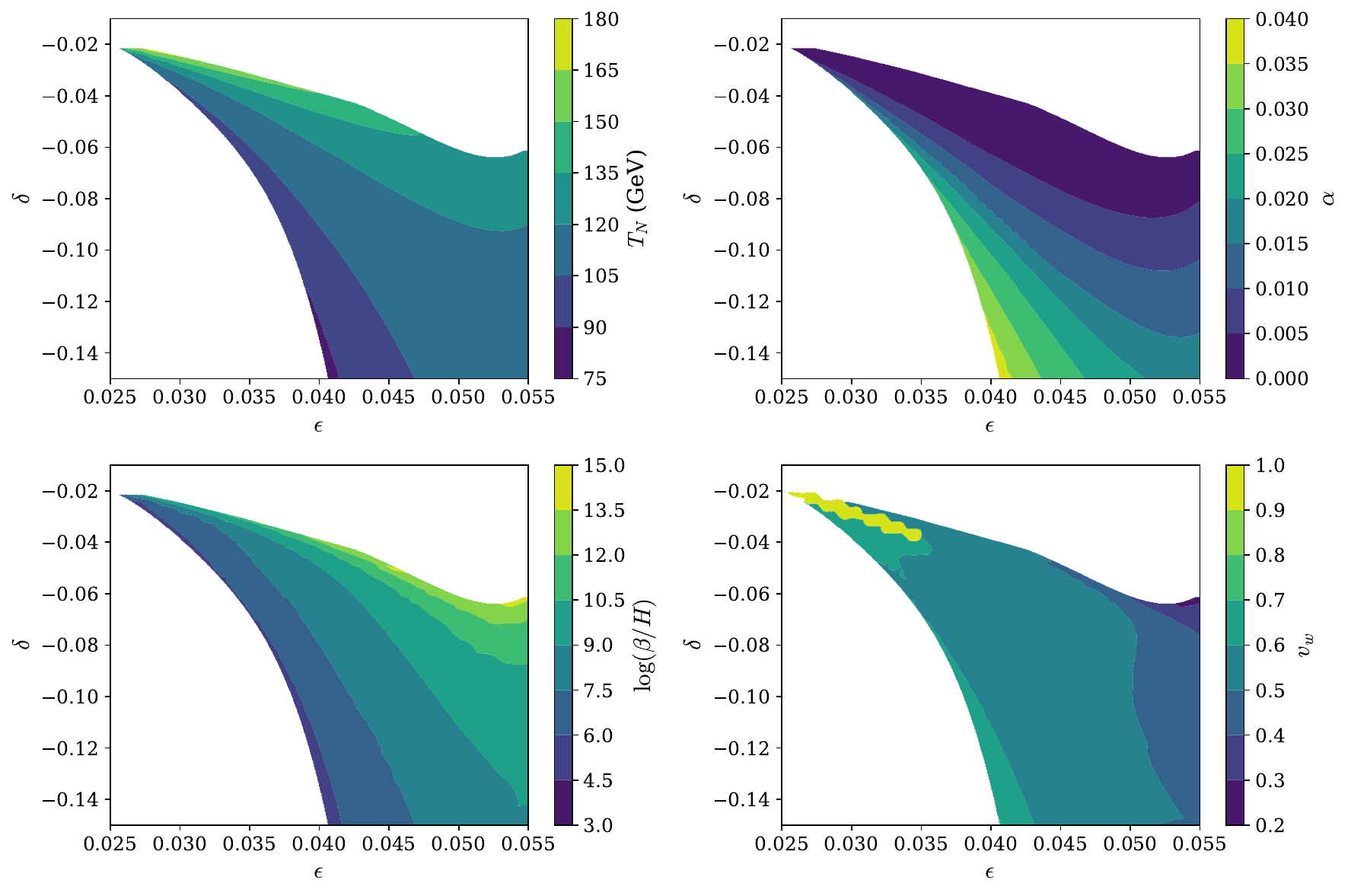}
    \caption{The thermal parameters $T_N,\alpha,\beta/H_*,v_w$ for a slice of $\epsilon,\delta$ parameter space with $\gamma_4 = 1.6$ and $\chi=\sqrt{0.1}$. The thermal parameters are defined respectively in Eqs.~\eqref{eq:Tn}, \eqref{eq:alpha}, and \eqref{eq:beta} with the wall velocity estimated using prescription outlined
    in \cite{Ai:2023see}. Note for a small region of parameter space, runaway bubbles with $v_w\rightarrow1$ are predicted. However, the region is numerically noisy as a result of finite sampling. Fig.~\ref{fig:daline} is included again on the top left plot.}
    \label{fig:dalinefull}
\end{figure}

As demonstrated in Sec.~\ref{sec:EWPT}, our non-linear theory model is capable of FOPTs that are qualitatively different than those accessible in SMEFT theories. In particular, we are interested in bubble nucleation that occurs in the parameter space identified in Fig.~\ref{fig:daline}. We now present in Fig.~\ref{fig:dalinefull} the behaviour of the associated thermal parameters $\alpha,\ \beta/H_*$ along with $T_N$ and $v_w$ as a function of $\epsilon,\delta$, using the same slice of parameter space as Fig.~\ref{fig:daline}. With these parameters associated with bubble nucleation, we can proceed to estimate the GW signal observable today.

Next, we discuss our calculation of the gravitational wave signal from the thermal parameters. The following discussion is largely pedagogical, and we note that our analysis follows standard techniques. There are three sources of gravitational wave production during bubble nucleation: the bubble collisions, plasma sound waves driven by the expanding bubble wall, and turbulence~\cite{Caprini:2015zlo}. The total gravitational wave energy density, $\Omega_{\text{GW}}$, is the sum of the signal from each of these sources is
	\begin{align}
	\Omega_\text{GW}
		&= \Omega_\text{col}+ \Omega_\text{sw} + \Omega_\text{tu} \,.
	\end{align}
The initial source for GW radiation occurs when the bubbles first collide, breaking spherical symmetry. The contribution from bubble collisions, however, is thought to be small (order percent) so long as the walls do not enter the runaway regime~\cite{Caprini:2019egz}, so here we do not include them in our estimation of the GW signal. 

As the bubbles expand, wave fronts emerge preceding the the bubble walls, forming acoustic sound waves propagating in the plasma. This is typically the dominant contribution to the GW signal near the peak frequency of the gravitational wave power spectrum.
Schematically, this power spectrum as measured today can be estimated by redshifting the GW signal at the source,
	\begin{align}
	\Omega_\text{sw}
		&= F_\text{redshift}
			\times \Omega_\text{amp}
			\times S_\text{sw}(f)\,,
	\end{align}
where $S_\text{sw}(f)$ is the spectral shape of the GW signal.	
Accounting for the redshift of GW radiation from the time of production introduces the factor (as outlined in, for example, Refs.~\cite{Fixsen:2009ug, Caprini:2019egz, Athron:2023xlk})
	\begin{align}
	F_\text{redshift}
		&= 3.57 \times 10^{-5} \left( \frac{ 100 }{ g_* } \right)^{1/3}\,.
	\end{align}
The GW energy density from the sound waves in the plasma is~\cite{Hindmarsh:2017gnf}:
	\begin{align}
	\Omega_{\text{sw}, 0}
		&=3  \Gamma^2 \bar U_f^4 (H_* R_*)  \tilde\Omega_\text{sw} \,,
	\end{align}
where
$\Gamma\sim4/3$ is the adiabatic index,
 $\bar U_f$ is the root mean square (RMS) fluid four-velocity, and $R_*$ is the average separation between bubbles $R_*= (8\pi)^{1/3} v_w /\beta$~\cite{Caprini:2019egz}. The efficiency factor $\tilde\Omega_\text{sw}$ stems from converting motion
  in the fluid to metric perturbations, estimated from simulation to be $\tilde\Omega_\text{gw}=0.012$~\cite{Hindmarsh:2015qta, Hindmarsh:2017gnf}.
The RMS velocity $\bar U_f$ is set by the strength of the phase transition $\alpha$ through
	\begin{align}
	\bar U_f^2 
		&= \frac34 \kappa_\text{sw} \frac\alpha{ 1 + \alpha}\,.
	\end{align}
The quantity $\kappa_\text{sw}$ controls how much of the vacuum energy is transferred into kinetic energy of the plasma. This has been studied in detail in Ref.~\cite{Espinosa:2010hh}, resulting in numerical fits of $\kappa_\text{sw}( v_w, \alpha)$ to approximate full solutions to the relativistic fluid equations of the plasma. The results of these fits are used here and checked against PTPlot~\cite{Caprini:2019egz}.

During the phase transition, shock waves will develop, at which point the motion of the plasma is better described by turbulence than acoustic sound waves. If the time scale of shock formation is small compared one Hubble time $\tau_\text{sh} < 1/H_*$, the resulting sound wave signal is reduced by a factor $ H_* \tau_\text{sh} = H_* R_* / \bar U_{fl}.$ The total $\Omega_\text{amp}$ is then\footnote{The numerical prefactor ensures the total GW power resulting from integrating the power spectrum $\Omega_\text{sw}$ reproduces the total power estimate $\Omega_{\text{sw}, 0}$~\cite{Hindmarsh:2017gnf}}
	\begin{align}
	\Omega_\text{amp} 
		&= 2.061 \, \Omega_{\text{sw}, 0} \text{Max}[H_* \tau_\text{sh}, 1]\,.
	\end{align}
Putting all of this together gives
	\begin{align}
	h^2 \Omega_{\text{sw}}(f) 
    &= 2.59 \times 10^{-6}
    	h^2  \left( \frac{ 100 }{ g_*} \right)^{1/3}
       \Gamma^2 \bar{U}^4_{fl}
        \left( \frac{H_* }{\beta} \right) 
         v_w
        \text{Max}[H_* \tau_\text{sh}, 1]
        S_\text{sw} (f)
         \,.
     	\end{align}
The spectral shape of the sound wave power spectrum is~\cite{Caprini:2015zlo}
	\begin{align}
	S_\text{sw}
		&= \left( \frac f{ f_\text{sw} } \right)^2 \left( \frac7{ 4 + 3 \left(  f/f_\text{sw}\right)^2 } \right)^{7/2}\,,
	\end{align}
written in terms of the peak frequency as observed today,
	\begin{align}
	f_\text{sw}
		&= \left( 8.876 \ \mu\text{ Hz} \right)
		\left( \frac{ g_\text{eff} }{100} \right)^{1/6}
		\left( \frac{ T_N }{ 100 \text{ GeV } } \right)
		\left( \frac1{ v_w } \right)
		\left( \frac\beta H \right) 
		\left( \frac{ z_p }{10} \right)\,.
	\end{align}
This results from red shifting the frequency profile obtained from numerical simulations, which predict $z_p\sim10$~\cite{Hindmarsh:2017gnf}.

\begin{figure}[h!]
    \centering

    \begin{subfigure}[b]{0.475\textwidth}
         \centering
         \includegraphics[width=\textwidth, clip=true, trim=.0cm .1cm .1cm .0cm]
         {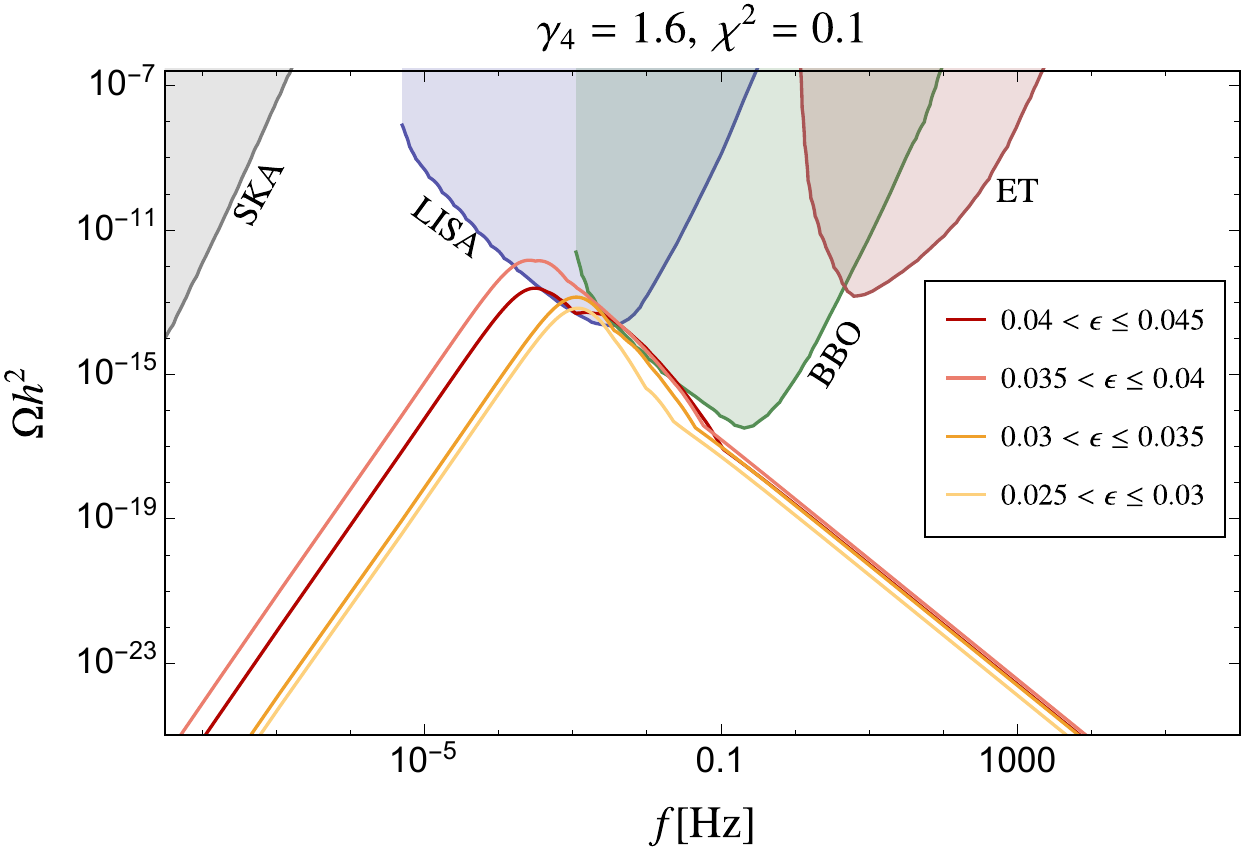}
         \caption{}
        \label{fig:GWepsilon}
     \end{subfigure}
     \hfill
     \begin{subfigure}[b]{0.475\textwidth}
         \centering
         \includegraphics[width=\textwidth, clip=true, trim=.0cm .1cm .1cm .0cm]
         {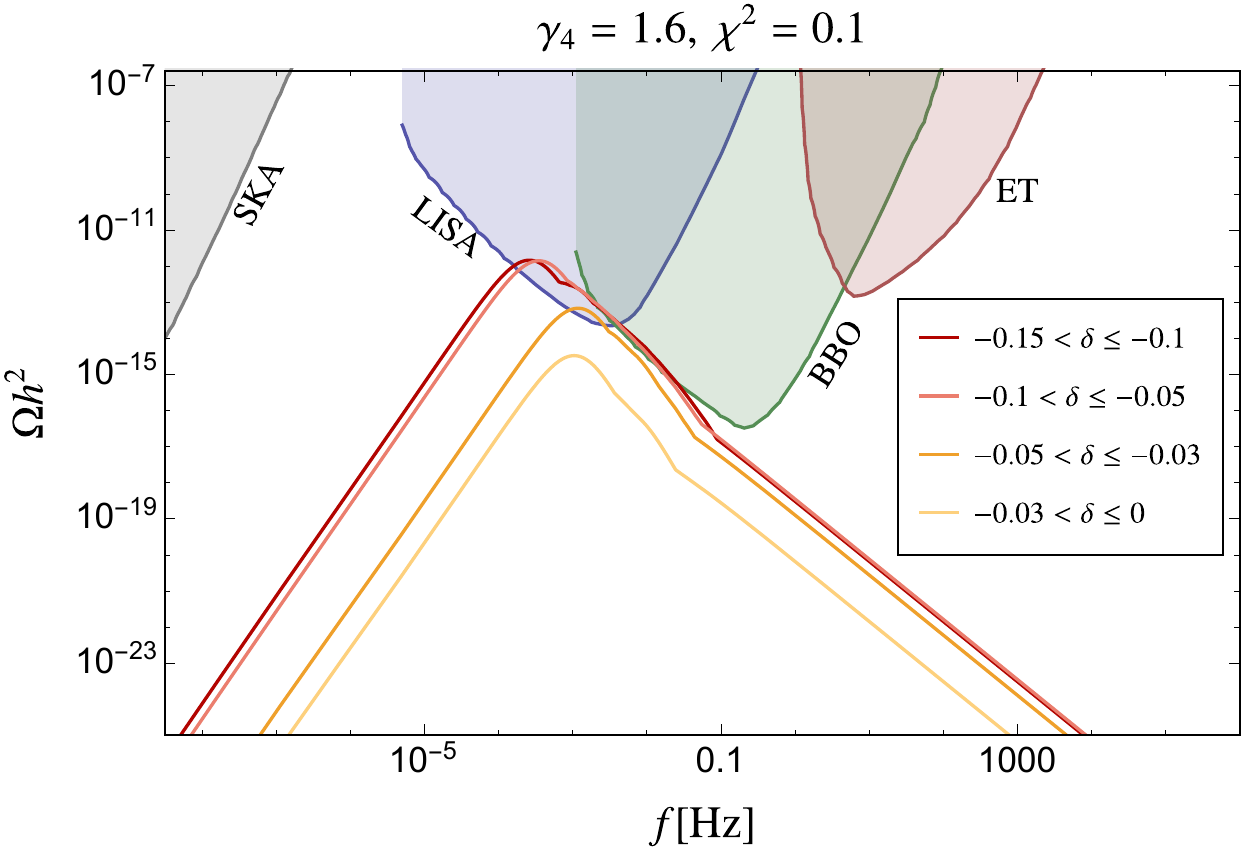}
         \caption{}
        \label{fig:GWdelta}
     \end{subfigure}
    \caption{The gravitational wave signal for $\gamma_4 = 1.6$, $\chi^2=0.1$. The GW signal curve is the maximal envelope of the power spectra obtained from varying $\epsilon$ and $\delta$. To give a sense of the $\epsilon$- and $\delta$-dependence, the parameter space is broken up into subsets, and a resulting maximal envelope power spectrum is drawn for each subset. In Fig.~\ref{fig:GWepsilon}, the $\epsilon$-dependence is emphasised by breaking the parameter space up into subsets of $\epsilon$ ranges as specified in the legend, while varying over all $\delta$ for each curve. In Fig.~\ref{fig:GWdelta}, the $\delta$-dependence is correspondingly emphasised, this time varying over all $\epsilon$ for each curve. The GW sensitivity curves are drawn for SKA~\cite{Weltman:2018zrl}, LISA~\cite{amaroseoane2017laser}, the Big Bang Observer (BBO)~\cite{Corbin_2006}, and the Einstein telescope (ET)~\cite{Sathyaprakash_2012}.}
    \label{fig:GW}
\end{figure}

The next important contribution to the overall signal is the turbulence term. In this case, the power of the gravitational wave signal at the source is now generated by turbulence 
in the fluid~\cite{Athron:2023xlk, RoperPol:2019wvy, Kolmogorov:91}. While our understanding of turbulence during cosmological phase transitions is still evolving~\cite{Caprini:2009yp, Kosowsky:2001xp, RoperPol:2018sap, Brandenburg:2021bvg,Athron:2023xlk}, the result used here is:
    \begin{align}
    h^2 \Omega_\text{tu} (f)
        &= 3.35 \times 10^{-4} \left( \frac{ H_* }\beta \right)
            \left( \frac{ \kappa_\text{tu} \alpha}{ 1 + \alpha }\right)^{3/2}
            \left( \frac{ 100 }{ g_* } \right)^{1/3}
            v_\text{w} S_\text{tu} (f) \,,
    \end{align}
where the efficiency factor $\kappa_\text{tu}$ parameterises how much of the kinetic energy is converted into turbulent motion. Here we use $\kappa_\text{tu} = 0.05 \,\kappa_\text{sw}$, as chosen in Ref.~\cite{Caprini:2015zlo}.
This approximation is based on simulated fluid motion~\cite{Hindmarsh:2015qta}, though this simulation does not last long enough to realistically capture turbulence effects. We emphasise that, as noted in the literature, this efficiency factor is not yet well-understood~\cite{Caprini:2019egz,Athron:2023xlk, Ellis:2019oqb, Alanne:2019bsm}.
Finally, the spectral shape of the turbulence signal is~\cite{Weir:2017wfa}:
    \begin{align}
    S_\text{tu} (f)
        &= \frac{ \left( f/f_\text{tu} \right)^3 }
            { \left[ 1 +  f/f_\text{tu} \right]^{11/3} \left( 1 + 8\pi f/h_N \right) }\,,
    \end{align}
  where
    \begin{align}
    f_\text{tu}
        &= 27 \mu\text{ Hz} \frac1{ v_\text{w}}
            \left( \frac\beta{ H_* } \right)
            \left( \frac{ T_N }{ 100 \text{ GeV } } \right)
            \left( \frac{ g_* }{ 100 } \right)^{1/6} \,,
       \\
    h_N &= 16.6\mu\text{ Hz} 
            \left( \frac{ T_N }{ 100 \text{ GeV } } \right)
            \left( \frac{ g_* }{ 100 } \right)^{1/6}\,.
    \end{align}

The total gravitational wave signal that results from both the sound wave and turbulence contributions is plotted in Fig.~\ref{fig:GW} for a benchmark choice of $\gamma_4$, $\chi$ in the nonlinear theory model. We also show the sensitivity curves for planned gravitational wave detectors. Given $\gamma_4$ and $\chi$, there is still a range of $\epsilon$ and $\delta$ that give bubble nucleation and generate a gravitational wave signal. This figure summarizes this parameter space by plotting the maximal envelope of the many power spectra that result from varying $\epsilon$ and $\delta$. While no single choice of $(\epsilon, \delta)$ will reproduce the whole curve, the maximal envelope overlapping with the GW detector sensitivity curves indicates at least one point in $(\epsilon, \delta)$ parameter space with a power spectrum that exceeds the sensitivity curve. In the remainder of this work, we estimate that a GW signal is within observable reach of a GW detector if there exist some frequencies for which the GW power spectrum exceeds the detector's sensitivity curve.

%%%%%%%%%%%%%%%%%%%%%%%%%%%%%%%%%%%%%%%%%%%%%%%%%%
\section{Complementarity  with LHC}\label{sec:LHC}
While measurements at the LHC have opened the door to the electroweak symmetry breaking mechanism and made formidable advances in its exploration, these measurements are also limited to only probe couplings of the electroweak sector around the vacuum. As this work has tried to underline, cosmology has the potential to reach where such scattering experiments cannot, to non-local effects in field space. Here we establish this complementarity quantitatively on a concrete case study by confronting our previous cosmological analysis with LHC bounds. % studying the phenomenology of a concrete case study, as in this letter. 

Local observables correspond to coefficients in a covariant expansion of fields of the action around the vacuum. These can be given in the scalar sector in terms of covariant derivatives of the cuvature tensor and scalar potential, reproducing here for simplicity the curvatures of Sec.~\ref{sec:Classical}:
\begin{align}\label{eq:curv}
v^2R_\varphi(0)&=1-c_\chi^4\gamma_a^2,
  & v^2R_h(0)&=-\gamma_a^2s_\chi^2c_\chi^2\,,
\end{align}
which are bounded by ATLAS
to a $95\%$ confidence level to be
\begin{align}\label{eq:LHCBounds}
    v^2R_\varphi(0)&=-0.080^{+0.12}_{-0.13},
  & v^2R_h(0)&=+0.080^{+1.0}_{-1.1} \,,
\end{align}
where these bounds are derived from \cite{ATLAS:2023qzf,ATLAS:2021vrm}.
 Also relevant will be the element of the third covariant derivative of the potential $(\mathcal \nabla^3V)_{hhh}=V'''(0)=3\lambda \gamma_4(1-\epsilon)v$, probed by measurements of triple Higgs coupling.
Bounds from ATLAS \cite{ATLAS:2021tyg} give the following constraints at $95\%$ confidence on the triple-Higgs coupling in the kappa framework (i.e. ratio of the triple-Higgs coupling to the SM expectation)
\begin{align}\label{eq:triplehbound}
    -1.0<\kappa_\lambda = \frac{V'''(0)}{V'''_{\text{SM}}(0)}=\gamma_4(1-\epsilon)<6.6 \,.
\end{align}

Here, we combine experimental input from the LHC with the cosmological analyses of Secs.~\ref{sec:Walls} and \ref{sec:PT} in order to present a final comprehensive phenomenological picture. We categorise these results into different different cosmological processes as follows:

\begin{itemize}
    \clearpage
    \item \textbf{Symmetry restoration.}
     The small $\epsilon,\delta$ limit allows for identification of $\chi>0.3$ as the region for high temperature symmetry restoration. This does not include the $\chi=0$ limit which returns SM-like couplings locally; the Standard Model itself presents high temperature symmetry restoration. The feature then arises in this non-linear theory that the non-local phenomenology of the SM is not recovered in the local SM coupling limit. A consequence of this is that the region for high temperature symmetry restoration in our non-linear theory lies a finite distance away in parameter space from the SM couplings limit and as such provides a target for collider experiments. This is illustrated in Fig.~\ref{FigRest} where said region for small $\delta,\epsilon$ is depicted and is partially ruled out by LHC bounds and would explored in full by future experiments such as the FCC.
     \begin{figure}
    \centering
        \includegraphics[width=.65\textwidth ]{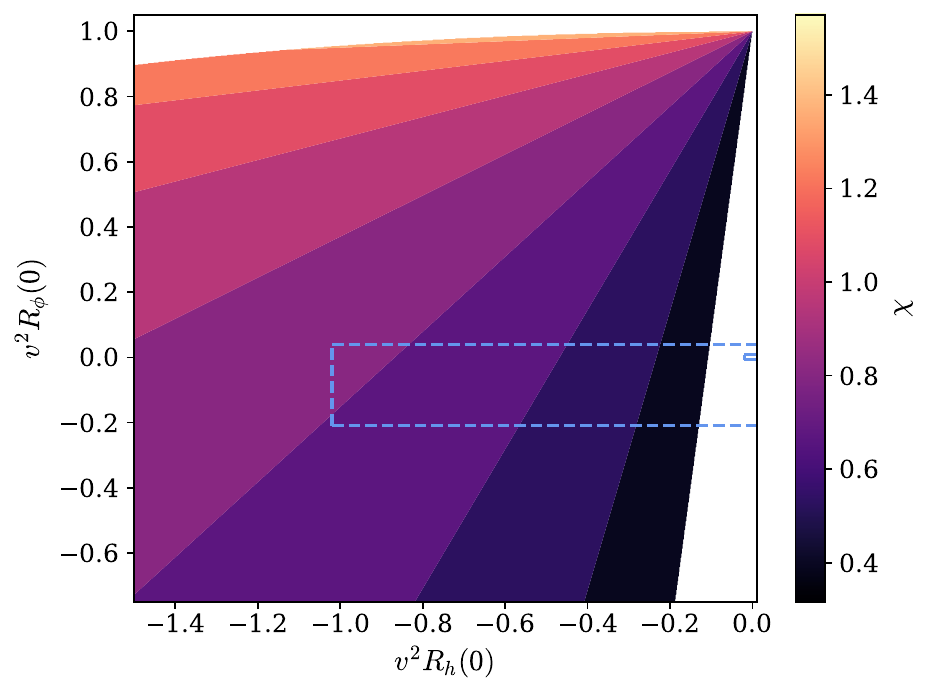}
        \caption{\label{FigRest} The curvature plane $v^2R_\varphi(0)$ and $v^2R_h(0)$ as defined in Eq.~\eqref{eq:curv}. In colour is the region for high-temperature symmetry restoration (see Sec.~\ref{sec:HighT}), i.e. $\pi/2>\chi>0.3$. We vary $\gamma_a^2$ independently to account for varying $\epsilon,\delta,\gamma_4$. We note for interest but do not show on the plot for clarity that increasing $\gamma_a^2$ decreases the curvatures radially originating from $(v^2R_h(0),v^2R_\varphi(0))=(0.0,1.0)$. The area outside the blue dashed box is excluded by LHC bounds from Eq.~\eqref{eq:LHCBounds}. The FCC would be expected to be sensitive, assuming SM-like couplings, up to the small, inner box \cite{deBlas:2019rxi,Bishara:2016kjn}.}
    \end{figure}

    \clearpage
    \item \textbf{Domain walls.}
    The wall formation scenario required $\epsilon,\delta<10^{-15}$, far smaller than collider limits can hope to compete with.
    LHC constraints therefore will have a relevant impact only in the remaining parameters of the theory, namely $\gamma_4, \chi$. In the case that $\chi>0.3$ the theory exhibits symmetry restoration and produces domain walls. As Fig.~\ref{fig:wps} shows, cosmological bounds alone (primarily in the form of wall annihilation before BBN) allow for potentially detectable GW signals at SKA. To produce a signal visible at SKA, one requires small $\gamma_4<0.1$ given the parametric dependence in Eq.~\eqref{WallsGW}. The triple gauge coupling constraint allows very small and even vanishing $\gamma_4$. However, the relevant $\epsilon,\delta\to0$ limit implies $\gamma_a=\gamma_4$, and curvature bounds of Eq.~\eqref{eq:LHCBounds} apply. These constrain $\gamma_4$ to a neighbourhood around $1/c_\chi^2$, so that there is a minimum attainable $\gamma_4$ well above 0.1. Plugging this value back into eq~\eqref{WallsGW} gives a maximum peak of the GW power spectrum resulting from domain walls: 
    \begin{align}
        \Omega_{\rm g.w.|peak}^{max}&=2.2\times10^{-17} \left(\frac{10^{-10}\textrm{ Hz}}{f_{peak}}\right)^4 & f_{\textrm{peak}}&>10^{-10}\textrm{ Hz}\label{MaxWalls} \,.
    \end{align}
    Figure \ref{fig:DWLHC}, shows the impact of LHC bounds and Fig. \ref{fig:DWSKA} shows the maximal prediction for the peak spectrum in this non-linear theory versus SKA sensitivity, see Eq.~\eqref{MaxWalls}. From this figure, it is clear that the parameter region which sources a domain wall GW signal that SKA is sensitive to is already ruled out by the LHC while the maximum attainable signal in spectrum lies five orders of magnitude below projected sensitivity.
\begin{figure}
    \centering

    \begin{subfigure}[b]{0.53\textwidth}
         \centering
         \includegraphics[width=\textwidth]{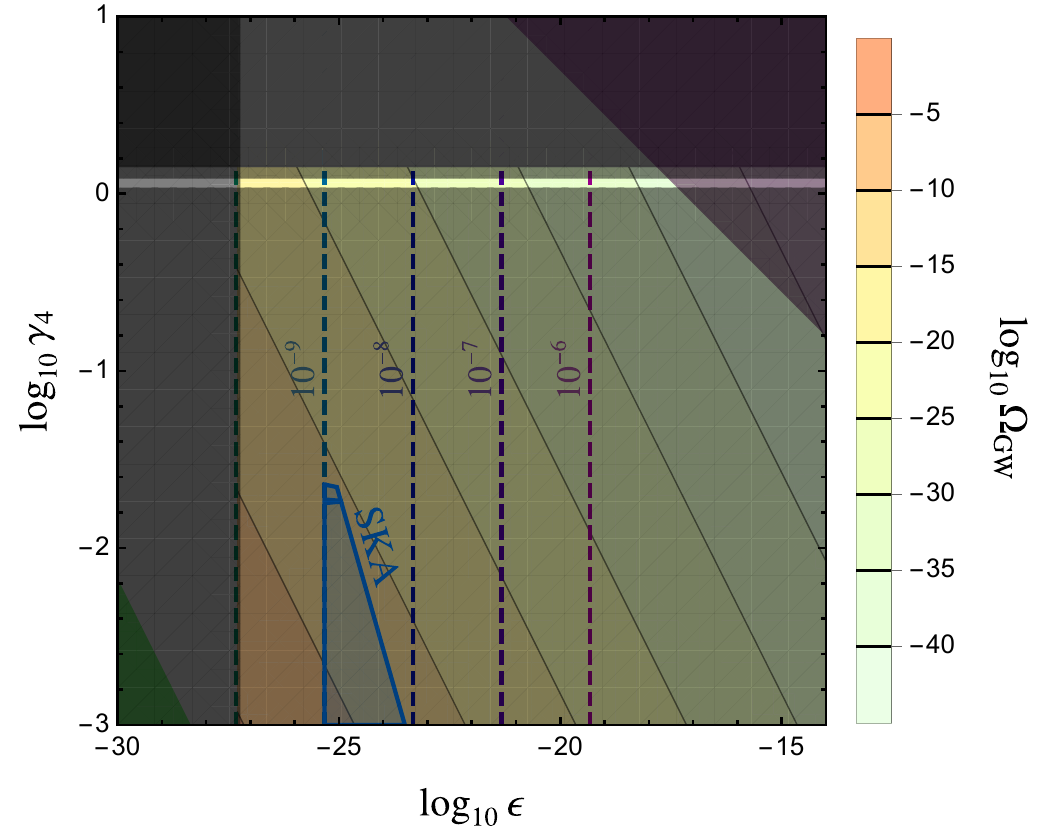}
         \caption{}
        \label{fig:DWLHC}
     \end{subfigure}
     \hfill
     \begin{subfigure}[b]{0.45\textwidth}
         \centering
         \includegraphics[width=\textwidth]{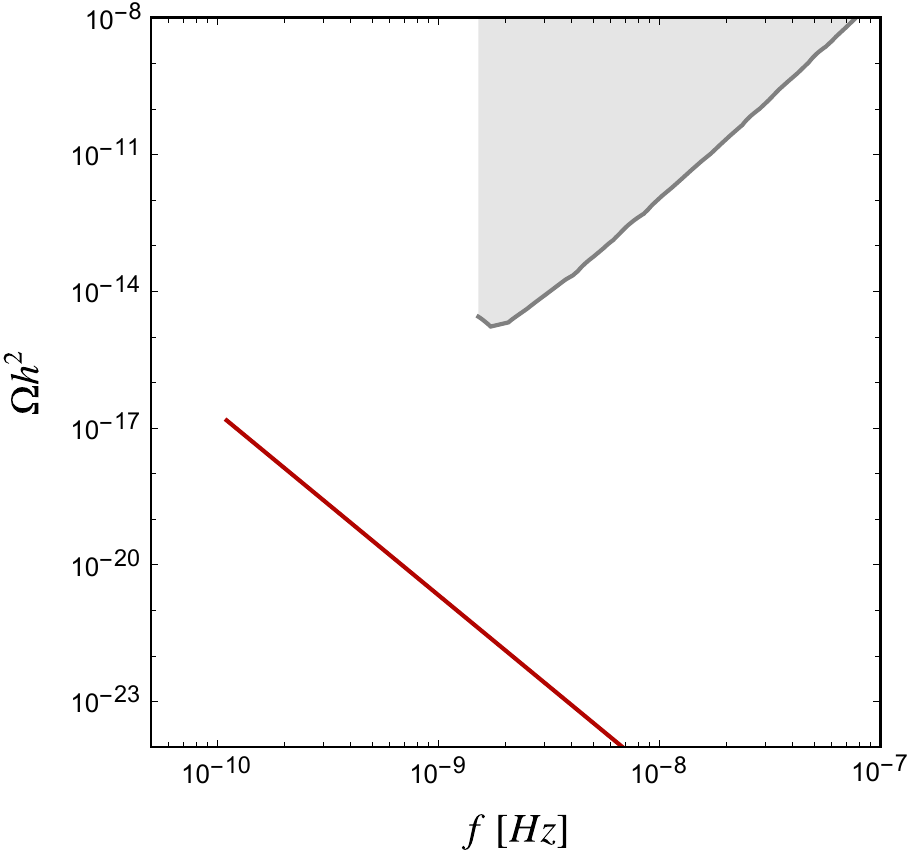}
         \caption{}
        \label{fig:DWSKA}
     \end{subfigure}
    \caption{(a) The ($\epsilon-\gamma_4$) plane delineating the wall formation region in parameter space as in Fig.~\ref{fig:wps} with the addition of LHC bounds, Eq.~\eqref{eq:LHCBounds}, in the vertical gray shaded region. (b) Curve for the value of the spectrum at the peak frequency for the allowed region of parameter space taking into account LHC bounds and the sensitivity of SKA~\cite{Weltman:2018zrl}.}
    
    \label{fig:WallsLHC}
\end{figure}

    \clearpage
    \item \textbf{First order phase transition.} 
For first order phase transitions, one requires couplings sizeably different from the SM, which would naively give collider constraints a more prominent role than in the walls case. The main LHC constraint arises again through the bound on $R_\varphi$, which constrains $c_\chi^2\gamma_a$ to lie close to 1.
On the other hand we require largish negative $\delta$ in this region. Recalling
\begin{align}
    \delta =\gamma_4^{-1}\gamma_\epsilon-\gamma_a^{-1}\label{eq:delforga}
\end{align}
together with $1<\gamma_\epsilon<\sqrt{2}$ for $\epsilon>0$ means $\delta<0$ would require $\gamma_4>\gamma_\epsilon \gamma_a\simeq \gamma_\epsilon/ c_\chi^2>1$. The effect of greater than one $\gamma_4$ is illustrated in Fig.~\ref{fig:changepar}; as $\gamma_4$ increases, the extrema of the $T=0$ potential is pushed closer together, which also facilitates nucleation. We plot $\gamma_4=1.6$ in Fig.~\ref{fig:bubble_summary} to provide an illustrative example of a region in parameter space allowed by current LHC bounds that also predicts a gravitational wave signal detectable at upcoming experiments LISA and BBO, for the region marked in orange and cyan respectively. To estimate the reach of upcoming experiments, we simply check that there exists some range of frequencies such that the GW power spectrum exceeds the detector's sensitivity curve. We leave a full signal-to-noise ratio analysis taking into account possible astrophysical GW backgrounds for later work. Note that, since $\gamma_4\neq1, \chi\neq0$, in the $(\epsilon, \delta)$ plane of Fig.~\ref{fig:bubble_summary} the LHC-allowed region does not go through the origin.
\begin{figure}[t]
    \centering

    \begin{subfigure}[b]{0.475\textwidth}
         \centering
         \includegraphics[width=\textwidth, clip=true, trim=.2cm .4cm .1cm .0cm]
         {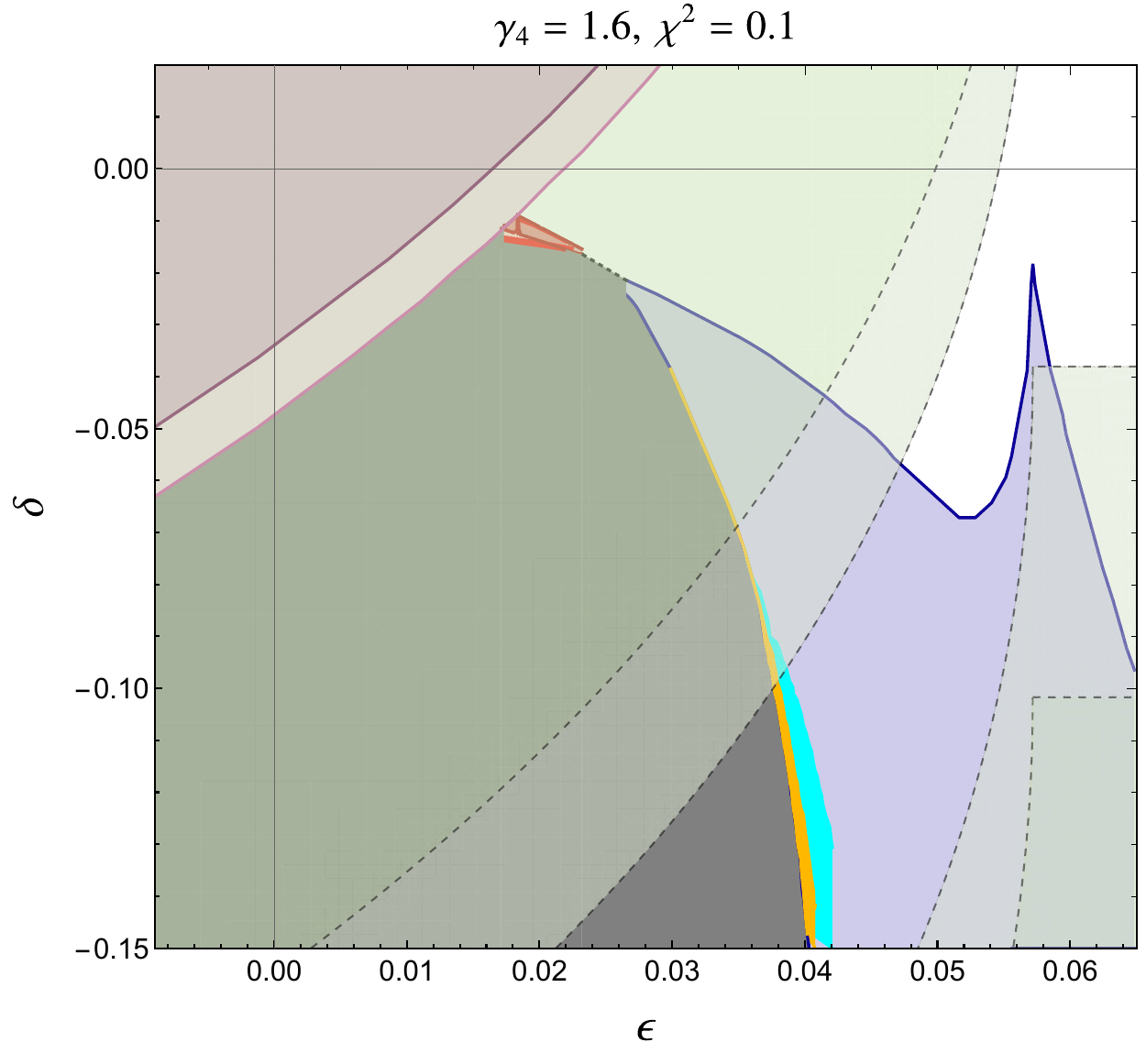}
         \caption{}
        \label{fig:bubble_summary_big}
     \end{subfigure}
     \hfill
     \begin{subfigure}[b]{0.475\textwidth}
         \centering
         \includegraphics[width=\textwidth, clip=true, trim=.2cm .4cm .1cm 0cm]
{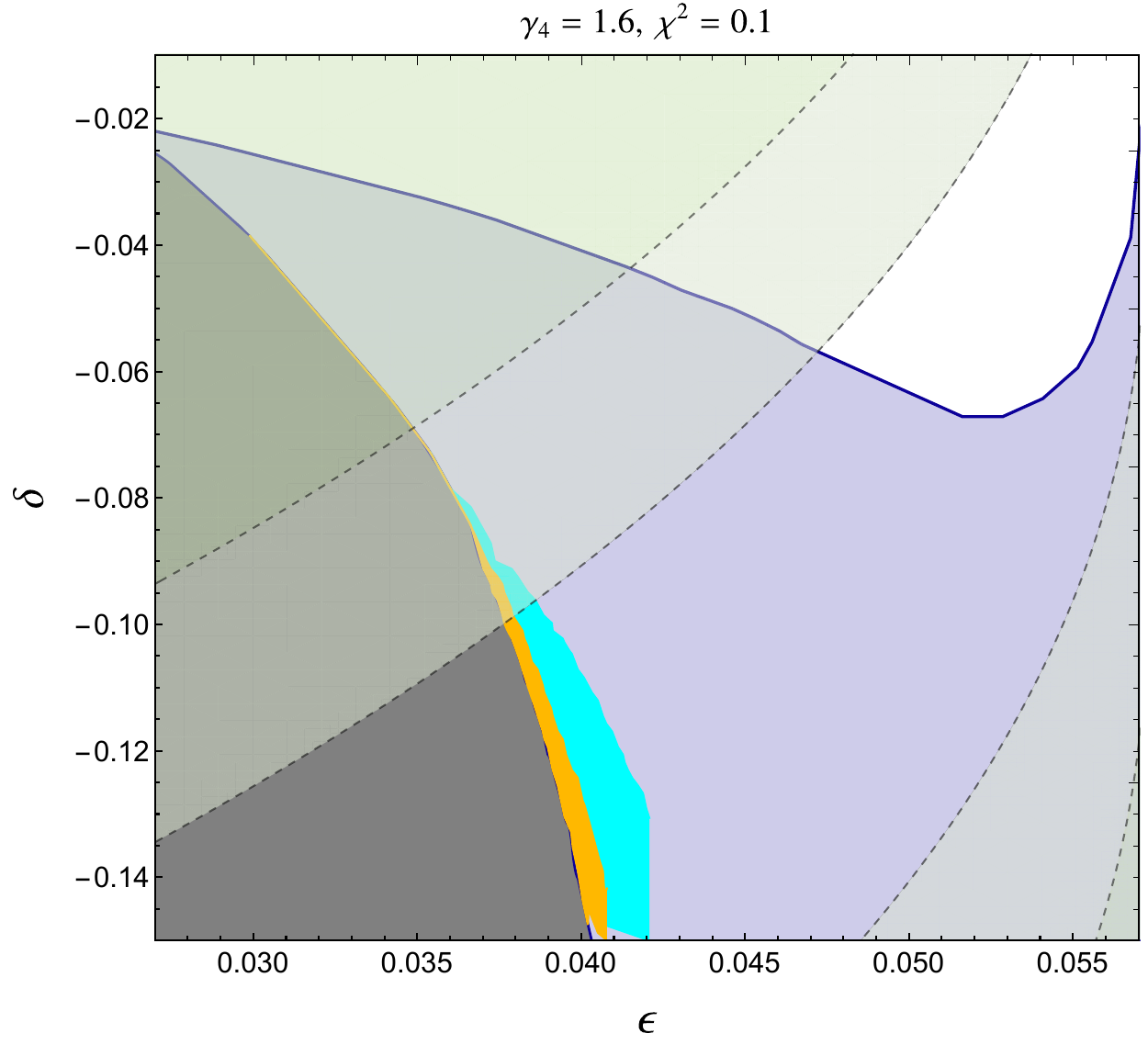}
         \caption{}
        \label{fig:bubble_summary_small}
     \end{subfigure}
    \caption{Summary of first order phase transition parameter space for $\gamma_4 = 1.6$, $\chi^2=0.1$. The region in blue shows the combinations of $(\epsilon, \delta)$ that we found to admit bubble nucleation while tunnelling from the false vacuum to the true vacuum today. The region in red (dark red) is excluded by the strong (weak) IR constraints. The pink (dark pink) regions are excluded by the boundedness from below (perturbativity) constraint. The light green (green) dashed lines show the LHC constraints on the curvature to the $1\sigma$ ($2\sigma)$ level. The gray region is unphysical, yielding the wrong vacuum today. The regions of first order phase transitions that give a GW signal detectable at LISA (BBO) are shown in orange (cyan). Fig.~\ref{fig:bubble_summary_small} magnifies the detectable region.}
    \label{fig:bubble_summary}
 
\end{figure}

\clearpage
\begin{figure}[h!]
\centering
\includegraphics[width=0.7\textwidth]{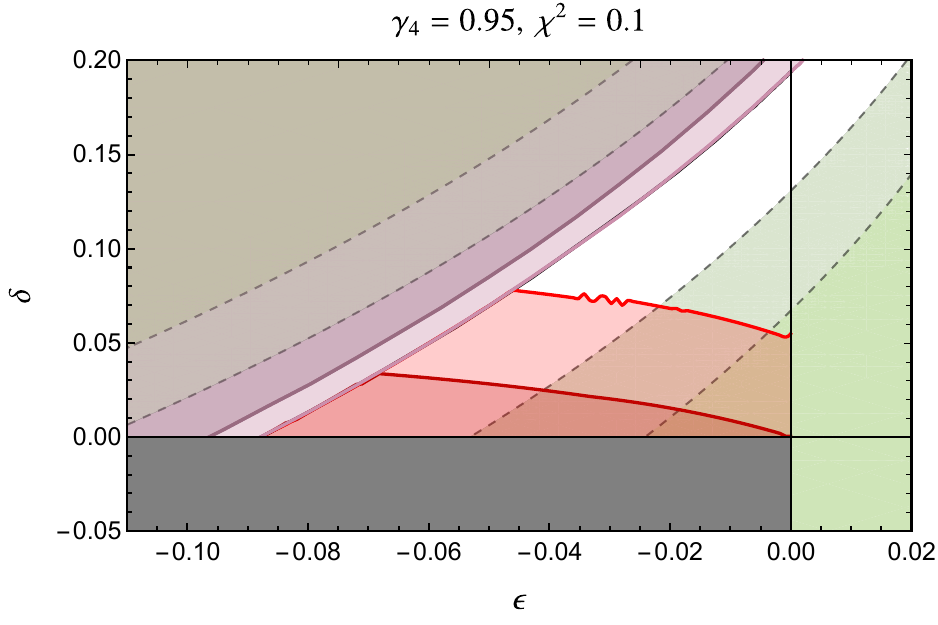}
    \includegraphics[width=0.32\textwidth]{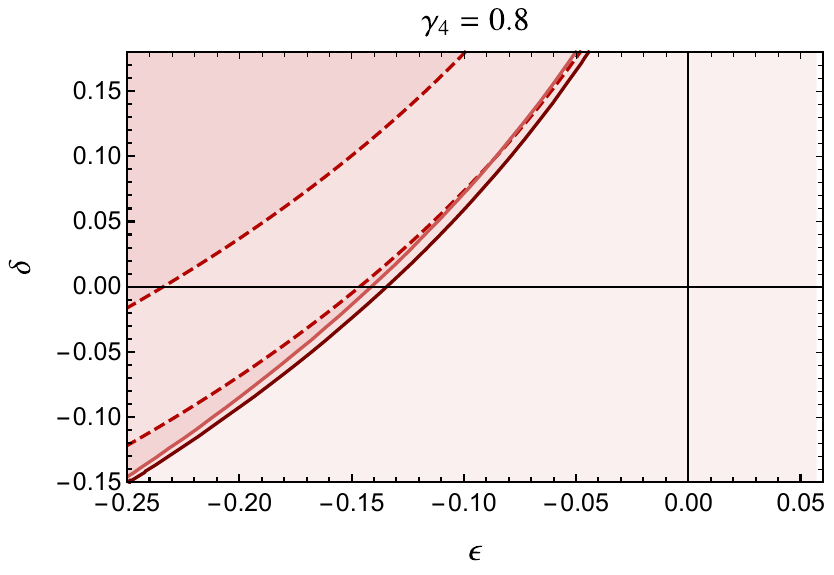}
    \includegraphics[width=0.32\textwidth]{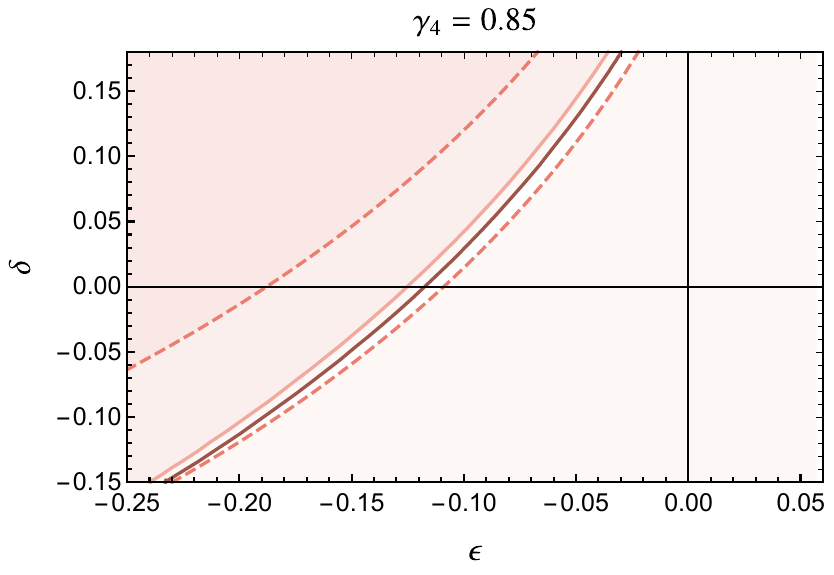}
    \includegraphics[width=0.32\textwidth]{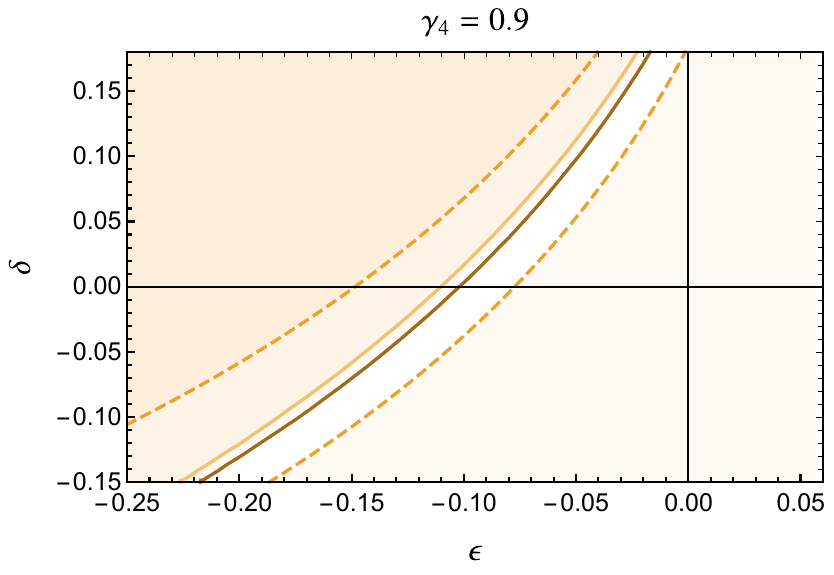}
    \includegraphics[width=0.32\textwidth]{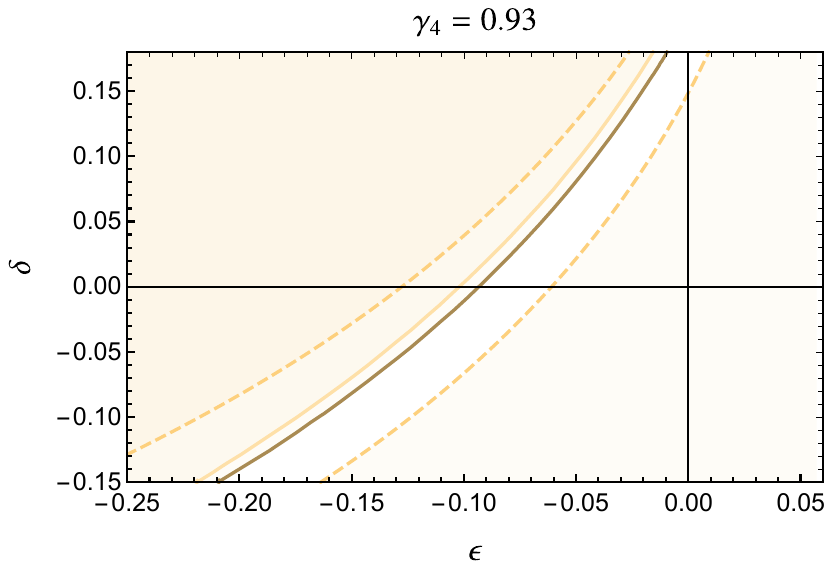}
    \includegraphics[width=0.32\textwidth]{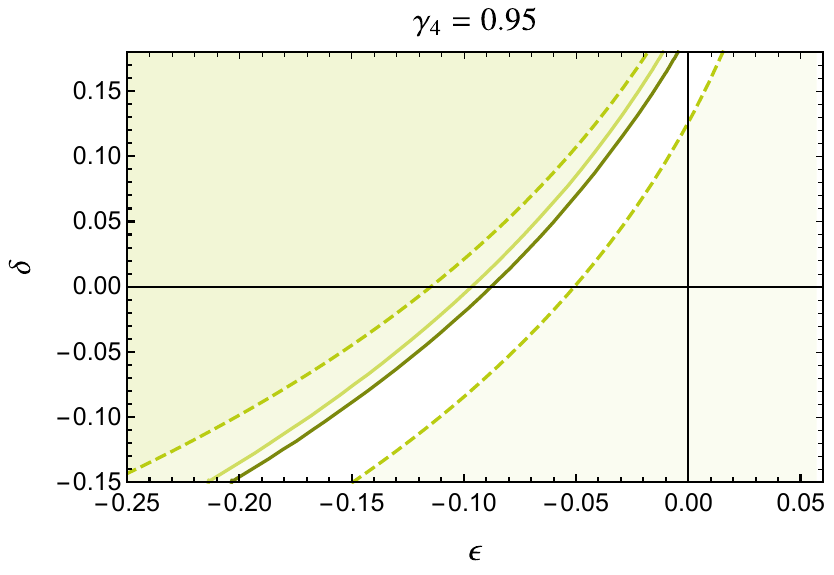}
    \includegraphics[width=0.32\textwidth]{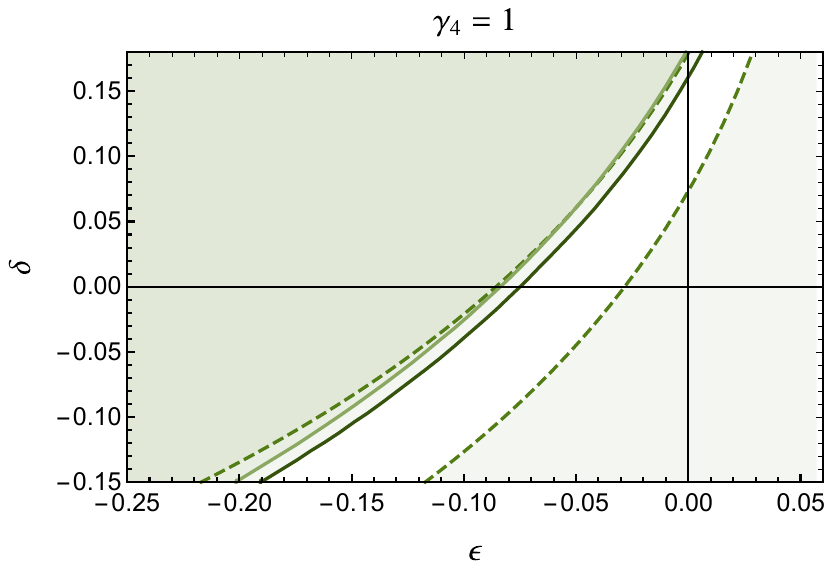}
    \caption{\label{DoomLHC} Top: Summary of the parameter space that leads to future vacuum decay (the doom scenario). Combination of Fig.~\ref{fig:Doom} with LHC bounds in light green ($1\sigma$) and darker green ($2\sigma$). Bottom: Series of plots with the combination of LHC exlusion regions (delineated with dashed lines) and perturbativity to illustrate that values of $\gamma_4<0.85$ are ruled out by a combination of LHC and perturbativity bounds, and therefore discarded in our analysis. \rm }
\end{figure}
\item \textbf{Doom.} 
Lastly the quadrant $\epsilon<0$, $\delta>0$ can lead to $h_0$ as the single high temperature vacuum while the true vacuum today, $h_-$, having appeared at a distance in field space large enough so that we are currently trapped in a false vacuum. The requirement of $\delta>0$ in conjunction with LHC bounds with the same reasoning as the one outlined around Eq.~\eqref{eq:delforga} but now in the opposite direction implies $\gamma_4<1$. This limit decreases the Higgs quartic coupling at tree level, increasing the relative contribution of loop effects so that a stronger perturbativity and boundedness from below constraint applies now. This is illustrated in the lower panel of Fig.~\ref{DoomLHC} where one can see that the combination of LHC and perturbativity bounds excludes $\gamma_4<0.8$. The reduced range for $\gamma_4$ still allows for a region, partially shown in Fig.~\ref{DoomLHC} upper panel, where the IR constraint, the LHC and perturbativity constraints are satisfied and we would be currently trapped in a false vacuum. This doom possibility has been found to lead to a lifetime for our universe in excess of $10^{2000}$ times the lifetime of the universe, see Fig.~\ref{fig:Saved}.

\end{itemize}

\section{Summary}

Today the question of whether electroweak symmetry is realised in Nature linearly or non-linearly is within experimental reach. The distinction is non-local in field space and thus calls for non-perturbative phenomena as the unambiguous probe. Such phenomena do arise in cosmology, and this paper has focused on studying the cosmological phenomenology of the non-linear realisation, termed here the \emph{non-linear theory space}, i.e. HEFT$\backslash$SMEFT which opens qualitatively new features with respect to SMEFT.

Our study has revealed non-decoupling deviations in the sphaleron energy, the possibility of domain wall formation, first order phase transitions at the electroweak scale and vacuum decay in the very distant future, as well as symmetry non-restoration at high temperature. These can be used to answer the EW realisation question because processes such as the formation of electroweak domain walls are, to the best of our current knowledge, exclusive to non-linear theories.

There is evidence to support the statement that non-linear theories are non-decoupling, meaning that no limit can be taken in such theories that returns the SM only. A non-decoupling feature, albeit localised a finite distance away from our vacuum, has been identified with the non-linear theory used here, which has wormhole-like space as the scalar manifold. This feature is to be found in the flat limit that corresponds to the SM locally; the closing of the throat of the wormhole produces a localised singularity in the effective potential that prevents symmetry restoration at high temperature and would likely lead to light new states. Conversely, away from this local SM limit, there is a region in parameter space that leads to symmetry restoration and presents local couplings around the vacuum which are a `finite distance' away from the SM ones. This minimum size of deviations provides a specific target, and we find that FCC would meet it and rule out the symmetry restoring region. Another non-decoupling effect has been found in the sphaleron energy, although this effect seems difficult to probe. The exploration of its implications for baryogenesis are hitherto unexplored.

The formation of domain walls requires minima that are approximately degenerate, but not exactly so. This requirement sets upper and lower cosmological bounds on $\epsilon$, which is related to the triple Higgs coupling. This window of $\epsilon$ is within a range too small to be probed directly by the LHC, although indirect LHC constraints on other parameters of the theory have implications on cosmology. Among them, there is an upper bound on the strength of gravitational waves from walls which lies a few orders of magnitude below the sensitivity of SKA. 

The case of a strong first order phase transition is realised in this theory in a way not possible in SMEFT; the extension of the Higgs field range with the wormhole topology considered here contains naturally a second minimum and a barrier such that no large potential corrections are needed.. This is qualitatively different from 1OPT's in the SMEFT, which require the addition of higher dimensional operators to generate a barrier. In order to achieve first order phase transitions in non-linear theories considered here, as opposed to walls, we require sizeable energy difference between minima and hence sizeable deviations from SM couplings. In fact, one has that the region in parameter space that leads to first order transitions is accessible experimentally via the LHC. However, the combination of LHC and current cosmological data does only probe this region partially. It is a promising possibility in fact that non-linear theories could have produced gravitational waves detectable at LISA {\it and}  will give rise to new signals at the HL-LHC.

Lastly, a scenario with vacuum decay in the future is difficult to probe with cosmological observation; the impending annihilation event has been found to be at least $10^{2000}$ years in the future. The LHC however can probe the possibility of vacuum decay through the determination of Higgs couplings.

This paper has explored a small fraction of the possible cosmological phenomenology of HEFT$\backslash$SMEFT. It is a fraction which nonetheless included new phenomena not possible in SMEFT and provides the elementary results to attempt a more comprehensive survey. Such exploration, to be complete, should include LHC and cosmological data, but also theory considerations such as perturbativity and the range of validity of an EFT at finite temperature and its relation to curvature. Indeed non-linear theories are constrained from all directions and hence, once these lines are drawn, they will
provide specific targets for experiments both on earth and in space, to answer the question of what type of electroweak symmetry realisation is present in Nature.

%Cosmology mate, it's reet.

\appendix
\section{UV model for non-linear type A theory.}

In this appendix we present a model that presents the geometry of a type A non-linear theory, i.e. a manifold which does not contain a fixed point, not even a singular one. This model is not meant to be phenomenologically viable or representative. In particular it is built out of linear representations and integration of heavy modes which leaves out the tantalizing but hitherto unrealised possibility of starting from non linear representations.

The model is built with an $SU(2)$ doublet $\phi$ and a singlet $h$, with vacuum expectation values given by $\left\langle|\phi|^2\right\rangle = v^2/2$ and $\left\langle h\right\rangle = 0$.
The Lagrangian is
\begin{equation}
  \mathcal{L}_{\text{UV}} =
  \frac{1}{2}(\partial_\mu h)^2
  + |\partial_\mu \phi|^2
  - V_h(h)
  - \frac{\lambda_\phi}{2} \left(|\phi|^2 - \frac{v^2}{2}\right)^2
  + v^2 G(h) |\phi|^2\,,
\end{equation}
with arbitrary functions $V_h(h)$ and $G(h)$ of the singlet $h$, with the minimum for $V_h(h)$ at $h=0$ and $G(0)=0$, without loss of generality.
We employ the usual parametrization of the doublet
\begin{equation}
  \phi = \frac{1}{\sqrt{2}} \; U \begin{pmatrix} 0 \\ v + H \end{pmatrix}
\end{equation}
with a radial coordinate $H$ such that $\left\langle H\right\rangle = 0$. The Lagrangian becomes
\begin{align}
  \mathcal{L}_{\text{UV}}
  &=
    \frac{1}{2}(\partial_\mu h)^2
    + \frac{1}{2} (\partial_\mu H)^2
    + \frac{1}{2} (v + H)^2 \, 
    \operatorname{Tr}\left[\partial^\mu U^\dagger \partial^\mu U^\dagger\right]
  \\
  &\phantom{=}
    - V_h(h)
    - V_H(H)
    + \frac{v^2}{2} G(h) (v + H)^2.
\end{align}
The $H$ scalar gets a mass $m_H^2 = \lambda_\phi v^2$.
Assuming a perturbative $\lambda_\phi \lesssim (4 \pi)^2$, we have $m_H^2 \lesssim (4\pi)^2 v^2$.
Here, we assume that $m_H^2 \simeq (4\pi)^2 v^2 \gg v^2$ and integrate out $H$ at tree level.
This can be done by plugging the solution to the equation of motion for $H$,
\begin{equation}
  H = \frac{1}{m_H^2} \Big\{v \operatorname{Tr}\left[\partial^\mu U^\dagger \partial^\mu U^\dagger\right] + v^3 G(h)\Big\}
  + O\left(\frac{1}{m_H^4}\right),
\end{equation}
into the UV Lagrangian, which gives:
\begin{align}
  \mathcal{L}_{\text{eff}}
  &=
    \frac{1}{2} (\partial_\mu h)^2
    + \frac{v^2}{2} \operatorname{Tr}\left[\partial^\mu U^\dagger \partial^\mu U^\dagger\right]
    - V_h(h)
    + \frac{1}{2 m_H^2} \Big\{
    v \operatorname{Tr}\left[\partial^\mu U^\dagger \partial^\mu U^\dagger\right]
    + v^3 G(h)
    \Big\}^2,
\end{align}
neglecting $O(1/m_H^4)$ terms.
In particular, we have
\begin{equation}
  F(h)^2 = 1 + \frac{v^2}{m_H^2} G(h),
  \qquad
  V(h) = V_h(h)
  - \frac{v^6}{2 m_H^2} G(h)^2.
\end{equation}
This means that non-linear theories with any $F(h)$ can be achieved with this model.
Both type A and B theories can be obtained with an appropriate choice of $G(h)$. In particular, a renormalizable UV completion with
\begin{align}
    G(h) &= \frac{c_\chi^2 m_H^2}{v^2} \left(
      2 \frac{h}{v_\star} + \frac{h^2}{v_\star^2}
    \right),
    \\
    V_h(h) &= V(h) + \frac{v^6}{m_H^2} G(h)^2
    \nonumber \\
    &= \left(\frac{m_h^2}{2} + \frac{2c_\chi^4v^6}{m_H^2 v_\star^2} \right)h^2
    +\left(\frac{m_h\sqrt{\lambda}}{2}\gamma_4(1-\epsilon) + \frac{2c_\chi^4v^6}{m_H^2 v_\star^3} \right)h^3
    +\left(\frac{\lambda}{8}\gamma_4^2+ \frac{2c_\chi^4v^6}{2m_H^2v_\star^4} \right)h^4,
\end{align}
gives the $F(h)$ and $V(h)$ functions we have used in this work, although additional terms of the form
\begin{equation}
    \mathcal{L}_{\text{eff}}
    \supset
    \frac{v^2}{2 m_H^2}
    \left\{\operatorname{Tr}\left[\partial^\mu U^\dagger \partial^\mu U^\dagger\right]\right\}^2
    + O\left(\frac{1}{m_H^4}\right),
\end{equation}
which we have not considered, are present in the Lagrangian. Corrections to this Lagrangian of order $m_H^{-4}$ can be computed systematically with a functional approach as outlined in \cite{Cohen:2020xca}.

\newpage
%%%%%%%%%%%%%%%%%%%%%%%%%%%%%%%%%%%%%%%%%%%%%%%%%%%%

%%%%%%%%%%%%%%%%%%%%%%%%%%%%%%%%%%%%%%%%%%%%%%%%%%%%%%%%%
\appendix
%%%%%%%%%%%%%%%%%%%%%%%%%%%%%%%%%%%%%%%%%%%%%%%%%%%

\acknowledgments

The authors would like to thank Djuna Croon, Philipp Schicho and Dave Sutherland for helpful comments. R. A., R. H. and M. W. are supported by the STFC under Grant No. ST/T001011/1. J.C.C. is supported by grant RYC2021-030842-I funded by\newline MCIN/AEI/10.13039/501100011033 and by the European Union NextGenerationEU/PRTR, and grant PID2022-139466NB-C22 funded by MCIN/AEI/10.13039/501100011033 and by ERDF A way of making Europe.

% The bibliography will probably be heavily edited during typesetting.
% We'll parse it and, using the arxiv number or the journal data, will
% query inspire, trying to verify the data (this will probalby spot
% eventual typos) and retrive the document DOI and eventual errata.
% We however suggest to always provide author, title and journal data:
% in short all the informations that clearly identify a document.

\normalem
\bibliography{COS_HEFt.bib}
\bibliographystyle{JHEP.bst}

\end{document}